\documentstyle[12pt]{article}
\newcommand{\nc}{\newcommand}
\nc{\rnc}{\renewcommand}
\rnc{\title}[1]{\Large\bf\mbox{}\\#1\bigskip\medskip\\}
\rnc{\author}[1]{\large #1\\ \smallskip}
\nc{\address}[1]{{\narrower\normalsize\it #1\\}\bigskip}
\nc{\e}[1]{{\em #1\/}}
\nc{\comment}[1]{}
\rnc{\thesection}{\arabic{section}\,.}
\rnc{\thesubsection}{\arabic{section}.\arabic{subsection}}
\rnc{\theequation}{\arabic{section}.\arabic{equation}}
\nc{\sect}[1]{\section{#1}\setcounter{equation}{0}}
\nc{\sub}[1]{\subsection{#1}}
\nc{\subsub}[1]{\subsubsection{#1}}
\nc{\beq}{\begin{equation}}
\nc{\beqa}{\begin{eqnarray}}
\nc{\eql}[1]{\label{Eqn#1}}
\nc{\bleq}[1]{\beq\eql{#1}}
\nc{\eeq}{\end{equation}}
\nc{\eeqa}{\end{eqnarray}}
\nc{\noeqno}{\nonumber\\}
\nc{\eqref}[1]{(\ref{Eqn#1})}
\nc{\sm}[1]{{\scriptstyle #1}}
\nc{\ssm}[1]{{\scriptscriptstyle #1}}
\nc{\D}{\mbox{\boldmath$D$}}
\nc{\DD}{{\cal D}}
\rnc{\L}{{\cal L}}
\nc{\R}{{\cal R}}
\nc{\W}[5]{W\!\left(\,\begin{array}{@{}cc|}#4&#3\\#1&#2\end{array}
\;#5\right)}
\nc{\Wf}[6]{W^{#1}\!\!\left(\,\begin{array}{@{}cc|}#5&#4\\#2&#3
\end{array}\;#6\right)}
\nc{\BL}[4]{B_{\mbox{\tiny L}}\!\!\left(\left.\!\!
\begin{array}{c}#3\\#1\end{array}\,#2\:\right|#4\right)}
\nc{\BR}[4]{B_{\mbox{\tiny R}}\!\!\left(\left.\!#2\,
\begin{array}{c}#3\\#1\end{array}\!\right|#4\right)}
\nc{\BtildeL}[4]{\tilde{B}_{\mbox{\tiny L}}\!\!\left(\left.\!\!
\begin{array}{c}#3\\#1\end{array}\,#2\:\right|#4\right)}
\nc{\BtildeR}[4]{\tilde{B}_{\mbox{\tiny R}}\!\!\left(
\left.\!#2\,\begin{array}{c}#3\\#1\end{array}\!\right|#4\right)}
\nc{\BLf}[5]{B^{#1}_{\mbox{\tiny L}}\!\!\left(\left.\!\!
\begin{array}{c}#4\\#2\end{array}\,#3\:\right|#5\right)}
\nc{\BRf}[5]{B^{#1}_{\mbox{\tiny R}}\!\!\left(\left.\!#3\,
\begin{array}{c}#4\\#2\end{array}\!\right|#5\right)}
\nc{\xiL}{\xi_{\mbox{\tiny L}}\!}
\nc{\xiR}{\xi_{\mbox{\tiny R}}\!}
\nc{\aL}{a_{\mbox{\tiny L}}}
\nc{\aR}{a_{\mbox{\tiny R}}}
\rnc{\l}{\lambda}
\nc{\m}{\mu}
\nc{\lm}{{\textstyle\frac{\l-\m}{2}}}
\rnc{\t}{\theta}
\nc{\T}[3]{\t^{#1}_{\!#2}\!(#3)}
\nc{\N}{{\scriptscriptstyle N}}
\rnc{\vec}[1]{\mbox{\boldmath$#1$}}
\nc{\pos}[2]{\makebox(0,0)[#1]{$#2$}}
\nc{\spos}[2]{\makebox(0,0)[#1]{$\sm{#2}$}}
\nc{\text}[6]{\begin{picture}(#1,#2)
\put(#3,#4){\pos{#5}{\displaystyle#6}}\end{picture}}
\nc{\dl}[3]{\put(#1,#2){\makebox(#3,0){\dotfill}}}
\rnc{\d}[2]{\put(#1,#2){\spos{}{\bullet}}}
\nc{\dd}[3]{\multiput(#1,#2)(0,1){#3}{\spos{}{\bullet}}}
\rnc{\u}{\begin{picture}(0,0)
\put(-0.23,0){\spos{r}{u}}\end{picture}}
\nc{\uq}{\begin{picture}(0,0)
\put(-0.8,0){\spos{l}{\;u\!+\!q\l}}\end{picture}}
\nc{\uql}{\begin{picture}(0,0)
\put(-0.8,0){\spos{l}{u\!+\!(\!q\!-\!1\!)\l}}\end{picture}}
\nc{\uqll}{\begin{picture}(0,0)
\put(-0.8,0){\spos{l}{u\!+\!(\!q\!-\!2\!)\l}}\end{picture}}
\nc{\uuq}{\begin{picture}(0,0)
\put(-0.06,0){\spos{}{-\!2u\!-\!q\l\!+\!\m}}\end{picture}}
\nc{\uuql}{\begin{picture}(0,0)
\put(-0.1,0.3){\spos{}{-\!2u}}
\put(-0.09,-0.1){\spos{}{-\!(\!q\!-\!1\!)\l}}
\put(-0.05,-0.5){\spos{}{+\!\m}}\end{picture}}
\nc{\uuqll}{\begin{picture}(0,0)
\put(-0.1,0.3){\spos{}{-\!2u}}
\put(-0.09,-0.1){\spos{}{-\!(\!q\!-\!2\!)\l}}
\put(-0.05,-0.5){\spos{}{+\!\m}}\end{picture}}
\nc{\uuqql}{\begin{picture}(0,0)
\put(-0.1,0.3){\spos{}{-\!2u}}
\put(-0.06,-0.1){\spos{}{-\!(\!2q\!-\!1\!)\l}}
\put(-0.05,-0.5){\spos{}{+\!\m}}\end{picture}}
\nc{\uuqqll}{\begin{picture}(0,0)\put(-0.1,0.3){\spos{}{-\!2u}}
\put(-0.06,-0.1){\spos{}{-\!(\!2q\!-\!2\!)\l}}
\put(-0.05,-0.5){\spos{}{+\!\m}}\end{picture}}
\nc{\uuqqlll}{\begin{picture}(0,0)
\put(-0.1,0.3){\spos{}{-\!2u}}
\put(-0.06,-0.1){\spos{}{-\!(\!2q\!-\!3\!)\l}}
\put(-0.05,-0.5){\spos{}{+\!\m}}\end{picture}}
\nc{\um}{\begin{picture}(0,0)
\put(0.8,0){\spos{r}{-\!u\!+\!\m\;}}\end{picture}}
\nc{\uqm}{\begin{picture}(0,0)
\put(0.8,0){\spos{r}{-\!u\!-\!q\l\!+\!\m}}\end{picture}}
\nc{\uqlm}{\begin{picture}(0,0)
\put(0.8,0){\spos{r}{-\!u\!-\!(\!q\!-\!1\!)\l\!+\!\m}}
\end{picture}}
\nc{\uqllm}{\begin{picture}(0,0)
\put(0.8,0){\spos{r}{-\!u\!-\!(\!q\!-\!2\!)\l\!+\!\m}}
\end{picture}}
\nc{\luuq}{\begin{picture}(0,0)
\put(0,0){\spos{}{2u\!+\!q\l\!-\!\m}}\end{picture}}
\nc{\luuql}{\begin{picture}(0,0)
\put(0,0.3){\spos{}{2u}}
\put(-0.09,-0.1){\spos{}{+\!(\!q\!-\!1\!)\l}}
\put(-0.05,-0.5){\spos{}{-\!\m}}\end{picture}}
\nc{\luuqll}{\begin{picture}(0,0)
\put(0,0.3){\spos{}{2u}}
\put(-0.09,-0.1){\spos{}{+\!(\!q\!-\!2\!)\l}}
\put(-0.05,-0.5){\spos{}{-\!\m}}\end{picture}}
\nc{\luuqql}{\begin{picture}(0,0)
\put(0,0.3){\spos{}{2u}}
\put(-0.06,-0.1){\spos{}{+\!(\!2q\!-\!1\!)\l}}
\put(-0.05,-0.5){\spos{}{-\!\m}}\end{picture}}
\nc{\luuqqll}{\begin{picture}(0,0)
\put(0,0.3){\spos{}{2u}}
\put(-0.06,-0.1){\spos{}{+\!(\!2q\!-\!2\!)\l}}
\put(-0.05,-0.5){\spos{}{-\!\m}}\end{picture}}
\nc{\luuqqlll}{\begin{picture}(0,0)
\put(0,0.3){\spos{}{2u}}
\put(-0.06,-0.1){\spos{}{+\!(\!2q\!-\!3\!)\l}}
\put(-0.05,-0.5){\spos{}{-\!\m}}
\end{picture}}
\nc{\fu}{\begin{picture}(0,0)\put(0.75,0.5){\spos{}{u}}
\end{picture}}
\nc{\qll}{\begin{picture}(0,0)
\put(0.75,0.5){\spos{}{u\!+\!(\!q\!-\!2\!)\l}}
\end{picture}}
\nc{\ql}{\begin{picture}(0,0)
\put(0.75,0.5){\spos{}{u\!+\!(\!q\!-\!1\!)\l}}
\end{picture}}
\nc{\q}{\begin{picture}(0,0)\put(0.75,0.5){\spos{}{u\!+\!q\l}}
\end{picture}}
\nc{\pl}{\begin{picture}(0,0)
\put(0.75,0.5){\spos{}{u\!-\!(\!p\!-\!1\!)\l}}
\end{picture}}
\nc{\qpl}{\begin{picture}(0,0)\put(0.75,0.7){\spos{}{u\!+}}
\put(0.75,0.3){\spos{}{(\!q\!-\!p\!-\!1\!)\l}}\end{picture}}
\nc{\qp}{\begin{picture}(0,0)
\put(0.75,0.5){\spos{}{u\!+\!(\!q\!-\!p\!)\l}}\end{picture}}
\nc{\uqpl}{\begin{picture}(0,0)\put(0.75,0.7){\spos{}{u\!+}}
\put(0.75,0.3){\spos{}{(\!q\!-\!p\!+\!1\!)\l}}\end{picture}}
\nc{\fum}{\begin{picture}(0,0)
\put(0.75,0.5){\spos{}{-\!u\!+\!\m}}\end{picture}}
\nc{\umqll}{\begin{picture}(0,0)
\put(0.75,0.7){\spos{}{-\!u\!+\!\m\!-}}
\put(0.75,0.3){\spos{}{(\!q\!-\!2\!)\l}}\end{picture}}
\nc{\umql}{\begin{picture}(0,0)
\put(0.75,0.7){\spos{}{-\!u\!+\!\m\!-}}
\put(0.75,0.3){\spos{}{(\!q\!-\!1\!)\l}}\end{picture}}
\nc{\umq}{\begin{picture}(0,0)
\put(0.75,0.5){\spos{}{-\!u\!+\!\m\!-\!q\l}}\end{picture}}
\nc{\umpl}{\begin{picture}(0,0)
\put(0.75,0.7){\spos{}{-\!u\!+\!\m\!-}}
\put(0.75,0.3){\spos{}{(\!p\!-\!1\!)\l}}\end{picture}}
\nc{\umqplll}{\begin{picture}(0,0)
\put(0.75,0.7){\spos{}{-\!u\!+\!\m\!-}}
\put(0.75,0.3){\spos{}{(\!q\!+\!p\!-\!3\!)\l}}\end{picture}}
\nc{\umqpll}{\begin{picture}(0,0)
\put(0.75,0.7){\spos{}{-\!u\!+\!\m\!-}}
\put(0.75,0.3){\spos{}{(\!q\!+\!p\!-\!2\!)\l}}\end{picture}}
\nc{\umqpl}{\begin{picture}(0,0)
\put(0.75,0.7){\spos{}{-\!u\!+\!\m\!-}}
\put(0.75,0.3){\spos{}{(\!q\!+\!p\!-\!1\!)\l}}\end{picture}}
\rnc{\v}{\begin{picture}(0,0)\put(-0.23,0){\spos{r}{v}}
\end{picture}}
\nc{\vrl}{\begin{picture}(0,0)
\put(-0.8,0){\spos{l}{v\!+\!(\!r\!-\!1\!)\l}}
\end{picture}}
\nc{\vvrl}{\begin{picture}(0,0)\put(-0.1,0.3){\spos{}{-\!2v}}
\put(-0.09,-0.1){\spos{}{-\!(\!r\!-\!1\!)\l}}
\put(-0.05,-0.5){\spos{}{+\!\m}}
\end{picture}}
\nc{\uv}{\begin{picture}(0,0)\put(0,0){\spos{}{u\!-\!v}}
\end{picture}}
\nc{\uvq}{\begin{picture}(0,0)\put(0,0.25){\spos{}{u\!-\!v}}
\put(-0.05,-0.15){\spos{}{+\!(\!q\!-\!1\!)\l}}\end{picture}}
\nc{\uvr}{\begin{picture}(0,0)\put(0,0.25){\spos{}{u\!-\!v}}
\put(-0.05,-0.15){\spos{}{-\!(\!r\!-\!1\!)\l}}\end{picture}}
\nc{\uvqr}{\begin{picture}(0,0)\put(0,0.25){\spos{}{u\!-\!v}}
\put(-0.05,-0.15){\spos{}{+\!(\!q\!-\!r\!)\l}}\end{picture}}
\nc{\uvll}{\begin{picture}(0,0)\put(0,0.25){\spos{}{u\!-\!v}}
\put(-0.05,-0.15){\spos{}{+\!(\!q\!-\!2\!)\l}}\end{picture}}
\nc{\uvm}{\begin{picture}(0,0)
\put(-0.05,0){\spos{}{-\!u\!-\!v\!+\!\m}}
\end{picture}}
\nc{\uvqm}{\begin{picture}(0,0)
\put(-0.1,0.3){\spos{}{-\!u\!-\!v}}
\put(-0.06,-0.1){\spos{}{-\!(\!q\!-\!1\!)\l}}
\put(-0.05,-0.5){\spos{}{+\!\m}}
\end{picture}}
\nc{\uvrm}{\begin{picture}(0,0)
\put(-0.1,0.3){\spos{}{-\!u\!-\!v}}
\put(-0.06,-0.1){\spos{}{-\!(\!r\!-\!1\!)\l}}
\put(-0.05,-0.5){\spos{}{+\!\m}}
\end{picture}}
\nc{\uvqrm}{\begin{picture}(0,0)
\put(-0.1,0.4){\spos{}{-\!u\!-\!v}}
\put(-0.06,0){\spos{}{-\!(\!q\!+\!r\!-\!2\!)\l}}
\put(-0.05,-0.4){\spos{}{+\!\m}}
\end{picture}}
\nc{\uvllm}{\begin{picture}(0,0)
\put(-0.1,0.3){\spos{}{-\!u\!-\!v}}
\put(-0.06,-0.1){\spos{}{-\!(\!q\!-\!2\!)\l}}
\put(-0.05,-0.5){\spos{}{+\!\m}}
\end{picture}}
\nc{\face}[5]{\begin{picture}(1.6,1.6)
\multiput(0.3,0.3)(1,0){2}{\line(0,1){1}}
\multiput(0.3,0.3)(0,1){2}{\line(1,0){1}}
\put(0.26,0.26){\spos{tr}{#1}}\put(1.34,0.26){\spos{tl}{#2}}
\put(1.34,1.34){\spos{bl}{#3}}
\put(0.26,1.34){\spos{br}{#4}}
\put(0.8,0.8){\spos{}{#5}}\end{picture}}
\nc{\dface}[5]{\begin{picture}(1.6,1.6)
\multiput(0.3,0.8)(0.5,0.5){2}{\line(1,-1){0.5}}
\multiput(0.3,0.8)(0.5,-0.5){2}{\line(1,1){0.5}}
\put(0.24,0.8){\spos{r}{#1}}\put(0.8,0.24){\spos{t}{#2}}
\put(1.36,0.8){\spos{l}{#3}}
\put(0.8,1.36){\spos{b}{#4}}
\put(0.8,0.8){\spos{}{#5}}\end{picture}}
\nc{\lefttri}[4]{\begin{picture}(1.1,1.6)
\put(0.3,0.3){\line(0,1){1}}\put(0.3,0.3){\line(1,1){0.5}}
\put(0.3,1.3){\line(1,-1){0.5}}
\put(0.26,0.26){\spos{tr}{#1}}\put(0.86,0.8){\spos{l}{#2}}
\put(0.26,1.34){\spos{br}{#3}}
\put(0.38,0.8){\spos{l}{#4}}\end{picture}}
\nc{\righttri}[4]{\begin{picture}(1.1,1.6)
\put(0.8,0.3){\line(0,1){1}}\put(0.8,0.3){\line(-1,1){0.5}}
\put(0.8,1.3){\line(-1,-1){0.5}}
\put(0.84,0.26){\spos{tl}{#1}}\put(0.24,0.8){\spos{r}{#2}}
\put(0.84,1.34){\spos{bl}{#3}}
\put(0.72,0.8){\spos{r}{#4}}\end{picture}}
\nc{\bdoubface}[8]{\begin{picture}(1.6,2.6)
\multiput(0.3,0.3)(1,0){2}{\line(0,1){2}}
\multiput(0.3,0.3)(0,1){3}{\line(1,0){1}}\d{1.3}{1.3}
\put(0.26,0.26){\spos{tr}{#1}}\put(1.34,0.26){\spos{tl}{#2}}
\put(1.4,1.3){\spos{l}{#3}}\put(1.34,2.34){\spos{bl}{#4}}
\put(0.26,2.34){\spos{br}{#5}}\put(0.24,1.3){\spos{r}{#6}}
\put(0.8,0.8){\spos{}{#7}}\put(0.8,1.8){\spos{}{#8}}
\end{picture}}
\nc{\tdoubface}[8]{\begin{picture}(1.6,2.6)
\multiput(0.3,0.3)(1,0){2}{\line(0,1){2}}
\multiput(0.3,0.3)(0,1){3}{\line(1,0){1}}\d{0.3}{1.3}
\put(0.26,0.26){\spos{tr}{#1}}\put(1.34,0.26){\spos{tl}{#2}}
\put(1.36,1.3){\spos{l}{#3}}\put(1.34,2.34){\spos{bl}{#4}}
\put(0.26,2.34){\spos{br}{#5}}\put(0.2,1.3){\spos{r}{#6}}
\put(0.8,0.8){\spos{}{#7}}\put(0.8,1.8){\spos{}{#8}}
\end{picture}}
\nc{\sdoubface}[8]{\begin{picture}(2.6,1.6)
\multiput(0.3,0.3)(1,0){3}{\line(0,1){1}}
\multiput(0.3,0.3)(0,1){2}{\line(1,0){2}}\d{1.3}{1.3}
\put(0.26,0.26){\spos{tr}{#1}}\put(1.3,0.24){\spos{t}{#2}}
\put(2.34,0.26){\spos{tl}{#3}}\put(2.34,1.34){\spos{bl}{#4}}
\put(1.3,1.4){\spos{b}{#5}}\put(0.26,1.34){\spos{br}{#6}}
\put(0.8,0.8){\spos{}{#7}}\put(1.8,0.8){\spos{}{#8}}
\end{picture}}
\nc{\leftdoubtri}[7]{\begin{picture}(2.2,3.8)
\put(0.3,0.3){\line(0,1){3.2}}\put(1.9,1.9){\line(-1,-1){1.6}}
\put(1.9,1.9){\line(-1,1){1.6}}\put(0.3,1.9){\line(1,-1){0.8}}
\put(0.3,1.9){\line(1,1){0.8}}\d{1.1}{1.1}
\put(0.26,0.26){\spos{tr}{#1}}\put(0.24,1.9){\spos{r}{#1}}
\put(0.26,3.54){\spos{br}{#1}}
\put(1.14,2.74){\spos{bl}{#2}}\put(1.96,1.9){\spos{l}{#3}}
\put(1.14,1.06){\spos{tl}{#4}}
\put(0.36,1.1){\spos{l}{#5}}\put(0.36,2.7){\spos{l}{#6}}
\put(1.1,1.9){\spos{}{#7}}
\end{picture}}
\nc{\rightdoubtri}[7]{\begin{picture}(2.2,3.8)
\put(1.9,0.3){\line(0,1){3.2}}\put(0.3,1.9){\line(1,-1){1.6}}
\put(0.3,1.9){\line(1,1){1.6}}\put(1.9,1.9){\line(-1,-1){0.8}}
\put(1.9,1.9){\line(-1,1){0.8}}\d{1.1}{2.7}
\put(1.94,0.26){\spos{tl}{#1}}\put(1.96,1.9){\spos{l}{#1}}
\put(1.94,3.54){\spos{bl}{#1}}\put(1.06,2.74){\spos{br}{#2}}
\put(0.24,1.9){\spos{r}{#3}}\put(1.06,1.06){\spos{tr}{#4}}
\put(1.84,1.1){\spos{r}{#5}}\put(1.84,2.7){\spos{r}{#6}}
\put(1.1,1.9){\spos{}{#7}}\end{picture}}
\nc{\longlowblock}[2]{\begin{picture}(4,1)
\multiput(0,0)(0.6,0){2}{\line(0,1){1}}
\multiput(3.4,0)(0.6,0){2}{\line(0,1){1}}
\multiput(0,0)(0,0.5){3}{\line(1,0){4}}
\multiput(0.3,0.25)(3.4,0){2}{\spos{}{#1}}
\multiput(0.3,0.75)(3.4,0){2}{\spos{}{#2}}
\multiput(0,0.5)(0.6,0){2}{\spos{}{\bullet}}
\multiput(3.4,0.5)(0.6,0){2}{\spos{}{\bullet}}
\end{picture}}
\nc{\shortlowblock}[2]{\begin{picture}(2,1)
\multiput(0,0)(0.6,0){2}{\line(0,1){1}}
\multiput(1.4,0)(0.6,0){2}{\line(0,1){1}}
\multiput(0,0)(0,0.5){3}{\line(1,0){2}}
\multiput(0.3,0.25)(1.4,0){2}{\spos{}{#1}}
\multiput(0.3,0.75)(1.4,0){2}{\spos{}{#2}}
\multiput(0,0.5)(0.6,0){2}{\spos{}{\bullet}}
\multiput(1.4,0.5)(0.6,0){2}{\spos{}{\bullet}}
\end{picture}}
\nc{\leftlowend}[1]{\begin{picture}(0.5,1)
\put(0,0){\line(0,1){1}}\put(0,0){\line(1,1){0.5}}
\put(0,1){\line(1,-1){0.5}}
\put(0.04,0.5){\spos{l}{#1}}
\multiput(0,0)(0,1){2}{\makebox(0.5,0){\dotfill}}
\end{picture}}
\nc{\rightlowend}[1]{\begin{picture}(0.5,1)
\put(0,0.5){\line(1,-1){0.5}}\put(0,0.5){\line(1,1){0.5}}
\put(0.5,0){\line(0,1){1}}
\put(0.46,0.5){\spos{r}{#1}}
\multiput(0,0)(0,1){2}{\makebox(0.5,0){\dotfill}}\end{picture}}
\nc{\leftbu}[2]{\begin{picture}(1.5,1)
\put(0,0){\line(0,1){1}}\put(0,0){\line(1,1){1}}
\put(0,1){\line(1,-1){1}}\put(1,0){\line(1,1){0.5}}
\put(1,1){\line(1,-1){0.5}}
\put(0.04,0.5){\spos{l}{#1}}\put(1,0.5){\spos{}{#2}}
\multiput(0,0)(0,1){2}{\makebox(1.5,0){\dotfill}}
\put(0.5,0.5){\spos{}{\bullet}}\end{picture}}
\nc{\rightbu}[2]{\begin{picture}(1.5,1)
\put(0,0.5){\line(1,-1){0.5}}\put(0,0.5){\line(1,1){0.5}}
\put(0.5,1){\line(1,-1){1}}\put(0.5,0){\line(1,1){1}}
\put(1.5,0){\line(0,1){1}}
\put(0.5,0.5){\spos{}{#1}}\put(1.46,0.5){\spos{r}{#2}}
\multiput(0,0)(0,1){2}{\makebox(1.5,0){\dotfill}}
\put(1,0.5){\spos{}{\bullet}}\end{picture}}
\nc{\longblock}[4]{\begin{picture}(4,2)
\multiput(0,0)(0.6,0){2}{\line(0,1){2}}
\multiput(3.4,0)(0.6,0){2}{\line(0,1){2}}
\multiput(0,0)(0,0.5){5}{\line(1,0){4}}
\multiput(0.3,0.25)(3.4,0){2}{\spos{}{#1}}
\multiput(0.3,0.75)(3.4,0){2}{\spos{}{#2}}
\multiput(0.3,1.25)(3.4,0){2}{\spos{}{#3}}
\multiput(0.3,1.75)(3.4,0){2}{\spos{}{#4}}
\multiput(0,0.5)(0,0.5){3}{\spos{}{\bullet}}
\multiput(0.6,0.5)(0,0.5){3}{\spos{}{\bullet}}
\multiput(3.4,0.5)(0,0.5){3}{\spos{}{\bullet}}
\multiput(4,0.5)(0,0.5){3}{\spos{}{\bullet}}
\end{picture}}
\nc{\shortblock}[4]{\begin{picture}(2,2)
\multiput(0,0)(0.6,0){2}{\line(0,1){2}}
\multiput(1.4,0)(0.6,0){2}{\line(0,1){2}}
\multiput(0,0)(0,0.5){5}{\line(1,0){2}}
\multiput(0.3,0.25)(1.4,0){2}{\spos{}{#1}}
\multiput(0.3,0.75)(1.4,0){2}{\spos{}{#2}}
\multiput(0.3,1.25)(1.4,0){2}{\spos{}{#3}}
\multiput(0.3,1.75)(1.4,0){2}{\spos{}{#4}}
\multiput(0,0.5)(0,0.5){3}{\spos{}{\bullet}}
\multiput(0.6,0.5)(0,0.5){3}{\spos{}{\bullet}}
\multiput(1.4,0.5)(0,0.5){3}{\spos{}{\bullet}}
\multiput(2,0.5)(0,0.5){3}{\spos{}{\bullet}}
\end{picture}}
\nc{\inv}[2]{\begin{picture}(2,1)
\put(0.5,0){\line(1,1){1}}\put(0.5,1){\line(1,-1){1}}
\multiput(0,0.5)(1.5,0.5){2}{\line(1,-1){0.5}}
\multiput(0,0.5)(1.5,-0.5){2}{\line(1,1){0.5}}
\put(0.5,0.5){\spos{}{#1}}\put(1.5,0.5){\spos{}{#2}}
\put(1,0.5){\spos{}{\bullet}}
\multiput(0,0)(0,1){2}{\makebox(2,0){\dotfill}}
\end{picture}}
\nc{\lowinv}[2]{\begin{picture}(2,2)
\put(0,0){\inv{#1}{#2}}
\multiput(0,1.5)(0,0.5){2}{\makebox(2,0){\dotfill}}
\multiput(0.5,1)(1,0){2}{\spos{}{\bullet}}
\end{picture}}
\nc{\midinv}[2]{\begin{picture}(2,2)
\put(0,0.5){\inv{#1}{#2}}
\multiput(0,0)(0,2){2}{\makebox(2,0){\dotfill}}
\multiput(0.5,0.5)(1,0){2}{\spos{}{\bullet}}
\multiput(0.5,1.5)(1,0){2}{\spos{}{\bullet}}
\end{picture}}
\nc{\highinv}[2]{\begin{picture}(2,2)
\put(0,1){\inv{#1}{#2}}
\multiput(0,0)(0,0.5){2}{\makebox(2,0){\dotfill}}
\multiput(0.5,1)(1,0){2}{\spos{}{\bullet}}
\end{picture}}
\nc{\leftend}[2]{\begin{picture}(0.5,2)
\put(0,0){\line(0,1){2}}\multiput(0,0)(0,1){2}{\line(1,1){0.5}}
\multiput(0,1)(0,1){2}{\line(1,-1){0.5}}
\put(0.04,0.5){\spos{l}{#1}}\put(0.04,1.5){\spos{l}{#2}}
\multiput(0,0)(0,1){3}{\makebox(0.5,0){\dotfill}}
\put(0,1){\spos{}{\bullet}}
\end{picture}}
\nc{\rightend}[2]{\begin{picture}(0.5,2)
\put(0.5,0){\line(0,1){2}}\multiput(0.5,0)(0,1){2}{\line(-1,1){0.5}}
\multiput(0.5,1)(0,1){2}{\line(-1,-1){0.5}}
\put(0.46,0.5){\spos{r}{#1}}\put(0.46,1.5){\spos{r}{#2}}
\multiput(0,0)(0,1){3}{\makebox(0.5,0){\dotfill}}
\put(0.5,1){\spos{}{\bullet}}
\end{picture}}
\nc{\lefttriangle}[3]{\begin{picture}(1,2)
\put(0,0){\line(0,1){2}}\put(0,1){\line(1,1){0.5}}
\put(0,1){\line(1,-1){0.5}}\put(1,1){\line(-1,1){1}}
\put(1,1){\line(-1,-1){1}}
\put(0.04,0.5){\spos{l}{#1}}\put(0.04,1.5){\spos{l}{#2}}
\put(0.5,1){\spos{}{#3}}
\multiput(0,0)(0,2){2}{\makebox(1,0){\dotfill}}
\multiput(0.5,0.5)(0,1){2}{\makebox(0.5,0){\dotfill}}
\put(0,1){\spos{}{\bullet}}
\multiput(0.5,0.5)(0,1){2}{\spos{}{\bullet}}
\end{picture}}
\nc{\righttriangle}[3]{\begin{picture}(1,2)
\put(1,0){\line(0,1){2}}\put(1,1){\line(-1,1){0.5}}
\put(1,1){\line(-1,-1){0.5}}\put(0,1){\line(1,1){1}}
\put(0,1){\line(1,-1){1}}
\put(0.96,0.5){\spos{r}{#1}}\put(0.96,1.5){\spos{r}{#2}}
\put(0.5,1){\spos{}{#3}}
\multiput(0,0)(0,2){2}{\makebox(1,0){\dotfill}}
\multiput(0,0.5)(0,1){2}{\makebox(0.5,0){\dotfill}}
\put(1,1){\spos{}{\bullet}}
\multiput(0.5,0.5)(0,1){2}{\spos{}{\bullet}}
\end{picture}}
\nc{\botleftref}[4]{\begin{picture}(1.5,2)
\put(0,0){\line(0,1){2}}\put(0,1){\line(1,1){0.5}}
\put(0,1){\line(1,-1){1}}\put(0,0){\line(1,1){1}}
\put(0,2){\line(1,-1){1.5}}\put(1.5,0.5){\line(-1,-1){0.5}}
\put(0.04,0.5){\spos{l}{#1}}\put(0.04,1.5){\spos{l}{#2}}
\put(1,0.5){\spos{}{#3}}\put(0.5,1){\spos{}{#4}}
\multiput(0,0)(0,2){2}{\makebox(1.5,0){\dotfill}}
\put(1,1){\makebox(0.5,0){\dotfill}}
\put(0.5,1.5){\makebox(1,0){\dotfill}}
\multiput(0,1)(1,0){2}{\spos{}{\bullet}}
\multiput(0.5,0.5)(0,1){2}{\spos{}{\bullet}}
\end{picture}}
\nc{\topleftref}[4]{\begin{picture}(1.5,2)
\put(0,0){\line(0,1){2}}\put(0,1){\line(1,1){1}}
\put(0,1){\line(1,-1){0.5}}\put(0,0){\line(1,1){1.5}}
\put(0,2){\line(1,-1){1}}\put(1.5,1.5){\line(-1,1){0.5}}
\put(0.04,0.5){\spos{l}{#1}}\put(0.04,1.5){\spos{l}{#2}}
\put(0.5,1){\spos{}{#3}}\put(1,1.5){\spos{}{#4}}
\multiput(0,0)(0,2){2}{\makebox(1.5,0){\dotfill}}
\put(1,1){\makebox(0.5,0){\dotfill}}
\put(0.5,0.5){\makebox(1,0){\dotfill}}
\multiput(0,1)(1,0){2}{\spos{}{\bullet}}
\multiput(0.5,0.5)(0,1){2}{\spos{}{\bullet}}
\end{picture}}
\nc{\botrightref}[4]{\begin{picture}(1.5,2)
\put(1.5,0){\line(0,1){2}}\put(1.5,1){\line(-1,1){0.5}}
\put(1.5,1){\line(-1,-1){1}}\put(1.5,0){\line(-1,1){1}}
\put(1.5,2){\line(-1,-1){1.5}}\put(0,0.5){\line(1,-1){0.5}}
\put(1.46,0.5){\spos{r}{#1}}\put(1.46,1.5){\spos{r}{#2}}
\put(0.5,0.5){\spos{}{#3}}\put(1,1){\spos{}{#4}}
\multiput(0,0)(0,2){2}{\makebox(1.5,0){\dotfill}}
\put(0,1){\makebox(0.5,0){\dotfill}}
\put(0,1.5){\makebox(1,0){\dotfill}}
\multiput(0.5,1)(1,0){2}{\spos{}{\bullet}}
\multiput(1,0.5)(0,1){2}{\spos{}{\bullet}}
\end{picture}}
\nc{\toprightref}[4]{\begin{picture}(1.5,2)
\put(1.5,0){\line(0,1){2}}\put(1.5,1){\line(-1,1){1}}
\put(1.5,1){\line(-1,-1){0.5}}\put(1.5,0){\line(-1,1){1.5}}
\put(1.5,2){\line(-1,-1){1}}\put(0,1.5){\line(1,1){0.5}}
\put(1.46,0.5){\spos{r}{#1}}\put(1.46,1.5){\spos{r}{#2}}
\put(1,1){\spos{}{#3}}\put(0.5,1.5){\spos{}{#4}}
\multiput(0,0)(0,2){2}{\makebox(1.5,0){\dotfill}}
\put(0,1){\makebox(0.5,0){\dotfill}}
\put(0,0.5){\makebox(1,0){\dotfill}}
\multiput(0.5,1)(1,0){2}{\spos{}{\bullet}}
\multiput(1,0.5)(0,1){2}{\spos{}{\bullet}}
\end{picture}}

\begin{document}
\begin{flushright}hep-th/9507118\end{flushright}
\begin{center}
\title{Interaction-Round-a-Face Models with Fixed Boundary
Conditions: The ABF Fusion Hierarchy}
\author{Roger E. Behrend\footnote{E-mail: reb@maths.mu.oz.au},
Paul A. Pearce\footnote{E-mail: pap@maths.mu.oz.au} and
David L. O'Brien\footnote{E-mail: dlo@maths.mu.oz.au}}
\address{Mathematics Department, University of Melbourne,\\Parkville,
Victoria 3052, Australia}

\begin{abstract}
We use boundary weights and reflection equations to obtain families of
commuting double-row transfer matrices for interaction-round-a-face
models with fixed boundary conditions. In particular, we consider the
fusion hierarchy of the Andrews-Baxter-Forrester models, for which we
find that the double-row transfer matrices satisfy functional
equations with an $su(2)$ structure.
\end{abstract}
{\small July 24, 1995}
\end{center}

\sect{Introduction}
\sub{Overview}
Two-dimensional lattice spin models in statistical mechanics have
traditionally been solved by imposing periodic boundary conditions on
the rows of the lattice.  The Yang-Baxter equation, together with such
boundary conditions, then leads to families of commuting row transfer
matrices and hence solvability~\cite{Bax82}.  However, the work of
Sklyanin~\cite{Skl88} shows that, by using reflection equations, it is
also possible to construct commuting double-row transfer matrices for
vertex models with open boundary conditions. In this paper, we
present a scheme, motivated by Sklyanin's formalism for open
boundaries and Baxter's correspondence between vertex and
interaction-round-a-face (IRF) models~\cite{Bax73,Bax82}, for
obtaining solvable IRF models with fixed boundary conditions.

Although the usual bulk quantities of physical interest are
independent of the boundary conditions in the thermodynamic limit,
there are many surface properties, such as the interfacial tension,
which are also important. Moreover, at criticality, the conformal
spectra of lattice models do depend on the boundary
conditions~\cite{Car86b}. For these reasons it is of interest to study
lattice models with non-periodic boundary conditions.

The layout of this paper is as follows. In the remainder of this
section, we discuss related work on solvability with open boundaries
and outline the formalism for vertex models.  In Section~2, we present
the general procedure for obtaining commuting double-row transfer
matrices for IRF models with fixed boundary conditions and then
specialise to the case of Andrews-Baxter-Forrester
(ABF)~\cite{AndBaxFor84} models.  In Section~3, we consider fusion of
IRF models with fixed boundary conditions and concentrate on the ABF
fusion hierarchy.  We conclude with a discussion of future work and
three appendices, in which we prove some of the results used in the
main text.

\sub{Background}
Reflection equations were introduced by Cherednik~\cite{Che84} as a
means of obtaining factorisable scattering matrices for particles on a
semi-infinite line.  Sklyanin~\cite{Skl88} then considered these
equations in the context of one-dimensional quantum spin chains and
showed that they could be used to obtain integrable systems with
non-periodic boundary conditions.  When translated into the context of
two-dimensional lattice models in statistical mechanics, Sklyanin's
formalism provides a scheme for obtaining exactly solvable vertex
models with open boundary conditions.  More specifically, this
procedure uses left and right boundary weights, represented by
$K$~matrices, in addition to the usual bulk vertex weights,
represented by a spectral-parameter-dependent $R$~matrix.  {}From
Sklyanin's work, it follows that if a particular $R$ matrix satisfies
the Yang-Baxter equation, the first and second inversion relations,
and certain symmetries, and if corresponding $K$ matrices can be found
which satisfy the reflection equations, then families of commuting,
open-boundary transfer matrices can be constructed.  Each such
transfer matrix involves vertex weights from two adjacent rows of the
lattice as well as a left and right boundary weight.

Various modifications of the reflection equations have been considered
since Sklyanin's original work.  Mezincescu and
Nepomechie~\cite{MezNep91a,MezNep92a} and Yue and
Chen~\cite{YueChe93b} have generalised the formalism to allow for
$R$~matrices satisfying less restrictive symmetries, while recently,
Kulish~\cite{Kul95} has, independently from us, obtained reflection
equations for IRF models.  Another variation is
spectral-parameter-independent reflection equations, which have been
studied by Kulish and Sklyanin~\cite{KulSkl92} and Kulish, Sasaki and
Schweibert~\cite{KulSasSch93}.

The reflection equations have been solved to give $K$~matrices
corresponding to the $R$~matrices of a number of models.
Cherednik~\cite{Che84} and Yue and Chen~\cite{YueChe93b} found
diagonal, elliptic solutions for the $Z_n\times Z_n$ Belavin model,
which is related to the Lie algebra $A^{(1)}_{n\!-\!1}$. Sklyanin used
Cherednik's solution for $n=2$ to obtain trigonometric, diagonal
$K$~matrices for the six-vertex model, or XXZ chain, while
non-diagonal $K$~matrices for the six-vertex model have been obtained
by de~Vega and Gonz\'{a}lez-Ruiz~\cite{DevGon93} and Ghoshal and
Zamolodchikov~\cite{GhoZam94a}. For the eight-vertex model, elliptic,
diagonal $K$~matrices have been found by Cuerno and
Gonz\'{a}lez-Ruiz~\cite{CueGon93}, while non-diagonal $K$~matrices
have been found by de~Vega and Gonz\'{a}lez-Ruiz~\cite{DevGon94b}, Hou
and Yue~\cite{HouYue93} and Inami and Konno~\cite{InaKon94}.

The original six-vertex model corresponds to the spin-$\frac{1}{2}$
representation of $A^{(1)}_1$.  Mezincescu, Nepomechie and
Rittenberg~\cite{MezNepRit90} have also found diagonal, trigonometric
$K$~matrices for the Zamolodchikov-Fateev 19-vertex model, which
corresponds to the spin-$1$ representation of $A^{(1)}_1$.
Furthermore, diagonal, trigonometric $K$~matrices for the
$n(2n\!-\!1)$-vertex models which correspond to the fundamental
representation of $A^{(1)}_{n\!-\!1}$ have been found by de~Vega and
Gonz\'{a}lez-Ruiz~\cite{DevGon93}, while non-diagonal $K$~matrices for
these models have been found by Abad and Rios~\cite{AbaRio95}.

The models mentioned so far have all been based on non-exceptional,
untwisted affine Lie algebras and we now outline the models based on
other Lie algebras, for which $K$~matrices are known.  For the
19-vertex Izergin-Korepin model, which corresponds to the fundamental
representation of the twisted affine Lie algebra $A^{(2)}_2$,
diagonal, trigonometric $K$~matrices have been obtained by Mezincescu
and Nepomechie~\cite{MezNep91c} and non-diagonal $K$~matrices have
been obtained by Kim~\cite{Kim94}.  For the 175-vertex model based on
the exceptional Lie algebra $G^{(1)}_2$, diagonal $K$~matrices have
been obtained by Yung and Batchelor~\cite{YunBat95b}, and for the
$spl(2,1)$-related 15-vertex $t$-$J$ model, diagonal $K$~matrices have
been obtained by Gonz\'{a}lez-Ruiz~\cite{Gon94}.  Finally, it should
be noted that Mezincescu and Nepomechie~\cite{MezNep91c,MezNep92a}
have shown that the identity matrix satisfies the right reflection
equation for vertex models corresponding to the fundamental
representations of the algebras $A^{(1)}_n$, $A^{(2)}_n$, $B^{(1)}_n$,
$C^{(1)}_n$ and $D^{(1)}_n$.

Having found $K$~matrices for a particular model, it remains to obtain
the eigenspectra of the corresponding double-row transfer matrices.
This has been done, using various forms of the Bethe ansatz, by Kulish
and Sklyanin~\cite{Skl88,KulSkl91}, Artz, Mezincescu, Nepomechie and
Rittenberg~\cite{MezNepRit90,MezNep92b,ArtMezNep94,ArtMezNep95},
Destri, De~Vega and
Gonz\'{a}lez-Ruiz~\cite{DesDev92a,DesDev92b,DevGon94a,DevGon94c,Gon94},
Yung and Batchelor~\cite{YunBat95a,YunBat95b}, Foerster and
Karowski~\cite{FoeKar93}, and Zhou~\cite{Zho95}.  In all of this work,
constant or trigonometric, diagonal $K$-matrices were used.  In many
cases, the important property of quantum algebra invariance---that is,
commutation of the transfer matrix or Hamiltonian with elements of the
associated quantum algebra---was obtained, by choosing specific,
usually constant, $K$~matrices.  Indeed, the fact that quantum
algebra invariance can not be achieved using standard periodic
boundary conditions has been a strong motivation for much of the work
on open boundary conditions.  We also note that
in~\cite{DesDev92a,DesDev92b,YunBat95a}, diagonal-to-diagonal
open-boundary transfer matrices were considered.  These were obtained
by setting to zero the spectral parameters of alternate vertex
weights in the double-row transfer matrices.

Other directions in which recent work with open boundaries has
proceeded include the consideration of loop models by Yung and
Batchelor~\cite{YunBat95a}, and the use of vertex operators by Jimbo,
Kedem, Kojima, Konno and Miwa~\cite{JimKedKojKonMiw95}.  There has
also been a substantial return to the use of reflection equations in
the theory of scattering on finite or semi-infinite lines, which has
been initiated by the work of Ghoshal and
Zamolodchikov~\cite{GhoZam94a,GhoZam94b} and Fring and
K\"{o}berle~\cite{FriKob94b}.  Of particular relevance here
is the boundary crossing equation introduced in~\cite{GhoZam94a},
since we use corresponding equations in our treatment of IRF models.

In this paper, we apply the fusion procedure to IRF models with fixed
boundary conditions and obtain functional equations satisfied by the
fixed-boundary double-row transfer matrices of the ABF fusion
hierarchy.  Fusion was introduced by Kulish, Reshetikhin and
Sklyanin~\cite{KulResSkl81} as a means of obtaining new solutions to
the Yang-Baxter equation, by combining $R$~matrices from a known
solution of the Yang-Baxter equation.  In terms of the associated Lie
algebra, if the original $R$~matrices correspond to a particular
representation, then the fused $R$~matrices correspond to
higher-dimensional representations.

Fusion was first applied to IRF models, and in particular to the ABF
models, by Date, Jimbo, Kuniba, Miwa and Okado
\cite{DatJimMiwOka86,DatJimMiwOka87a,DatJimKunMiwOka87,DatJimKunMiwOka88}.
Bazhanov and Reshetikhin~\cite{BazRes89} then
obtained functional equations satisfied by the periodic-boundary row
transfer matrices of the ABF fusion hierarchy and used these equations
to derive Bethe ansatz solutions for the eigenspectra of the transfer
matrices.  These functional equations were also used by Kl\"{u}mper
and Pearce~\cite{KluPea92} to obtain a generalised inversion
identity.  Fusion of other IRF models, leading to functional equations
satisfied by periodic-boundary transfer matrices, has been implemented
by Jimbo, Kuniba, Miwa and Okado~\cite{JimKunMiwOka88} and Bazhanov
and Reshetikhin~\cite{BazRes90} for the $A^{(1)}_n$ IRF models, by Zhou
and Pearce~\cite{ZhoPea94a} for the $A$--$D$--$E$ models, which
include the critical ABF models, and by Zhou, Pearce and
Grimm~\cite{ZhoPeaGri95} for the dilute~$A$ models.

In the case of vertex models with open boundaries, the aim of fusion
is to construct new solutions to the reflection equation, by combining
$K$ and $R$~matrices from known solutions of the Yang-Baxter and
reflection equations.  A general formalism for this procedure has been
presented by Mezincescu and Nepomechie~\cite{MezNep92c} and has been
applied to the eight-vertex model by Yue~\cite{Yue94}. Furthermore,
the boundary bootstrap equations, introduced in the context of
scattering by Ghoshal and Zamolodchikov~\cite{GhoZam94a} and Fring and
K\"{o}berle~\cite{FriKob94b}, correspond to the process of fusion.  In
each of these cases only level~2 fusion, in which two $K$~matrices are
combined with one $R$~matrix, was considered explicitly, although it
is clear that the process can be extended to higher fusion levels.  In
fact, Zhou~\cite{Zho95} has recently applied fusion at arbitrary
levels to the six-vertex model with open boundary conditions.

Finally, we note that the work of Saleur and Bauer~\cite{SalBau89} and
of Destri and de~Vega~\cite{DesDev92b} has involved the consideration
of ABF models with fixed boundary conditions, however in each of these
cases the fixed boundary conditions were applied along diagonal rows
of the lattice.

\sub{Vertex Models with Open Boundary Conditions}
We now schematically outline the formalism for vertex models on which
our treatment of IRF models is based.  We note that the main
differences between our formalism and that originally presented by
Sklyanin~\cite{Skl88} are that here only the first inversion relation
is assumed, no $R$~matrix symmetries are assumed, and the top row of
the transfer matrix has the form $\vec{T}(\m-u)^{t_0}$ rather than the
form $\vec{T}(-u)^{-1}$, where $\vec{T}(u)$ is the form of the bottom
row, $\m$ is arbitrary and $t_0$ is transposition on the auxiliary
space.

We are considering a vertex model with Boltzmann vertex weights
%
%  Vertex Weights
%
\setlength{\unitlength}{12mm}\begin{center}
\begin{picture}(1,1)\put(0.5,0){\line(0,1){1}}
\put(0,0.5){\line(1,0){1}}
\put(0.42,0.42){\spos{tr}{u}}\end{picture}\end{center}
where $u$ is a spectral parameter. We assume that these
satisfy the Yang-Baxter equation
%
%  Vertex Yang-Baxter Equation
%
\setlength{\unitlength}{8mm}
\begin{center}\begin{picture}(4,4)\put(4,0.5){\line(-2,1){4}}
\put(4,3.5){\line(-2,-1){4}}\put(3,0){\line(0,1){4}}
\put(0.65,2){\spos{r}{u\!-\!v}}\put(2.9,0.95){\spos{tr}{u}}
\put(2.9,2.75){\spos{tr}{v}}\end{picture}
\text{2}{4}{1}{2}{}{=}
\begin{picture}(4,4)\put(0,0.5){\line(2,1){4}}
\put(0,3.5){\line(2,-1){4}}\put(1,0){\line(0,1){4}}
\put(2.65,2){\spos{r}{u\!-\!v}}\put(0.9,0.75){\spos{tr}{v}}
\put(0.9,2.95){\spos{tr}{u}}\end{picture}\end{center}
and the inversion relation
%
%  Vertex Inversion Relation
%
\setlength{\unitlength}{15mm}
\begin{center}
\begin{picture}(2,1)\multiput(0,0)(1,0){2}{\line(1,1){1}}
\multiput(0,1)(1,0){2}{\line(1,-1){1}}
\put(0.4,0.5){\spos{r}{u}}\put(1.4,0.5){\spos{r}{-u}}\end{picture}
\text{0.8}{1.2}{0.4}{0.5}{}{=}
\text{1.4}{1}{0.7}{0.5}{}{\rho(u)\,\rho(-\!u)}
\begin{picture}(1,1)\put(0,0.2){\line(1,0){1}}\put(0,0.8){\line(1,0){1}}
\end{picture}
\end{center}
where $\rho$ is a model-dependent scalar function.
We now introduce left and right boundary weights
%
%  Vertex Boundary Weights
%
\setlength{\unitlength}{10mm}\begin{center}
\begin{picture}(0.5,1)\put(0,0.5){\line(1,1){0.5}}
\put(0,0.5){\line(1,-1){0.5}}
\put(0.08,0.1){\spos{l}{u}}\multiput(0,0)(0,0.15){7}{\line(0,1){0.1}}
\end{picture}
\text{4}{1}{2}{0.5}{}{\mbox{and}}
\begin{picture}(0.5,1)\put(0,0){\line(1,1){0.5}}
\put(0,1){\line(1,-1){0.5}}
\put(0.42,0.1){\spos{r}{u}}\multiput(0.5,0)(0,0.15){7}{\line(0,1){0.1}}
\end{picture}\end{center}
which satisfy left and right reflection equations
%
%  Vertex Left Reflection Equation
%
\setlength{\unitlength}{10mm}
\begin{center}\begin{picture}(3,4)
\put(0,2){\line(1,-1){2}}\put(0,2){\line(1,1){2}}
\put(0,1){\line(4,-1){3}}
\put(0,1){\line(4,1){3}}
\put(0.08,0.78){\spos{l}{u}}\put(0.08,1.6){\spos{l}{v}}
\put(0.9,1.62){\vector(-1,-2){0.2}}
\put(1.1,1.67){\spos{b}{\m\!-\!u\!-\!v}}
\put(1.75,1.05){\vector(-2,-1){0.6}}\put(1.8,1.07){\spos{l}{u\!-\!v}}
\multiput(0,0)(0,0.15){28}{\line(0,1){0.1}}\end{picture}
\text{3}{4}{1.5}{2}{}{=}
\begin{picture}(3,4)
\put(0,2){\line(1,-1){2}}\put(0,2){\line(1,1){2}}
\put(0,3){\line(4,-1){3}}\put(0,3){\line(4,1){3}}
\put(0.08,1.6){\spos{l}{v}}\put(0.08,2.78){\spos{l}{u}}
\put(0.9,2.37){\vector(-1,2){0.2}}\put(1.1,2.32){\spos{t}{\m\!-\!u\!-\!v}}
\put(1.75,2.95){\vector(-2,1){0.6}}\put(1.8,2.95){\spos{l}{u\!-\!v}}
\multiput(0,0)(0,0.15){28}{\line(0,1){0.1}}\end{picture}\end{center}
and
%
%  Vertex Right Reflection Equation
%
\setlength{\unitlength}{10mm}
\begin{center}\begin{picture}(3,4)
\put(3,2){\line(-1,-1){2}}\put(3,2){\line(-1,1){2}}
\put(3,1){\line(-4,-1){3}}\put(3,1){\line(-4,1){3}}
\put(2.92,0.78){\spos{r}{u}}\put(2.92,1.6){\spos{r}{v}}
\put(1.2,0.22){\spos{br}{u\!-\!v}}\put(2,1.15){\spos{r}{\m\!-\!u\!-\!v}}
\multiput(3,0)(0,0.15){28}{\line(0,1){0.1}}\end{picture}
\text{3}{4}{1.5}{2}{}{=}
\begin{picture}(3,4)
\put(3,2){\line(-1,-1){2}}\put(3,2){\line(-1,1){2}}
\put(3,3){\line(-4,-1){3}}\put(3,3){\line(-4,1){3}}
\put(2.92,1.6){\spos{r}{v}}\put(2.92,2.78){\spos{r}{u}}
\put(1.2,3.6){\spos{br}{u\!-\!v}}\put(2,2.9){\spos{r}{\m\!-\!u\!-\!v}}
\multiput(3,0)(0,0.15){28}{\line(0,1){0.1}}\end{picture}\end{center}
where $\m$ is an arbitrary parameter. If we define a double-row transfer matrix
as
%
%  Vertex Double Row Transfer Matrix
%
\setlength{\unitlength}{6.5mm}\begin{center}
\text{5}{4}{2.5}{2}{}{\D(u)\;\;=}
\begin{picture}(11,4)(0,0)
\multiput(2,0)(2,0){2}{\line(0,1){4}}\put(9,0){\line(0,1){4}}
\multiput(1,1)(0,2){2}{\line(1,0){9}}
\multiput(0,2)(10,1){2}{\line(1,-1){1}}
\multiput(0,2)(10,-1){2}{\line(1,1){1}}
\multiput(0,0.7)(0,0.3){9}{\line(0,1){0.2}}
\multiput(11,0.7)(0,0.3){9}{\line(0,1){0.2}}
\put(0.04,0.96){\spos{l}{\m\!-\!u}}\put(10.88,1.37){\spos{r}{u}}
\multiput(1.9,0.9)(2,0){2}{\spos{tr}{u}}
\put(8.9,0.9){\spos{tr}{u}}
\multiput(1.9,2.9)(2,0){2}{\spos{tr}{\m\!-\!u}}
\put(8.9,2.9){\spos{tr}{\m\!-\!u}}
\end{picture}\end{center}
then it can be shown that, for any fixed value of $\m$, these matrices form a
commuting family,
\beq\D(u)\:\D(v)=\D(v)\:\D(u)\eeq
We also note that if we regard each value, $u$, of the spectral parameter as an
effective angle
\[\t(u)=\frac{\pi u}{\m}\]
then the geometric angles in the diagrammatic Yang-Baxter and reflection
equations correspond exactly with the effective angles given by values of the
spectral parameter.
This is due to the interpretation of these equations in terms of scattering.

\sect{IRF Models with Fixed Boundary Conditions}
\sub{Boltzmann Weights and Transfer Matrices}
We now present our formalism for interaction-round-a-face (IRF) models, which
was motivated by the preceding formalism for vertex models and Baxter's
vertex-face
correspondence~\cite{Bax73,Bax82}.  Our use of boundary crossing equations was
also motivated by the work of Ghoshal and Zamolodchikov~\cite{GhoZam94a}.

We are considering an IRF model with Boltzmann face weights
\setlength{\unitlength}{13mm}
\[\text{2}{1.6}{1}{0.8}{}{\W{a}{b}{c}{d}{u}}
\text{1}{1.6}{0.5}{0.8}{}{=}
\begin{picture}(1.14,1.6)
\put(0.57,0.8){\setlength{\unitlength}{0.71\unitlength}
\makebox(0,0){\face{a}{b}{c}{d}{u}}}\end{picture}
\text{1}{1.6}{0.5}{0.8}{}{=}
\dface{a}{b}{c}{d}{u}\]
Here, the spins $a$, $b$, $c$, $d$ take values from a discrete set and the
spectral parameter $u$ is a complex variable.

In order to accommodate fixed boundary conditions, we introduce left and right
boundary weights
\setlength{\unitlength}{13mm}
\[\text{1.6}{1.6}{0.8}{0.8}{}{\BL{a}{b}{c}{u}}
\text{1}{1.6}{0.5}{0.8}{}{=}\lefttri{a}{b}{c}{u}
\text{2.4}{1.6}{1.2}{0.8}{}{\mbox{and}}
\text{1.6}{1.6}{0.8}{0.8}{}{\BR{a}{b}{c}{u}}
\text{1}{1.6}{0.5}{0.8}{}{=}\righttri{a}{b}{c}{u}\]
We now consider a lattice of width $N$ and use these weights to construct a
double-row transfer matrix.
If $a_1,\ldots,a_{\N\!+\!1}$  and $b_1,\ldots,b_{\N\!+\!1}$ are two rows of
spins, and $\m$ is an arbitrary parameter,
then the corresponding entry of the double-row transfer matrix is defined by
%
% Fixed Boundary Double Row Transfer Matrix
%
\setlength{\unitlength}{12mm}
\begin{flushleft}$\displaystyle\langle a_1,\ldots,a_{\N\!+\!1}|
\D(u)|b_1,\ldots,b_{\N\!+\!1}\rangle\;=$\end{flushleft}
$$\sum_{c_1\ldots c_{N\!+\!1}}\;\BL{a_1}{c_1}{b_1}{\m\!-\!u}\,
\left[\prod_{j=1}^N\;\W{a_j}{a_{j\!+\!1}}{c_{j\!+\!1}}{c_j}{u}\:
\W{c_j}{c_{j\!+\!1}}{b_{j\!+\!1}}{b_j}{\m\!-\!u}\right]\,
\BR{a_{\N\!+\!1}}{c_{\N\!+\!1}}{b_{\N\!+\!1}}{u}$$
\beq=\raisebox{-1.4\unitlength}[1.6\unitlength][
1.4\unitlength]{\begin{picture}(4.5,3)
\multiput(0.5,0.5)(6,0){2}{\line(0,1){2}}
\multiput(1,0.5)(1,0){3}{\line(0,1){2}}
\multiput(5,0.5)(1,0){2}{\line(0,1){2}}
\multiput(1,0.5)(0,1){3}{\line(1,0){5}}
\put(1,1.5){\line(-1,2){0.5}}\put(1,1.5){\line(-1,-2){0.5}}
\put(6,1.5){\line(1,2){0.5}}\put(6,1.5){\line(1,-2){0.5}}
\put(0.5,0.45){\spos{t}{a_1}}\put(1,0.45){\spos{t}{a_1}}
\put(2,0.45){\spos{t}{a_2}}\put(3,0.45){\spos{t}{a_3}}
\put(5,0.45){\spos{t}{a_N}}\put(6,0.45){\spos{t}{a_{N\!+\!1}}}
\put(6.7,0.45){\spos{t}{a_{N\!+\!1}}}
\put(0.5,2.6){\spos{b}{b_1}}\put(1,2.6){\spos{b}{b_1}}
\put(2,2.6){\spos{b}{b_2}}\put(3,2.6){\spos{b}{b_3}}
\put(5,2.6){\spos{b}{b_N}}\put(6,2.6){\spos{b}{b_{N\!+\!1}}}
\put(6.7,2.6){\spos{b}{b_{N\!+\!1}}}
\put(1.05,1.45){\spos{tl}{c_1}}\put(2.05,1.45){\spos{tl}{c_2}}
\put(3.05,1.45){\spos{tl}{c_3}}\put(4.99,1.45){\spos{tr}{c_N}}
\put(5.99,1.45){\spos{tr}{c_{N\!+\!1}}}
\multiput(1.5,1)(1,0){2}{\spos{}{u}}\put(5.5,1){\spos{}{u}}
\multiput(1.5,2)(1,0){2}{\spos{}{\m\!-\!u}}
\put(5.5,2){\spos{}{\m\!-\!u}}
\put(0.71,1.5){\spos{}{\m\!-\!u}}\put(6.29,1.5){\spos{}{u}}
\multiput(0.5,0.5)(0,2){2}{\makebox(0.5,0){\dotfill}}
\multiput(6,0.5)(0,2){2}{\makebox(0.5,0){\dotfill}}
\multiput(1,1.5)(1,0){3}{\spos{}{\bullet}}
\multiput(5,1.5)(1,0){2}{\spos{}{\bullet}}
\end{picture}}\qquad\qquad\qquad
\qquad\qquad\eql{DRTM}\eeq
In this and all subsequent diagrams, we use solid circles to indicate spins
which are summed over and dotted lines to connect identical spins.

In general, there will be restrictions on the spins allowed on any neighbouring
lattice sites, as specified by an adjacency matrix
\[A_{ab}=\left\{\begin{array}{ll}0\;,&\mbox{$a$ and $b$ may not be adjacent}\\
1\;,&\mbox{$a$ and $b$ may be adjacent}\end{array}\right.\]
We assume that the face and boundary weights satisfy the adjacency condition as
follows:
\beqa\rule[-3.5ex]{0ex}{3.5ex}\W{a}{b}{c}{d}{u}&=&A_{ab}
\:A_{bc}\:A_{cd}\:A_{da}\;\W{a}{b}{c}{d}{u}\noeqno
\rule[-3.5ex]{0ex}{3.5ex}\BL{a}{b}{c}{u}&=&A_{ab}\:A_{bc}
\;\BL{a}{b}{c}{u}\eql{AC}\\
\BR{a}{b}{c}{u}&=&A_{ab}\:A_{bc}\;
\BR{a}{b}{c}{u}\nonumber\eeqa

\sub{Local Relations}
The face weights and boundary weights are assumed to satisfy the following
local relations:\\
the Yang-Baxter equation
%
%  Yang-Baxter Equation
%
\beqa\lefteqn{\rule[-3.5ex]{0ex}{3.5ex}\sum_g\;
\W{a}{b}{g}{f}{u\!-\!v}\;\W{b}{c}{d}{g}{u}\;
\W{g}{d}{e}{f}{v}=}\qquad\qquad\noeqno
&&\sum_g\;\W{b}{c}{g}{a}{v}\;\W{a}{g}{e}{f}{u}\;
\W{g}{c}{d}{e}{u\!-\!v}\eql{YBE}\eeqa
\setlength{\unitlength}{9mm}\begin{center}
\begin{picture}(4.5,3)\multiput(0.5,1.5)(1,-1){2}{\line(1,1){1}}
\multiput(0.5,1.5)(1,1){2}{\line(1,-1){1}}
\multiput(2.5,0.5)(0,1){3}{\line(1,0){1.5}}
\multiput(2.5,0.5)(1.5,0){2}{\line(0,1){2}}
\put(1.5,1.5){\spos{}{u-v}}\put(3.25,1){\spos{}{u}}
\put(3.25,2){\spos{}{v}}
\put(0.4,1.5){\spos{r}{a}}\put(1.5,0.4){\spos{t}{b}}
\put(2.5,0.4){\spos{t}{b}}\put(4.05,0.45){\spos{tl}{c}}
\put(4.1,1.5){\spos{l}{d}}\put(4.05,2.55){\spos{bl}{e}}
\put(1.5,2.6){\spos{b}{f}}\put(2.5,2.6){\spos{b}{f}}
\put(2.58,1.42){\spos{tl}{g}}
\multiput(1.5,0.5)(0,2){2}{\makebox(1,0){\dotfill}}
\put(2.5,1.5){\spos{}{\bullet}}\end{picture}
\text{1}{3}{0.5}{1.5}{}{=}
\begin{picture}(4.5,3)
\multiput(2,1.5)(1,-1){2}{\line(1,1){1}}
\multiput(2,1.5)(1,1){2}{\line(1,-1){1}}
\multiput(0.5,0.5)(0,1){3}{\line(1,0){1.5}}
\multiput(0.5,0.5)(1.5,0){2}{\line(0,1){2}}
\put(3,1.5){\spos{}{u-v}}\put(1.25,1){\spos{}{v}}
\put(1.25,2){\spos{}{u}}
\put(0.4,1.5){\spos{r}{a}}\put(0.45,0.45){\spos{tr}{b}}
\put(2,0.4){\spos{t}{c}}\put(3,0.4){\spos{t}{c}}
\put(4.1,1.5){\spos{l}{d}}\put(2,2.6){\spos{b}{e}}
\put(3,2.6){\spos{b}{e}}\put(0.45,2.55){\spos{br}{f}}
\put(1.92,1.42){\spos{tr}{g}}
\multiput(2,0.5)(0,2){2}{\makebox(1,0){\dotfill}}
\put(2,1.5){\spos{}{\bullet}}\end{picture}\end{center}
the inversion relation
%
%   Inversion Relation
%
\beq\eql{IR}\sum_e\;\W{a}{b}{e}{d}{u}\;
\W{e}{b}{c}{d}{\!\!-\!u}=\rho(u)\,\rho(-u)\:
\delta_{ac}\:A_{ab}\:A_{ad}\eeq
\setlength{\unitlength}{9mm}\begin{center}
\begin{picture}(5,3)\multiput(0.5,1.5)(3,1){2}{\line(1,-1){1}}
\multiput(0.5,1.5)(3,-1){2}{\line(1,1){1}}\put(1.5,0.5){\line(1,1){2}}
\put(1.5,2.5){\line(1,-1){2}}
\put(1.5,1.5){\spos{}{u}}\put(3.4,1.5){\spos{}{-u}}
\put(1.5,0.4){\spos{t}{b}}\put(3.5,0.4){\spos{t}{b}}
\put(1.5,2.6){\spos{b}{d}}\put(3.5,2.6){\spos{b}{d}}
\put(0.4,1.5){\spos{r}{a}}\put(4.6,1.5){\spos{l}{c}}
\put(2.5,1.35){\spos{t}{e}}
\multiput(1.5,0.5)(0,2){2}{\makebox(2,0){\dotfill}}
\put(2.5,1.5){\spos{}{\bullet}}\end{picture}
\text{1}{3}{0.5}{1.5}{}{=}
\text{5}{3}{2.5}{1.5}{}{\rho(u)\,\rho(-u)\:\delta_{ac}
\:A_{ab}\:A_{ad}}\end{center}
\newpage
\noindent the left reflection equation
%
% Left Reflection Equation
%
\beqa\lefteqn{\rule[-3.5ex]{0ex}{3.5ex}\sum_{f g}\;
\W{f}{a}{b}{c}{u\!-\!v}\;\W{g}{f}{c}{d}{\m\!-\!u\!-\!v}\;
\BL{a}{f}{g}{u}\;\BL{g}{d}{e}{v}=}\qquad\noeqno
&&\sum_{fg}\;\W{f}{c}{d}{e}{u\!-\!v}\;\W{g}{b}{c}{f}{\m\!-\!u\!-\!v}\;
\BL{g}{f}{e}{u}\;\BL{a}{b}{g}{v}\eql{LRE}\eeqa
\nc{\botleftrefnodots}[4]{\begin{picture}(1.5,2)\put(0,0){\line(0,1){2}}
\put(0,1){\line(1,1){0.5}}\put(0,1){\line(1,-1){1}}\put(0,0){\line(1,1){1}}
\put(0,2){\line(1,-1){1.5}}\put(1.5,0.5){\line(-1,-1){0.5}}
\put(0.1,0.5){\spos{l}{#1}}\put(0.1,1.5){\spos{l}{#2}}\put(1,0.5){\spos{}{#3}}
\put(0.5,1){\spos{}{#4}}\end{picture}}
\nc{\topleftrefnodots}[4]{\begin{picture}(1.5,2)\put(0,0){\line(0,1){2}}
\put(0,1){\line(1,1){1}}\put(0,1){\line(1,-1){0.5}}\put(0,0){\line(1,1){1.5}}
\put(0,2){\line(1,-1){1}}\put(1.5,1.5){\line(-1,1){0.5}}
\put(0.1,0.5){\spos{l}{#1}}\put(0.1,1.5){\spos{l}{#2}}\put(0.5,1){\spos{}{#3}}
\put(1,1.5){\spos{}{#4}}\end{picture}}
\setlength{\unitlength}{18mm}\begin{center}
\begin{picture}(2.5,2)\put(0.5,0){\botleftrefnodots{\,u}{\,v}{u-v}{\m-u-v}}
\put(0.47,2){\spos{r}{e}}\put(1.02,1.52){\spos{bl}{d}}
\put(1.52,1.02){\spos{bl}{c}}\put(2.02,0.5){\spos{l}{b}}
\put(1.5,-0.02){\spos{t}{a}}\put(0.47,0){\spos{r}{a}}
\put(0.98,0.43){\spos{t}{f}}\put(0.44,1){\spos{r}{g}}
\put(0.5,0){\makebox(1,0){\dotfill}}
\put(1,0.5){\spos{}{\bullet}}
\put(0.5,1){\spos{}{\bullet}}\end{picture}
\text{0.4}{2}{0.05}{1}{}{=}
\begin{picture}(2.5,2)
\put(0.5,0){\topleftrefnodots{\,v}{\,u}{\m-u-v}{u-v}}
\put(0.47,2){\spos{r}{e}}\put(1.5,2.02){\spos{b}{e}}
\put(2.02,1.5){\spos{l}{d}}\put(1.52,0.98){\spos{tl}{c}}
\put(1.02,0.48){\spos{tl}{b}}\put(0.47,0){\spos{r}{a}}
\put(1.02,1.6){\spos{b}{f}}\put(0.44,1){\spos{r}{g}}
\put(0.5,2){\makebox(1,0){\dotfill}}\put(1,1.5){\spos{}{\bullet}}
\put(0.5,1){\spos{}{\bullet}}\end{picture}\end{center}
the right reflection equation
%
% Right Reflection Equation
%
\beqa\lefteqn{\rule[-3.5ex]{0ex}{3.5ex}\sum_{f
g}\;\W{b}{a}{f}{c}{u\!-\!v}\;
\W{c}{f}{g}{d}{\m\!-\!u\!-\!v}\;\BR{a}{f}{g}{u}\;\BR{g}{d}{e}{v}=}
\qquad\noeqno
&&\sum_{fg}\;\W{d}{c}{f}{e}{u\!-\!v}\;\W{c}{b}{g}{f}{\m\!-\!u\!-\!v}\;
\BR{g}{f}{e}{u}\;\BR{a}{b}{g}{v}\eql{RRE}\eeqa
\nc{\botrightrefnodots}[4]{\begin{picture}(1.5,2)
\put(1.5,0){\line(0,1){2}}
\put(1.5,1){\line(-1,1){0.5}}
\put(1.5,1){\line(-1,-1){1}}\put(1.5,0){\line(-1,1){1}}
\put(1.5,2){\line(-1,-1){1.5}}\put(0,0.5){\line(1,-1){0.5}}
\put(1.4,0.5){\spos{r}{#1}}\put(1.4,1.5){\spos{r}{#2}}
\put(0.5,0.5){\spos{}{#3}}
\put(1,1){\spos{}{#4}}\end{picture}}
\nc{\toprightrefnodots}[4]{\begin{picture}(1.5,2)\put(1.5,0){\line(0,1){2}}
\put(1.5,1){\line(-1,1){1}}\put(1.5,1){\line(-1,-1){0.5}}
\put(1.5,0){\line(-1,1){1.5}}\put(1.5,2){\line(-1,-1){1}}
\put(0,1.5){\line(1,1){0.5}}
\put(1.4,0.5){\spos{r}{#1}}\put(1.4,1.5){\spos{r}{#2}}
\put(1,1){\spos{}{#3}}\put(0.5,1.5){\spos{}{#4}}\end{picture}}
\setlength{\unitlength}{18mm}\begin{center}
\begin{picture}(2.5,2)
\put(0.5,0){\botrightrefnodots{u\,}{v\,}{u-v}{\m-u-v}}
\put(2.03,2){\spos{l}{e}}\put(1.48,1.52){\spos{br}{d}}
\put(0.98,1.02){\spos{br}{c}}\put(0.48,0.5){\spos{r}{b}}
\put(1,-0.02){\spos{t}{a}}\put(2.03,0){\spos{l}{a}}
\put(1.48,0.43){\spos{t}{f}}\put(2.06,1){\spos{l}{g}}
\put(1,0){\makebox(1,0){\dotfill}}\put(1.5,0.5){\spos{}{\bullet}}
\put(2,1){\spos{}{\bullet}}\end{picture}
\text{0.4}{2}{0.2}{0.95}{}{=}
\begin{picture}(2.5,2)\put(0.5,0){\toprightrefnodots{v\,}{u\,}{\m-u-v}{u-v}}
\put(2.03,2){\spos{l}{e}}\put(1,2.02){\spos{b}{e}}
\put(0.48,1.5){\spos{r}{d}}
\put(0.98,0.98){\spos{tr}{c}}\put(1.48,0.48){\spos{tr}{b}}
\put(2.03,0){\spos{l}{a}}\put(1.52,1.6){\spos{b}{f}}
\put(2.06,1){\spos{l}{g}}
\put(1,2){\makebox(1,0){\dotfill}}\put(1.5,1.5){\spos{}{\bullet}}
\put(2,1){\spos{}{\bullet}}\end{picture}\end{center}
the left boundary crossing equation
%
%  Left Boundary Crossing
%
\beq\sum_{d}\;\BL{a}{d}{c}{u}\;\W{d}{a}{b}{c}{2u\!-\!\m}=
-\rho(\m\!-\!2u)\;\BL{a}{b}{c}{\m\!-\!u}\eql{LBC}\eeq
\setlength{\unitlength}{9mm}
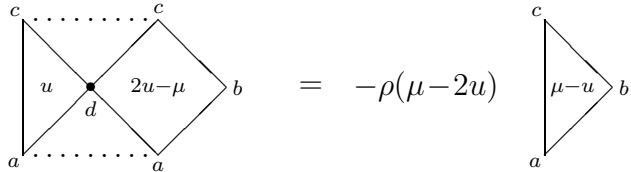
\begin{center}
\begin{picture}(4,3)\put(0.5,0.5){\line(0,1){2}}
\put(0.5,0.5){\line(1,1){2}}\put(0.5,2.5){\line(1,-1){2}}
\put(2.5,0.5){\line(1,1){1}}\put(2.5,2.5){\line(1,-1){1}}
\put(0.7,1.5){\spos{l}{\,u}}\put(2.5,1.5){\spos{}{2u-\m}}
\put(0.45,0.45){\spos{tr}{a}}\put(2.5,0.4){\spos{t}{a}}
\put(3.6,1.5){\spos{l}{b}}\put(0.45,2.55){\spos{br}{c}}
\put(2.5,2.6){\spos{b}{c}}\put(1.5,1.3){\spos{t}{d}}
\multiput(0.5,0.5)(0,2){2}{\makebox(2,0){\dotfill}}
\put(1.5,1.5){\spos{}{\bullet}}\end{picture}
\text{1.2}{3}{0.6}{1.5}{}{=}\text{2.2}{3}{1.1}{1.5}{}{-\rho(\m\!-\!2u)}
\begin{picture}(2,3)\put(0.5,0.5){\line(0,1){2}}
\put(0.5,0.5){\line(1,1){1}}\put(0.5,2.5){\line(1,-1){1}}
\put(0.7,1.5){\spos{l}{\!\!\m-u}}
\put(0.45,0.45){\spos{tr}{a}}\put(1.6,1.5){\spos{l}{b}}
\put(0.45,2.55){\spos{br}{c}}\end{picture}\end{center}
and the right boundary crossing equation
%
%  Right Boundary Crossing
%
\beq\sum_{d}\;\W{b}{a}{d}{c}{2u\!-\!\m}\;\BR{a}{d}{c}{u}=
-\rho(\m\!-\!2u)\;\BR{a}{b}{c}{\m\!-\!u}\eql{RBC}\eeq
\setlength{\unitlength}{9mm}\begin{center}
\begin{picture}(4,3)\put(0.5,1.5){\line(1,-1){1}}
\put(0.5,1.5){\line(1,1){1}}\put(1.5,0.5){\line(1,1){2}}
\put(1.5,2.5){\line(1,-1){2}}\put(3.5,0.5){\line(0,1){2}}
\put(1.5,1.5){\spos{}{2u-\m}}\put(3.3,1.5){\spos{r}{u\,}}
\put(1.5,0.4){\spos{t}{a}}\put(3.55,0.45){\spos{tl}{a}}
\put(0.4,1.5){\spos{r}{b}}\put(1.5,2.6){\spos{b}{c}}
\put(3.55,2.55){\spos{bl}{c}}\put(2.5,1.3){\spos{t}{d}}
\multiput(1.5,0.5)(0,2){2}{\makebox(2,0){\dotfill}}
\put(2.5,1.5){\spos{}{\bullet}}\end{picture}
\text{1.2}{3}{0.6}{1.5}{}{=}
\text{2.2}{3}{1.1}{1.5}{}{-\rho(\m\!-\!2u)}
\begin{picture}(2,3)\put(0.5,1.5){\line(1,-1){1}}
\put(0.5,1.5){\line(1,1){1}}\put(1.5,0.5){\line(0,1){2}}
\put(1.3,1.5){\spos{r}{\m-u\!\!}}
\put(1.55,0.45){\spos{tl}{a}}\put(0.4,1.5){\spos{r}{b}}
\put(1.55,2.55){\spos{bl}{c}}\end{picture}\end{center}
These equations are to be satisfied for all values of the external spins and
all values of the spectral parameters.
The function $\rho$ is model-dependent and $\m$ is the same fixed parameter as
in~\eqref{DRTM}.

We note that these local relations are consistent with the initial condition
\beq\eql{SIC}\W{a}{b}{c}{d}{0}=-\rho(0)\:\delta_{ac}\:A_{ab}\:A_{ad}\eeq
(The minus sign is used here, and in the boundary crossing equations, in order
to provide consistency with subsequent fusion equations.)
More specifically, with this intitial condition we see that~\eqref{YBE} holds
for $u=v$ or $v=0$, that~\eqref{IR} holds
for $u=0$, that~\eqref{LRE} and~\eqref{RRE} hold for $u=v$, and
that~\eqref{LBC} and~\eqref{RBC} hold for $u=\m/2$.
Furthermore, we find that~\eqref{YBE} holds for $u=0$, due to~\eqref{IR},
while~\eqref{LRE} and~\eqref{RRE} hold for $v=\m-u$ due to~\eqref{LBC}
and~\eqref{RBC}. Indeed, the inversion relation can be motivated by the
Yang-Baxter equation together with the initial condition,
while the boundary crossing equations can be motivated by the reflection
equations together with the initial condition and the inversion relation.

It can also be seen that if the face weights satisfy certain reflection and
rotation symmetries, then there are correspondences
between solutions of the reflection equations.  More specifically, if
\[\BL{a}{b}{c}{u}=\BtildeL{a}{b}{c}{u}\quad\mbox{and}\quad
\BR{a}{b}{c}{u}=\BtildeR{a}{b}{c}{u}\]
satisfy~\eqref{LRE} and~\eqref{RRE},
then the symmetry \beq\W{a}{b}{c}{d}{u}=\W{a}{d}{c}{b}{u}
\eql{refsym1}\eeq implies that so too do
\[\BL{a}{b}{c}{u}=\BtildeL{c}{b}{a}{u}\quad\mbox{and}\quad
\BR{a}{b}{c}{u}=\BtildeR{c}{b}{a}{u}\;,\]
the symmetry \beq\W{a}{b}{c}{d}{u}=\W{c}{b}{a}{d}{u}\eql{refsym2}\eeq implies
that so too do
\[\BL{a}{b}{c}{u}=\BtildeR{a}{b}{c}{u}\quad\mbox{and}
\quad\BR{a}{b}{c}{u}=\BtildeL{a}{b}{c}{u}\;,\]
and the symmetry \beq\W{a}{b}{c}{d}{u}=\W{c}{d}{a}{b}{u}
\eql{rotsym}\eeq
(which occurs if \eqref{refsym1} and \eqref{refsym2} both occur) implies that
so too do
\[\BL{a}{b}{c}{u}=\BtildeR{c}{b}{a}{u}\quad\mbox{and}\quad
\BR{a}{b}{c}{u}=\BtildeL{c}{b}{a}{u}\;.\]

\sub{Crossing Symmetry of Double-Row Transfer Matrices}
The double-row transfer matrices satisfy crossing symmetry,
\beq\D(u)=\D(\m\!-\!u)\eeq
We prove this by considering an entry of $\D(u)$, applying the inversion
relation at
an arbitrary point, then using the Yang-Baxter equation $N$ times, and finally
applying both boundary
crossing equations:
%
%  Crossing Symmetry Proof
%
\nc{\equals}{&\text{0.6}{1.2}{0.3}{0.5}{}{=}&}
\nc{\shfac}{\text{1}{1}{0.4}{0.5}{}{\frac{1}{\eta(u)}}}
\setlength{\unitlength}{15mm}
\begin{eqnarray*}
\text{0.6}{1}{0.3}{0.5}{}{\D(u)}\equals\leftlowend{\m-u}
\longlowblock{u}{\m-u}\rightlowend{u\;}\\
\equals\shfac\leftlowend{\m-u}\shortlowblock{u}{\m-u}
\inv{\m-2u}{2u-\m}\shortlowblock{u}{\m-u}\rightlowend{u\;}\\
\equals\shfac\leftbu{\m-u}{\m-2u}\longlowblock{\m-u}{u}
\rightbu{2u-\m}{u\;}\\
\equals\rule[-1.5ex]{0ex}{1.5ex}\leftlowend{\;u}
\longlowblock{\m-u}{u}\rightlowend{\m-u\!}\\
&=&\D(\m\!-\!u)\end{eqnarray*}
where $\eta(u)=\rho(\m\!-\!2u)\rho(2u\!-\m)$.

\sub{Commutation of Double-Row Transfer Matrices}
The double-row transfer matrices form a commuting family,
\beq\D(u)\:\D(v)=\D(v)\:\D(u)\eeq
We prove this by the following steps, in each of which we use either the
inversion relation, the Yang-Baxter equation $N$ times, or the reflection
equations:
%\comment{
%
%
%  Commutation Proof
%
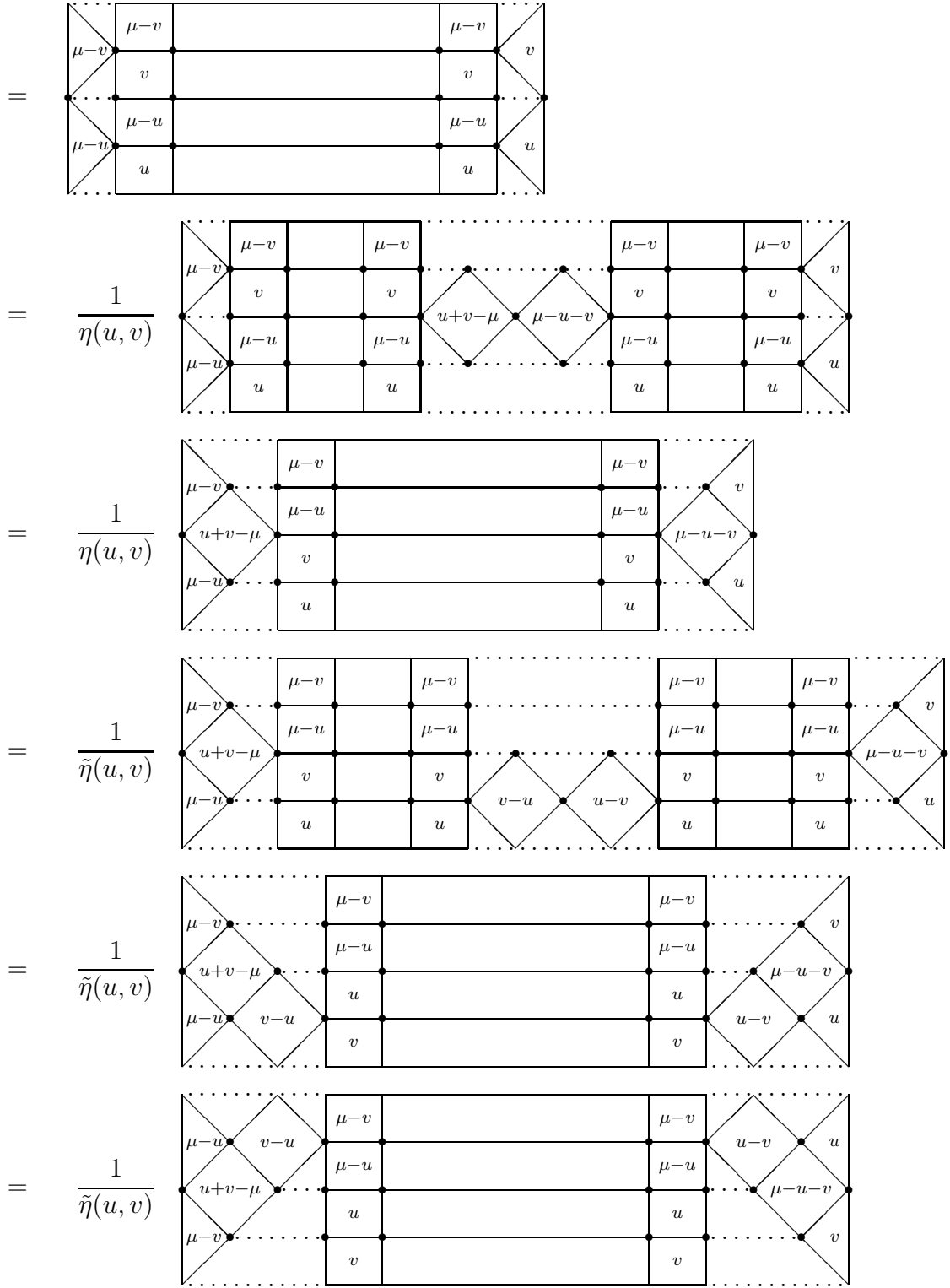
\rnc{\equals}{&\text{0.6}{2.2}{0.3}{1}{}{=}&}
\rnc{\shfac}{\text{1.2}{2}{0.5}{1}{}{\frac{1}{\eta(u,v)}}}
\nc{\lonfac}{\text{1.2}{2}{0.5}{1}{}{\frac{1}{\tilde{\eta}(u,v)}}}
\setlength{\unitlength}{15mm}
\begin{eqnarray*}
\lefteqn{\rule[-1.5ex]{0ex}{1.5ex}\D(u)\:\D(v)}\\
\equals\leftend{\m-u}{\m-v}\longblock{u}{\m-u}{v}{\m-v}
\rightend{u\;}{v\;}\\
\equals\shfac\leftend{\m-u}{\m-v}\shortblock{u}{\m-u}{v}{\m-v}
\midinv{u+v-\m}{\m-u-v}\shortblock{u}{\m-u}{v}{\m-v}
\rightend{u\;}{v\;}\\
\equals\shfac\lefttriangle{\m-u}{\m-v}{u+v-\m}
\longblock{u}{v}{\m-u}{\m-v}\righttriangle{u\;}{v\;}{\m-u-v}\\
\equals\lonfac\lefttriangle{\m-u}{\m-v}{u+v-\m}
\shortblock{u}{v}{\m-u}{\m-v}\lowinv{v-u}{u-v}
\shortblock{u}{v}{\m-u}{\m-v}\righttriangle{u\;}{v\;}{\m-u-v}\\
\equals\lonfac\botleftref{\m-u}{\m-v}{v-u}{u+v-\m}
\longblock{v}{u}{\m-u}{\m-v}\botrightref{u\;}{v\;}{u-v}{\m-u-v}\\
\equals\lonfac\topleftref{\m-v}{\m-u}{u+v-\m}{v-u}
\longblock{v}{u}{\m-u}{\m-v}\toprightref{v\;}{u\;}{\m-u-v}{u-v}
\end{eqnarray*}
\newpage
\begin{eqnarray*}
\equals\lonfac\lefttriangle{\m-v}{\m-u}{u+v-\m}
\shortblock{v}{u}{\m-v}{\m-u}\highinv{v-u}{u-v}
\shortblock{v}{u}{\m-v}{\m-u}
\righttriangle{v\;}{u\;}{\m-u-v}\\
\equals\shfac\lefttriangle{\m-v}{\m-u}{u+v-\m}
\longblock{v}{u}{\m-v}{\m-u}
\righttriangle{v\;}{u\;}{\m-u-v}\\
\equals\shfac\leftend{\m-v}{\m-u}
\shortblock{v}{\m-v}{u}{\m-u}\midinv{u+v-\m}{\m-u-v}
\shortblock{v}{\m-v}{u}{\m-u}\rightend{v\;}{u\;}\\
\equals\rule[-1.5ex]{0ex}{1.5ex}\leftend{\m-v}{\m-u}
\longblock{v}{\m-v}{u}{\m-u}\rightend{v\;}{u\;}\\
&=&\D(v)\:\D(u)\end{eqnarray*}
%
%
%
%}
where $\eta(u,v)=\rho(u\!+\!v\!-\!\m)\rho(\m\!-\!u\!-\!v)$ and
$\tilde{\eta}(u,v)=
\rho(v\!-\!u)\rho(u\!-\!v)\rho(u\!+\!v\!-\!\m)\rho(\m\!-\!u\!-\!v)$.

\sub{ABF Models}
We now consider the particular case of Andrews-Baxter-Forrester (ABF)
restricted solid-on-solid models~\cite{AndBaxFor84}.
There is one such model for each integer $L\geq3$. The spins---sometimes known
also as heights---in the model labelled by $L$ take the values
\[a\in\{1,2,\ldots,L\}\]
and adjacent spins must differ by $1$,
\beq A_{ab}=\delta_{a,b\!-\!1}+\delta_{a,b\!+\!1}\eql{ABFAC}\eeq
For the model labelled by $L$, there is a fixed crossing parameter
\beq\l=\frac{\pi}{L+1}\eql{lambda}\eeq
and the non-zero face weights are given by
\beqa\rule[-3.5ex]{0ex}{3.5ex}
\W{a}{a\!\mp\!1}{a}{a\!\pm\!1}{u}&=&
\frac{\t(\l\!-\!u)}{\t(\l)}\noeqno
\rule[-3.5ex]{0ex}{3.5ex}\W{a\!\mp\!1}{a}{a\!\pm\!1}{a}{u}&=&
\frac{\sqrt{\t((a\!-\!1)\l)\,\t((a\!+\!1)\l)}}{\t(a\l)}\;
\frac{\t(u)}{\t(\l)}\eql{FW}\\
\W{a\!\pm\!1}{a}{a\!\pm\!1}{a}{u}&=&\frac{\t(a\l\!\pm\!u)}{\t(a\l)}
\nonumber\eeqa
Here $\t$ is a standard elliptic theta-$1$ function of fixed nome
$\hat{q}$, with $-1<\hat{q}^2<1$,
\beq\t(u)=\t_1(u,\hat{q})=2\hat{q}^{1/4}\sin u\:\prod_{n=1}^{\infty}
\left(1-2\hat{q}^{2n}\cos2u+\hat{q}^{4n}\right)\left(1-\hat{q}^{2n}
\right)\eql{theta}\eeq
At criticality, $\hat{q}=0$ and we can take $\t(u)=\sin u$.
The main properties of $\t$ which we shall use are that it is odd
\beq\t(u)=-\t(-u)\eeq
that it is periodic
\beq\t(u)=-\t(u\!+\!\pi)\eeq
and that it satisfies the identity
\beq\begin{array}{c}\t(s+x)\;\t(s-x)\;\t(t+y)\;\t(t-y)-\t(s+y)\;
\t(s-y)\;\t(t+x)\;\t(t-x)\\
\qquad\qquad\qquad=\;\t(s+t)\;\t(s-t)\;\t(x+y)\;\t(x-y)\end{array}
\eql{thetaid}\eeq
It can be seen that the ABF face weights satisfy various simple relations:
reflection and rotation symmetries,~\eqref{refsym1}--\eqref{rotsym},
\beq\W{a}{b}{c}{d}{u}=\W{c}{b}{a}{d}{u}=\W{a}{d}{c}{b}{u}=
\W{c}{d}{a}{b}{u}\eeq
crossing symmetry
\beq\eql{CS}\W{a}{b}{c}{d}{u}=\sqrt{\frac{\t(a\l)\,\t(c\l)}{\t(b\l)\,\t(d\l)}}
\;\W{d}{c}{b}{a}{\l\!-\!u}\eeq
full height reversal symmetry
\beq\eql{FHRS}\W{a}{b}{c}{d}{u}=
\W{L\!+\!1\!-\!a}{L\!+\!1\!-\!b}{L\!+\!1\!-\!c}{L\!+\!1\!-\!d}{u}
\eeq
and the initial condition
\beq\eql{IC}\W{a}{b}{c}{d}{0}=\delta_{ac}\:A_{ab}\:A_{ad}\eeq
It is well-known that, essentially due to~\eqref{thetaid},
the face weights also satisfy the Yang-Baxter equation,~\eqref{YBE},
and the inversion
relation,~\eqref{IR}, with the function $\rho$ given by
\beq\eql{rho}\rho(u)=\frac{\t(u\!-\!\l)}{\t(\l)}\eeq
We now define, as the only non-zero ABF boundary weights,
\beqa\rule[-3.5ex]{0ex}{3.5ex}\BL{a}{a\!\mp\!1}{a}{u}&=&
\sqrt{\frac{\t((a\!\mp\!1)\l)}{\t(a\l)}}\;\;\frac{\t(u\!+\!\lm\!\mp\!\xiL(a))
\,\t(u\!\pm\!a\l\!+\!\lm\!\pm\!\xiL(a))}{\t(\l)^2}\noeqno
\eql{BW}\\
\BR{a}{a\!\mp\!1}{a}{u}&=&
\sqrt{\frac{\t((a\!\mp\!1)\l)}{\t(a\l)}}\;\;\frac{\t(u\!+\!\lm\!\mp\!\xiR(a))
\,\t(u\!\pm\!a\l\!+\!\lm\!\pm\!\xiR(a))}{\t(\l)^2}\nonumber\eeqa
where $\xiL(a)$ and $\xiR(a)$ are arbitrary parameters.
In Appendix A, we show that the reflection equations,~\eqref{LRE}
and~\eqref{RRE}, are satisfied by the ABF face and boundary weights.
We also show that the ABF weights, together with $\rho$ given
by~\eqref{rho},
satisfy the boundary crossing equations,~\eqref{LBC} and~\eqref{RBC}.

If $\xiL$ and $\xiR$ satisfy
\beq\eql{xicond}\xiL(L\!+\!1\!-\!a)=-\xiL(a)\,,\qquad\xiR(L\!+\!1\!-\!a)
=-\xiR(a)\eeq
then the ABF boundary weights satisfy full height reversal symmetry
\beqa\BL{a}{b}{c}{u}&=&-
\BL{L\!+\!1\!-\!a}{L\!+\!1\!-\!b}{L\!+\!1\!-\!c}{u}\noeqno
\\
\BR{a}{b}{c}{u}&=&-
\BR{L\!+\!1\!-\!a}{L\!+\!1\!-\!b}{L\!+\!1\!-\!c}{u}
\nonumber\eeqa
The boundary weights~\eqref{BW} have the diagonal form
\beq\BL{a}{b}{c}{u}=\BL{a}{b}{c}{u}\:\delta_{ac}\,,
\qquad\BR{a}{b}{c}{u}=\BR{a}{b}{c}{u}\:\delta_{ac}\eeq
This form implies equality of the boundary spins at
 each end of the double-row $\D(u)$ (and hence at each end of the entire
lattice).
It is convenient to regard these spins as labels for the fixed boundaries and
only the internal spins as matrix indices. We therefore
define the ABF double-row transfer matrix with fixed left and right boundary
spins $\aL$ and $\aR$, $\D(\aL\aR|u)$, by
\beq\langle
a_2,\ldots,a_\N\,|\,\D(\aL\aR|u)\,|\,b_2,\ldots,b_\N\rangle=
\langle
\aL,a_2,\ldots,a_\N,\aR\,|\,\D(u)\,|\,\aL,b_2,\ldots,b_\N,\aR\rangle
\eeq
It is natural in these models to take $\m$ as \beq\m=\l\eeq
With this choice, crossing symmetry of the face weights,~\eqref{CS}, implies
that $\D(\aL\aR|u)$ is symmetric
\beq\D(\aL\aR|u)=\D(\aL\aR|u)^t\eeq
Ultimately, we will be interested in the isotropic point, $u=\l/2$,
 at which we now show it is possible to achieve a completely homogeneous
lattice,
with pure, fixed boundary conditions.  If we set $\xiL(\aL)=\pm\l/2$,
$\xiR(\aR)=\pm\l/2$ and $\m=\l$, then
\[\BL{\aL}{\aL\!\mp\!1}{\aL}{\l/2}=
\BR{\aR}{\aR\!\mp\!1}{\aR}{\l/2}=0\]
so that the transfer matrix $\D(\aL\aR|\l/2)$ is simply proportional to the
matrix product of two rows of face weights, all with spectral parameter $\l/2$,
with the three spins on the left boundary fixed to $\aL$,~$\aL\!\pm\!1$,~$\aL$
and the three spins on the right boundary fixed to
$\aR$,~$\aR\!\pm\!1$,~$\aR$.  Similarly, if we set  $\xiL(\aL)=\pm\l/2$ and
$\xiR(\aR)=\mp\l/2$,
then $\D(\aL\aR|\l/2)$ has the spins on the left boundary fixed to
$\aL$,~$\aL\!\pm\!1$,~$\aL$ and the spins on the right boundary fixed to
$\aR$,~$\aR\!\mp\!1$,~$\aR$.

\sect{Fusion of IRF Models with Fixed Boundary\protect\\Conditions}
\sub{General Formalism}
We now extend our formalism to cover models which have a fusion hierarchy.
For these models there is a discrete set of fusion levels,
and we assume that each of these is labelled by a single
integer, with the original, unfused model corresponding to fusion level~$1$.

The fused face weights, $W^{pq}$, are associated with two fusion levels---a
horizontal level,~$p$, and a
vertical level, $q$---and the fused boundary weights, $K^q$, are associated
with one fusion level,~$q$.
There is now an adjacency matrix, $A^q$, for each fusion level~$q$, with the
adjacency conditions on the fused weights being
\beqa\rule[-3.5ex]{0ex}{3.5ex}\Wf{pq}{a}{b}{c}{d}{u}&=&A^p_{ab}\:A^q_{bc}
\:A^p_{cd}\:A^q_{da}\;\Wf{pq}{a}{b}{c}{d}{u}\eql{genAC}\\
\rule[-3.5ex]{0ex}{3.5ex}\BLf{q}{a}{b}{c}{u}&=&A^q_{ab}\:A^q_{bc}\;
\BLf{q}{a}{b}{c}{u}\\
\BRf{q}{a}{b}{c}{u}&=&A^q_{ab}\:A^q_{bc}\;\BRf{q}{a}{b}{c}{u}\eeqa
The fused double-row transfer matrices are also associated with two fusion
levels, and are defined by
\newpage
\beqa\rule[-1.5ex]{0ex}{1.5ex}\lefteqn{\langle
a_1,\ldots,a_{\N\!+\!1}|
\D^{pq}\!(u)|b_1,\ldots,b_{\N\!+\!1}\rangle\;=}&&\noeqno
&&\rule[-3.5ex]{0ex}{3.5ex}\sum_{c_1\ldots c_{N\!+\!1}}\;
\BLf{q}{a_1}{c_1}{b_1}{\!-\!u\!-\!(q\!-\!1)\l\!+\!\m}\;\times\eql{genDRTM}\\
&&\qquad\qquad\left[\prod_{j=1}^N\;
\Wf{pq}{a_j}{a_{j\!+\!1}}{c_{j\!+\!1}}{c_j}{u}\:
\Wf{pq}{c_j}{c_{j\!+\!1}}{b_{j\!+\!1}}{b_j}{\!\!-\!u\!-\!(q\!-\!1)\l\!+\!\m}
\right]\;
\BRf{q}{a_{\N\!+\!1}}{c_{\N\!+\!1}}{b_{\N\!+\!1}}{u}\nonumber
\eeqa
where $\l$ and $\m$ are arbitrary fixed parameters.
In this generalised framework, the fused Yang-Baxter equation is
\beqa\lefteqn{\rule[-3.5ex]{0ex}{3.5ex}\sum_g\;
\Wf{rq}{a}{b}{g}{f}{u\!-\!v}\;
\Wf{pq}{b}{c}{d}{g}{u}\;\Wf{pr}{g}{d}{e}{f}{v}=}
\qquad\qquad\qquad\eql{genYBE}\\
&&\sum_g\;\Wf{pr}{b}{c}{g}{a}{v}\;\Wf{pq}{a}{g}{e}{f}{u}
\;\Wf{rq}{g}{c}{d}{e}{u\!-\!v}\nonumber\eeqa
the fused inversion relation is
\beq\sum_e\;\Wf{qr}{a}{b}{e}{d}{u}\;
\Wf{rq}{e}{b}{c}{d}{\!\!-\!u}
=\rho^{qr}\!(u)\:\rho^{rq}\!(\!-u)\:\delta_{ac}\:A^q_{ab}
\:A^r_{ad}\eql{genIR}\eeq
the fused left reflection equation is
\begin{flushleft}$\rho^{rq}\!(u\!-\!v)\:
\rho^{rq}\!(\!-\!u\!-\!v\!-\!(q\!-\!1)\l\!+\!\m)\;
\times$\end{flushleft}
$$\sum_{fg}\:\Wf{qr}{f}{a}{b}{c}{u\!-\!v+\!(q\!-\!r)\l}\:
\Wf{qr}{g}{f}{c}{d}{\!\!-\!u\!-\!v\!-\!(r\!-\!1)\l\!+\!\m}\:
\BLf{q}{a}{f}{g}{u}\:\BLf{r}{g}{d}{e}{v}$$
\begin{flushleft}$=\quad\rho^{qr}\!(u\!-\!v\!+\!(q\!-\!r)\l)\;
\rho^{qr}\!(\!-\!u\!-\!v\!-\!(r\!-\!1)\l\!+\!\m)\;\times$
\end{flushleft}
\beq\sum_{fg}\;\Wf{rq}{f}{c}{d}{e}{u\!-\!v}\;
\Wf{rq}{g}{b}{c}{f}{\!\!-\!u\!-\!v\!-\!(q\!-\!1)\l\!+\!\m}\;
\BLf{q}{g}{f}{e}{u}\;\BLf{r}{a}{b}{g}{v}\eql{genLRE}\eeq
the fused right reflection equation is
\begin{flushleft}$\rho^{qr}\!(u\!-\!v\!+\!(q\!-\!r)\l)\:
\rho^{qr}\!(\!-\!u\!-\!v\!-\!(r\!-\!1)\l\!+\!\m)\;\times$
\end{flushleft}
\beq\sum_{fg}\;\Wf{rq}{b}{a}{f}{c}{u\!-\!v}\;
\Wf{rq}{c}{f}{g}{d}{\!\!-\!u\!-\!v\!-\!(q\!-\!1)\l\!+\!\m}\;
\BRf{q}{a}{f}{g}{u}\;\BRf{r}{g}{d}{e}{v}\eql{genRRE}\eeq
\begin{flushleft}$=\quad\rho^{rq}\!(u\!-\!v)\;
\rho^{rq}\!(\!-\!u\!-\!v\!-\!(q\!-\!1)\l\!+\!\m)\;\times$
\end{flushleft}
$$\sum_{fg}\;\Wf{qr}{d}{c}{f}{e}{u\!-\!v\!+\!(q\!-\!r)\l}\;
\Wf{qr}{c}{b}{g}{f}{\!\!-\!u\!-\!v\!-\!(r\!-\!1)\l\!+\!\m}\;
\BRf{q}{g}{f}{e}{u}\;\BRf{r}{a}{b}{g}{v}$$
\newpage
\noindent the fused left boundary crossing equation is
\beqa\lefteqn{\sum_{d}\;\BLf{q}{a}{d}{c}{u}\;
\Wf{qq}{d}{a}{b}{c}{2u\!+\!(q\!-\!1)\l\!-\!\m}=}
\qquad\qquad\qquad\qquad\eql{genLBC}\\
&&(-1)^q\;\rho^{qq}(\!-\!2u\!-\!(q\!-\!1)\l\!+\!\m)\;
\BLf{q}{a}{b}{c}{\!-\!u\!-\!(q\!-\!1)\l\!+\!\m}\nonumber
\eeqa
and the fused right boundary crossing equation is
\beqa\lefteqn{\sum_{d}\;\Wf{qq}{b}{a}{d}{c}{2u\!+\!(q\!-\!1)\l\!-\!\m}
\;\BRf{q}{a}{d}{c}{u}=}\qquad\qquad\qquad\qquad\eql{genRBC}\\
&&(-1)^q\;\rho^{qq}(\!-\!2u\!-\!(q\!-\!1)\l\!+\!\m)\;
\BRf{q}{a}{b}{c}{\!-\!u\!-\!(q\!-\!1)\l\!+\!\m}\nonumber
\eeqa
where $\rho^{rq}$ are model-dependent functions.
The fused local relations are consistent with the fused initial condition
\beq\Wf{qq}{a}{b}{c}{d}{0}=(-1)^q\;\rho^{qq}(0)\;\delta_{ac}\;A^q_{ab}
\;A^q_{ad}\eeq
There are also correspondences between solutions of the fused reflection
equations if the fused face weights satisfy certain reflection and rotation
symmetries.
It can be seen that the fused adjacency conditions, double row transfer matrix,
local relations and initial condition reduce
to~\eqref{AC} and~\eqref{DRTM}--\eqref{SIC} for $p=q=r=1$.

By following a parallel sequence of steps to those of Section~2.3, but now
including the fusion levels $p$ and $q$, we can show that the fused inversion
relation and boundary crossing equations,~\eqref{genIR}, \eqref{genLBC} and
\eqref{genRBC}, imply that the fused double-row transfer matrices satisfy
crossing symmetry
\beq\eql{fusCS}\D^{pq}\!(u)=\D^{pq}\!(\!-\!u\!-\!(q\!-\!1)\l\!+\!\m)\eeq
Similarly, by following a parallel sequence of steps to those of Section~2.4,
we can show that the
fused Yang-Baxter equation, inversion relation, and reflection
equations,~\eqref{genYBE}--\eqref{genRRE}, imply that the fused double-row
transfer matrices form a commuting family
\beq\eql{fuscomm}\D^{pq}\!(u)\:\D^{pr}\!(v)=\D^{pr}\!(v)\:\D^{pq}\!(u)\eeq

\sub{ABF Models}
\subsub{Adjacency Conditions}
We now return to the case of ABF models and consider their fusion
hierarchy~\cite{DatJimMiwOka86,DatJimMiwOka87a,DatJimKunMiwOka88}.
These models are related to the Lie algebra $su(2)$, or more specifically
$A^{(1)}_1$.
The original ABF models are associated with the spin-$\frac{1}{2}$
representation
of $su(2)$ and the higher fusion levels are associated with higher-spin
representations of $su(2)$.

For each $L$, we have $L+2$ fusion levels, labelled $-1,0,\ldots,L$.
The level~$q$ adjacency matrix, $A^q$, is defined by the condition that $a$ and
$b$ are adjacent if and only if
\newpage
\beq a-b\:\in\:\{-q,\:-q\!+\!2,\:\ldots\:,\:q\!-\!2,\:q\}\eql{AC1}\eeq
and
\beq
a+b\:\in\:\{q\!+\!2,\:q\!+\!4,\:\ldots\:,\:2L\!-
\!q\!-\!2,\:2L\!-\!q\}\eql{AC2}\eeq
It can be seen that
\beq A^{-1}=0\;,\qquad A^{0}=I\;,\qquad A^{1}=A\;,\qquad
A^{L\!-\!2}=AY\;,
\qquad A^{L\!-\!1}=Y\;,\qquad A^{L}=0\eql{A}\eeq
where $I$ is the $L\times L$ identity matrix, $A$ is given
by~\eqref{ABFAC},
 and $Y$ is the $L\times L$ height reversal matrix
\beq Y_{ab}=\delta_{{\scriptscriptstyle L+1}\!-a\,,\,b}\eeq
It can be shown that the fused adjacency matrices satisfy full height reversal
symmetry
\beq A^q=YA^q\:Y\eeq
partial height reversal symmetry
\beq A^q=YA^{L-1-q}\eeq
and the $su(2)$ fusion rules
\beq A^qA\:=\:A^{q\!-\!1}+A^{q\!+\!1}\;,\qquad0\leq q\leq L\!-\!1
\eql{fusrule1}\eeq
\beq(A^q)^2\:=\:I+A^{q\!-\!1}\,A^{q\!+\!1}\;,\qquad0\leq q\leq L\!-\!1
\eql{fusrule2}\eeq
\beq(\tilde{A}^q)^2\:=\:(I+\tilde{A}^{q\!-\!1})\:(I+\tilde{A}^{q\!+\!1})
\;,\qquad1\leq q\leq L\!-\!2\eql{fusrule3}\eeq
where
\beq\tilde{A}^q=A^{q\!-\!1}\:A^{q\!+\!1}\;,\qquad0\leq q\leq L\!-\!1\eeq
\rnc{\thefootnote}{\fnsymbol{footnote}}\setcounter{footnote}{1}
For what follows, it is useful to define a set, $P^q_{ab}$, of
$q\!-\!1$-point
 paths between $a$ and $b$, as\footnote{It can be shown that, for $A^q_{ab}=1$,
the number of paths in $P^q_{ab}$ is
 $\displaystyle(A)^q{}_{ab}=\frac{q\,!}{\left(\!\frac{q+a-b}{2}\!\right)
\!\raisebox{-1ex}{\Large!}
\left(\!\frac{q+b-a}{2}\!\right)\!\raisebox{-1ex}{\Large!}}$}
\beq P^q_{ab}=\left\{\begin{array}{@{}c@{}l@{}}
\rule[-3ex]{0ex}{3ex}\{1,\ldots,L\}^{q\!-\!1}&,\,A^q_{ab}=0\\
\Big\{(c_1,\ldots,c_{q\!-\!1})\in\{1,\ldots,L\}^{q\!-\!1}\,
\Big|\,A_{ac_1}\:A_{c_1c_2}\,\ldots\,A_{c_{\!q\!-\!2}c_{\!q\!-\!1}}
\:A_{c_{\!q\!-\!1}b}=1\Big\}
&,\,A^q_{ab}=1\end{array}\right.\eeq

\subsub{Face and Boundary Weights}
We now define ABF fused face weights, $W^{pq}$, and fused boundary weights,
$K^q$.
These definitions will involve the fusion normalisation function
\beq\T{q}{k}{u}\quad=\quad\frac{\;\displaystyle\prod_{j=0}^{q\!-\!1}
\t(u\!+\!k\l\!-\!j\l)\;}{\rule{0ex}{2ex}\displaystyle\t(\l)^q}\eeq
and the fusion gauge factors
\beq
X^q_{ab}\quad=\quad\left\{\begin{array}{cl}\rule[-5ex]{0ex}{5ex}1&,\;
\;A^q_{ab}=0\\
\frac{\;\displaystyle\prod_{j=
\frac{a+b-q}{2}}^{\frac{a+b+q}{2}}\!\!\!\t(j\l)\;\;
\prod_{j=2}^{\frac{a-b+q}{2}}\!\!\!\t(j\l)\;\;
\prod_{j=2}^{\frac{b-a+q}{2}}\!\!\!\t(j\l)\;}
{\rule{0ex}{3ex}\displaystyle
\t(\l)^{2q\!-\!1}}&,\;\;A^q_{ab}=1\end{array}\right.
\eql{X}\eeq
and
\beq\rule[-4ex]{0ex}{4ex}G^q_{\!a_0,a_1,\ldots,a_{q\!-\!1},a_q}
\quad=\quad X^q_{a_0a_q}\;\;
\frac{\;\displaystyle\t(\l)^{q\!+\!1}}{\;\displaystyle
\prod_{j=0}^q\t(a_j\l)\;}\eeq
where, as before, $\l$ is given by~\eqref{lambda} and $\t$ is given
by~\eqref{theta}.
Throughout this section, a product $\displaystyle\prod_{j=j'}^{j''}P(j)$ is
taken to be $1$ if $j''<j'$.

For weights involving fusion level~$-1$, we must, in order to satisfy the
adjacency condition, define
\beq\Wf{p,-1}{a}{b}{c}{d}{u}=
\Wf{-1,q}{a}{b}{c}{d}{u}=\BLf{-1}{a}{b}{c}{u}=
\BRf{-1}{a}{b}{c}{u}=0\eql{fus-1wt}\eeq
For weights involving fusion level~$0$, we define
\beqa\rule[-3.5ex]{0ex}{3.5ex}\Wf{p,0}{a}{b}{c}{d}{u}
&=&\T{p}{-\!1}{u}\:\delta_{ad}\:\delta_{bc}\:A^p_{ab}\noeqno
\Wf{0,q}{a}{b}{c}{d}{u}&=&\delta_{ab}\:\delta_{cd}\:A^q_{ad}
\eql{fus0wt}\eeqa
$$\BLf{0}{a}{b}{c}{u}=\BRf{0}{a}{b}{c}{u}=\delta_{ab}\:\delta_{bc}$$
\rule{0ex}{2.5ex}For fusion level~$1$, the non-zero ABF weights are defined as
\beqa\rule[-3.5ex]{0ex}{3.5ex}\Wf{11}{a}{a\!\mp\!1}{a}{a\!\pm\!1}{u}
&=&\frac{\t(\l\!-\!u)}{\t(\l)}\noeqno
\rule[-3.5ex]{0ex}{3.5ex}\Wf{11}{a\!\mp\!1}{a}{a\!\pm\!1}{a}{u}&=&-
\;\frac{\t((a\!\pm\!1)\l)\:\t(u)}{\t(a\l)\:\t(\l)}\noeqno
\rule[-3.5ex]{0ex}{3.5ex}\Wf{11}{a\!\pm\!1}{a}{a\!\pm\!1}{a}{u}
&=&\frac{\t(a\l\!\pm\!u)}{\t(a\l)}\eql{fus11wt}\\
\rule[-3.5ex]{0ex}{3.5ex}\BLf{1}{a}{a\!\mp\!1}{a}{u}
&=&\mp\;\frac{\t((a\!\mp\!1)\l)\:\t(u\!+\!\lm\!\mp\!
\xiL(a))\:\t(u\!\pm\!a\l\!+\!\lm\!\pm\!\xiL(a))}{\t(\l)^3}
\noeqno
\rule[-3.5ex]{0ex}{3.5ex}\BRf{1}{a}{a\!\mp\!1}{a}{u}&=&\mp\;
\frac{\t(u\!+\!\lm\!\mp\!\xiR(a))\:\t(u\!\pm\!a\l\!+\!\lm\!
\pm\!\xiR(a))}{\t(a\l)\:\t(\l)}\nonumber\eeqa
where, as before, $\xiL(a)$ and $\xiR(a)$ are arbitrary constants.
These weights are related to the standard ABF weights,~\eqref{FW}
and~\eqref{BW}, by the gauge transformation
\beqa\rule[-3.5ex]{0ex}{3.5ex}\Wf{11}{a}{b}{c}{d}{u}&=&\epsilon_a\;
\epsilon_c\;\sqrt{\frac{\t(c\l)}{\t(a\l)}}\;\:\W{a}{b}{c}{d}{u}\noeqno
\rule[-3.5ex]{0ex}{3.5ex}\BLf{1}{a}{b}{c}{u}&=&\epsilon_{a\!+\!1}\;
\epsilon_b\;\frac{\sqrt{\t(a\l)\:\t(b\l)}}{\t(\l)}\;\:\BL{a}{b}{c}{u}
\eql{gauge}\\
\BRf{1}{a}{b}{c}{u}&=&\epsilon_{a\!+\!1}\;\epsilon_b\;
\frac{\t(\l)}{\sqrt{\t(a\l)\:\t(b\l)}}\;\:\BR{a}{b}{c}{u}
\nonumber\eeqa
where $\epsilon_a$ are factors whose required properties
are\setcounter{footnote}{1}\footnote{For example, one of the (four possible)
choices for $\epsilon$ is
$\epsilon_a=\left\{\begin{array}{@{}r@{}l@{}}1&,\;a=0\mbox{ or }1\mbox{ (mod }
4)\\-1&,\;a=2\mbox{ or }3\mbox{ (mod }4)\end{array}\right.$}
\beq(\epsilon_a)^2=1\;,\qquad\qquad\epsilon_a\;\epsilon_{a\!+\!2}=-1\eeq
The Yang-Baxter, inversion, reflection and boundary crossing
equations,~\eqref{YBE}--\eqref{RBC},
are still satisfied by these level~$1$ weights since the gauge factors
corresponding to the internal spins of these equations cancel,
while the gauge factors corresponding to the external spins are the same on
both sides of each equation.  However, the face weights no longer
satisfy the reflection symmetry~\eqref{refsym2}.
We note that the face and boundary weights which appear
in all subsequent diagrams are the level~$1$ ABF weights of~\eqref{fus11wt}.

We now proceed to ABF weights involving higher fusion levels, which are defined
in terms of sums of products of the level~$1$ weights of~\eqref{fus11wt} as
follows:
%
%  Fused Face Weight
%
\setlength{\unitlength}{11mm}\beq
\raisebox{
-3.5\unitlength}[3.5\unitlength][
3.5\unitlength]{\text{0}{7}{0}{6}{tl}{\Wf{pq}{a}{b}{c}{d}{u}\quad
=}\text{6.3}{7}{6.3}{3.2}{r}{\frac{A^p_{ab}\:A^q_{ad}}{\displaystyle
\prod_{j=0}^{q\!-\!2}\T{p}{j}{u}}\;
\sum_{e_1\ldots e_{p\!-\!1}}\;\sum_{h_1\ldots h_{\!q\!
-\!1}}}\begin{picture}(7,7)
\multiput(1,1)(1,0){2}{\line(0,1){5}}
\multiput(4,1)(1,0){3}{\line(0,1){5}}
\multiput(1,1)(0,1){3}{\line(1,0){5}}
\multiput(1,5)(0,1){2}{\line(1,0){5}}
\put(1.95,1.39){\spos{r}{(p\!-\!1)\l}}
\put(1.06,1.66){\spos{l}{u\!-}}
\put(4.5,1.5){\spos{}{u\!-\!\l}}
\put(5.5,1.5){\spos{}{u}}
\put(1.95,2.39){\spos{r}{(p\!-\!2)\l}}
\put(1.06,2.66){\spos{l}{u\!-}}
\put(4.5,2.5){\spos{}{u}}\put(5.5,2.5){\spos{}{u\!+\!\l}}
\put(1.95,5.39){\spos{r}{(q\!-\!p)\l}}
\put(1.06,5.66){\spos{l}{u\!+}}
\put(4.95,5.39){\spos{r}{(q\!-\!2)\l}}
\put(4.06,5.66){\spos{l}{u\!+}}
\put(5.95,5.39){\spos{r}{(q\!-\!1)\l}}
\put(5.06,5.66){\spos{l}{u\!+}}
\put(0.95,0.9){\pos{tr}{a}}
\put(6.1,0.9){\pos{tl}{b}}
\put(6.1,6.1){\pos{bl}{c}}
\put(0.9,6.1){\pos{br}{d}}
\put(2,0.92){\spos{t}{e_1}}
\put(4,0.92){\spos{t}{e_{p\!-\!2}}}
\put(5,0.92){\spos{t}{e_{p\!-\!1}}}
\put(6.08,2){\spos{l}{f_1}}\put(6.08,3){\spos{l}{f_2}}
\put(6.08,5){\spos{l}{f_{\!q\!-\!1}}}
\put(5,6.12){\spos{b}{g_1}}\put(4,6.12){\spos{b}{g_2}}
\put(2,6.12){\spos{b}{g_{p\!-\!1}}}
\put(0.94,5){\spos{r}{h_1}}\put(0.7,3){\spos{b}{h_{q\!-\!2}}}
\put(0.71,2){\spos{b}{h_{q\!-\!1}}}
\multiput(2,1)(0,1){3}{\spos{}{\bullet}}
\multiput(4,1)(0,1){3}{\spos{}{\bullet}}
\multiput(5,1)(0,1){3}{\spos{}{\bullet}}
\multiput(1,2)(0,1){2}{\spos{}{\bullet}}
\multiput(1,5)(1,0){2}{\spos{}{\bullet}}
\multiput(4,5)(1,0){2}{\spos{}{\bullet}}
\end{picture}}\eql{fusFW}\eeq
where $(f_1,\ldots,f_{\!q\!-\!1})\in P^q_{bc}$ and
$(g_1,\ldots,g_{p\!-\!1})\in P^p_{cd}$,
%
%  Fused Left Boundary Weight
%
\setlength{\unitlength}{7.3mm}\beq\BLf{q}{a}{b}{c}{u}\quad=
\quad\frac{\delta_{ac}\;A^q_{ab}}{\;\displaystyle
\prod_{j=0}^{q\!-\!2}\T{j\!+\!1}{2j\!+\!1\;}{2u\!-\!\m}\;}\;\;
\sum_{c_1\ldots c_{q\!-\!1}}\quad\raisebox{
-7\unitlength}[
7\unitlength][7\unitlength]{\begin{picture}(8,14)
\put(7,7){\line(-1,1){6}}
\put(6,6){\line(-1,1){5}}
\put(5,5){\line(-1,1){4}}\put(3,3){\line(-1,1){2}}
\put(2,2){\line(-1,1){1}}
\put(7,7){\line(-1,-1){6}}\put(6,8){\line(-1,-1){5}}
\put(4,10){\line(-1,-1){3}}\put(3,11){\line(-1,-1){2}}
\put(2,12){\line(-1,-1){1}}
\put(1,13){\line(0,-1){8.7}}\put(1,3.7){\line(0,-1){1.4}}
\put(1,1.7){\line(0,-1){0.7}}
\put(1.23,12){\spos{l}{u}}\put(1.06,10){\spos{l}{u\!+\!\l}}
\put(1.02,8.02){\spos{l}{u\!+\!2\l}}
\put(1.95,11){\spos{}{-\!2u\!-\!\l\!+\!\m}}
\put(1.95,9){\spos{}{-\!2u\!-\!3\l\!+\!\m}}
\put(2.95,10){\spos{}{-\!2u\!-\!2\l\!+\!\m}}
\put(5.7,7.3){\spos{}{-\!2u}}
\put(5.91,6.9){\spos{}{-\!(q\!-\!1)\l}}
\put(6.1,6.5){\spos{}{+\!\m}}
\put(4.95,6){\spos{}{-\!2u\!-\!q\l\!+\!\m}}
\put(1.7,3.3){\spos{}{-\!2u}}
\put(1.94,2.9){\spos{}{-\!(2q\!-\!3)\l}}
\put(2.15,2.5){\spos{}{+\!\m}}
\put(1.8,4){\spos{r}{u\!+\!(q\!-\!2)\l}}
\put(1.8,2){\spos{r}{u\!+\!(q\!-\!1)\l}}
\put(7.15,7.1){\pos{l}{b}}\put(0.9,0.9){\pos{tr}{a}}
\put(0.9,13.1){\pos{br}{a}}
\multiput(0.92,3)(0,2){2}{\spos{r}{a}}
\multiput(0.92,7)(0,2){3}{\spos{r}{a}}
\put(2.1,12.1){\spos{bl}{c_1}}\put(3.1,11.1){\spos{bl}{c_2}}
\put(4.1,10.1){\spos{bl}{c_3}}\put(6.1,8.1){\spos{bl}{c_{q\!-\!1}}}
\put(2.05,1.95){\spos{tl}{d_1}}\put(3.05,2.95){\spos{tl}{d_2}}
\put(5.05,4.95){\spos{tl}{d_{\!q\!-\!2}}}
\put(6.05,5.95){\spos{tl}{d_{\!q\!-\!1}}}
\multiput(2,4)(0,6){2}{\spos{}{\bullet}}
\multiput(4,6)(1,1){2}{\spos{}{\bullet}}
\multiput(2,8)(1,1){2}{\spos{}{\bullet}}
\d{2}{12}\d{3}{11}\d{4}{10}\d{6}{8}
\put(6,10.5){\pos{}{\displaystyle
\,G^q_{\!a,c_1,\ldots,c_{q\!-\!1},b}}}
\end{picture}}\eql{fusLBW}\eeq
where $(d_1,\ldots,d_{\!q\!-\!1})\in P^q_{ab}$, and
%
%  Fused Right Boundary Weight
%
\setlength{\unitlength}{7.3mm}\beq\BRf{q}{a}{b}{c}{u}\quad=
\quad\frac{\delta_{ac}\;A^q_{ab}}{\;\displaystyle
\prod_{j=0}^{q\!-\!2}\T{j\!+\!1}{2j\!+\!1}{2u\!-\!\m}\;}\;\;
\sum_{c_1\ldots c_{q\!-\!1}}
\quad\raisebox{
-7\unitlength}[7\unitlength][
7\unitlength]{\begin{picture}(8,14)
\put(1,7){\line(1,1){6}}\put(2,6){\line(1,1){5}}
\put(4,4){\line(1,1){3}}\put(5,3){\line(1,1){2}}
\put(6,2){\line(1,1){1}}
\put(1,7){\line(1,-1){6}}\put(2,8){\line(1,-1){5}}
\put(3,9){\line(1,-1){4}}\put(5,11){\line(1,-1){2}}
\put(6,12){\line(1,-1){1}}
\put(7,1){\line(0,1){8.7}}\put(7,10.3){\line(0,1){1.4}}
\put(7,12.3){\line(0,1){0.7}}
\put(6.77,2){\spos{r}{u}}\put(6.94,4){\spos{r}{u\!+\!\l}}
\put(6.98,6.02){\spos{r}{u\!+\!2\l}}
\put(5.95,3){\spos{}{-\!2u\!-\!\l\!+\!\m}}
\put(5.95,5){\spos{}{-\!2u\!-\!3\l\!+\!\m}}
\put(4.95,4){\spos{}{-\!2u\!-\!2\l\!+\!\m}}
\put(1.7,7.3){\spos{}{-\!2u}}
\put(1.91,6.9){\spos{}{-\!(q\!-\!1)\l}}
\put(2.1,6.5){\spos{}{+\!\m}}
\put(2.95,8){\spos{}{-\!2u\!-\!q\l\!+\!\m}}
\put(5.7,11.3){\spos{}{-\!2u}}
\put(5.94,10.9){\spos{}{-\!(2q\!-\!3)\l}}
\put(6.15,10.5){\spos{}{+\!\m}}
\put(6.2,10){\spos{l}{u\!+\!(q\!-\!2)\l}}
\put(6.2,12){\spos{l}{u\!+\!(q\!-\!1)\l}}
\put(0.85,7.1){\pos{r}{b}}\put(7.1,0.9){\pos{tl}{a}}
\put(7.1,13.1){\pos{bl}{a}}
\multiput(7.08,3)(0,2){3}{\spos{l}{a}}
\multiput(7.08,9)(0,2){2}{\spos{l}{a}}
\put(5.9,1.9){\spos{tr}{c_1}}\put(4.9,2.9){\spos{tr}{c_2}}
\put(3.9,3.9){\spos{tr}{c_3}}\put(1.9,5.9){\spos{tr}{c_{q\!-\!1}}}
\put(5.9,12.1){\spos{br}{d_1}}\put(4.9,11.1){\spos{br}{d_2}}
\put(2.9,9.1){\spos{br}{d_{\!q\!-\!2}}}
\put(1.9,8.1){\spos{br}{d_{\!q\!-\!1}}}
\put(1.9,10.9){\pos{}{\displaystyle
\frac{1}{\;G^q_{\!a,d_1,\ldots,d_{q\!-\!1},b}\;}}}
\multiput(3,7)(1,1){2}{\spos{}{\bullet}}
\multiput(5,5)(1,1){2}{\spos{}{\bullet}}
\multiput(6,4)(0,6){2}{\spos{}{\bullet}}
\d{2}{6}\d{4}{4}\d{5}{3}\d{6}{2}
\end{picture}}\eql{fusRBW}\eeq
where $(d_1,\ldots,d_{\!q\!-\!1})\in P^q_{ab}$.
It is shown in~\cite{DatJimKunMiwOka88} that
$\rule[-3ex]{0ex}{3ex}\Wf{pq}{a}{b}{c}{d}{u}$ is independent of the choice of
$(f_1,\ldots,f_{\!q\!-\!1})\in P^q_{bc}$
and $(g_1,\ldots,g_{p\!-\!1})\in P^p_{cd}$, and it can be shown similarly that
$\rule[-3ex]{0ex}{8ex}\BLf{q}{a}{b}{a}{u}$ and
$\rule[-3ex]{0ex}{7ex}\BRf{q}{a}{b}{a}{u}$ are
each independent of the choice of $(d_1,\ldots,d_{\!q\!-\!1})\in P^q_{ab}$.

It can also be shown that $\rule[-3ex]{0ex}{3ex}\Wf{pq}{a}{b}{c}{d}{u}$
satisfies~\eqref{genAC} even though the adjacency condition is only explicitly
applied on the
edges being summed.  This can be regarded as a push-through property
of the entries of the adjacency matrix,
\setlength{\unitlength}{11mm}
\beq\text{0}{5}{0}{3.3}{tl}{A^p_{ab}\:A^q_{bc}\:A^p_{cd}\:A^q_{da}}
\text{2.3}{5}{2.8}{2.1}{r}{\times\sum_{e_1\ldots e_{p\!-\!1}}\;
\sum_{h_1\ldots h_{\!q\!-\!1}}}
\begin{picture}(5,5)\multiput(1,1)(0,1){2}{\line(1,0){3}}
\multiput(1,3)(0,1){2}{\line(1,0){3}}
\multiput(1,1)(1,0){2}{\line(0,1){3}}
\multiput(3,1)(1,0){2}{\line(0,1){3}}
\put(0.95,0.9){\pos{tr}{a}}\put(4.1,0.9){\pos{tl}{b}}
\put(4.1,4.1){\pos{bl}{c}}\put(0.9,4.1){\pos{br}{d}}
\put(2,0.92){\spos{t}{e_1}}\put(3,0.92){\spos{t}{e_{p\!-\!1}}}
\put(4.08,2){\spos{l}{f_1}}\put(4.08,3){\spos{l}{f_{\!q\!-\!1}}}
\put(3,4.12){\spos{b}{g_1}}\put(2,4.12){\spos{b}{g_{p\!-\!1}}}
\put(0.94,3){\spos{r}{h_1}}\put(0.72,2){\spos{b}{h_{q\!-\!1}}}
\d{2}{2}\d{3}{2}\d{2}{3}\d{3}{3}\d{2}{1}\d{3}{1}\d{1}{2}\d{1}{3}
\put(3.5,1.5){\spos{}{u}}
\put(1.06,1.66){\spos{l}{u\!-}}
\put(1.95,1.39){\spos{r}{(p\!-\!1)\l}}
\put(1.06,3.66){\spos{l}{u\!+}}
\put(1.95,3.39){\spos{r}{(q\!-\!p)\l}}
\put(3.06,3.66){\spos{l}{u\!+}}
\put(3.95,3.39){\spos{r}{(q\!-\!1)\l}}
\end{picture}\text{0.5}{5}{0}{2.5}{}{=}
\text{0}{5}{0.2}{3.3}{tl}{A^p_{ab}\:A^q_{da}}
\text{2.1}{5}{2.6}{2.1}{r}{\times\sum_{e_1\ldots e_{p\!-\!1}}\;
\sum_{h_1\ldots h_{\!q\!-\!1}}}
\begin{picture}(5,5)\multiput(1,1)(0,1){2}{\line(1,0){3}}
\multiput(1,3)(0,1){2}{\line(1,0){3}}
\multiput(1,1)(1,0){2}{\line(0,1){3}}
\multiput(3,1)(1,0){2}{\line(0,1){3}}
\put(0.95,0.9){\pos{tr}{a}}\put(4.1,0.9){\pos{tl}{b}}
\put(4.1,4.1){\pos{bl}{c}}\put(0.9,4.1){\pos{br}{d}}
\put(2,0.92){\spos{t}{e_1}}\put(3,0.92){\spos{t}{e_{p\!-\!1}}}
\put(4.08,2){\spos{l}{f_1}}\put(4.08,3){\spos{l}{f_{\!q\!-\!1}}}
\put(3,4.12){\spos{b}{g_1}}\put(2,4.12){\spos{b}{g_{p\!-\!1}}}
\put(0.94,3){\spos{r}{h_1}}\put(0.72,2){\spos{b}{h_{q\!-\!1}}}
\d{2}{2}\d{3}{2}\d{2}{3}\d{3}{3}\d{2}{1}\d{3}{1}\d{1}{2}\d{1}{3}
\put(3.5,1.5){\spos{}{u}}
\put(1.06,1.66){\spos{l}{u\!-}}\put(1.95,1.39){\spos{r}{(p\!-\!1)\l}}
\put(1.06,3.66){\spos{l}{u\!+}}\put(1.95,3.39){\spos{r}{(q\!-\!p)\l}}
\put(3.06,3.66){\spos{l}{u\!+}}\put(3.95,3.39){\spos{r}{(q\!-\!1)\l}}
\end{picture}\eql{adjpushthrough}\eeq
Furthermore, it can be shown that the configuration of level~1 weights in
$\rule[-3ex]{0ex}{8ex}\Wf{pq}{a}{b}{c}{d}{u}$ can be re-oriented, as follows:
\setlength{\unitlength}{11mm}\beqa
\lefteqn{\left(\prod_{j=0}^{q\!-\!2}\T{p}{j}{u}\!\!\right)\;
\;\Wf{pq}{a}{b}{c}{d}{u}}\noeqno
&=&A^p_{ab}\:A^q_{da}\;\sum_{e_1\ldots e_{p\!-\!1}}\;
\sum_{h_1\ldots h_{\!q\!-\!1}}
\raisebox{-2.5\unitlength}[2.5\unitlength][
2.5\unitlength]{\begin{picture}(5,5)
\multiput(1,1)(0,1){2}{\line(1,0){3}}
\multiput(1,3)(0,1){2}{\line(1,0){3}}
\multiput(1,1)(1,0){2}{\line(0,1){3}}
\multiput(3,1)(1,0){2}{\line(0,1){3}}
\put(0.95,0.9){\pos{tr}{a}}\put(4.1,0.9){\pos{tl}{b}}
\put(4.1,4.1){\pos{bl}{c}}\put(0.9,4.1){\pos{br}{d}}
\put(2,0.92){\spos{t}{e_1}}\put(3,0.92){\spos{t}{e_{p\!-\!1}}}
\put(4.08,2){\spos{l}{f_1}}\put(4.08,3){\spos{l}{f_{\!q\!-\!1}}}
\put(3,4.12){\spos{b}{g_1}}\put(2,4.12){\spos{b}{g_{p\!-\!1}}}
\put(0.94,3){\spos{r}{h_1}}\put(0.72,2){\spos{b}{h_{q\!-\!1}}}
\d{2}{2}\d{3}{2}\d{2}{3}\d{3}{3}\d{2}{1}\d{3}{1}\d{1}{2}\d{1}{3}
\put(3.5,1.5){\spos{}{u}}
\put(1.06,1.66){\spos{l}{u\!-}}
\put(1.95,1.39){\spos{r}{(p\!-\!1)\l}}
\put(1.06,3.66){\spos{l}{u\!+}}
\put(1.95,3.39){\spos{r}{(q\!-\!p)\l}}
\put(3.06,3.66){\spos{l}{u\!+}}
\put(3.95,3.39){\spos{r}{(q\!-\!1)\l}}
\end{picture}}\eql{reorient0}\\
&=&A^p_{ab}\:A^q_{bc}\;\sum_{e_1\ldots e_{p\!-\!1}}\;
\sum_{f_1\ldots f_{\!q\!-\!1}}\;
\raisebox{-2.5\unitlength}[2.5\unitlength][
2.5\unitlength]{\begin{picture}(8,5)
\put(0,2.5){\pos{l}{\displaystyle\frac{1}{\;G^q_{\!d,h_1,
\ldots,h_{q\!-\!1},a}\;}}}
\put(8,2.5){\pos{r}{G^q_{\!b,f_1,\ldots,f_{\!q\!-\!1},c}}}
\put(1.7,0){\begin{picture}(5,5)\multiput(1,1)(0,1){2}{\line(1,0){3}}
\multiput(1,3)(0,1){2}{\line(1,0){3}}
\multiput(1,1)(1,0){2}{\line(0,1){3}}\multiput(3,1)(1,0){2}{\line(0,1){3}}
\put(0.95,0.9){\pos{tr}{a}}\put(4.1,0.9){\pos{tl}{b}}
\put(4.1,4.1){\pos{bl}{c}}\put(0.9,4.1){\pos{br}{d}}
\put(2,0.92){\spos{t}{e_1}}\put(3,0.92){\spos{t}{e_{p\!-\!1}}}
\put(4.08,2){\spos{l}{f_1}}\put(4.08,3){\spos{l}{f_{\!q\!-\!1}}}
\put(3,4.12){\spos{b}{g_1}}\put(2,4.12){\spos{b}{g_{p\!-\!1}}}
\put(0.94,3){\spos{r}{h_1}}\put(0.72,2){\spos{b}{h_{q\!-\!1}}}
\d{2}{2}\d{3}{2}\d{2}{3}\d{3}{3}\d{2}{1}\d{3}{1}\d{4}{2}\d{4}{3}
\put(3.5,3.5){\spos{}{u}}
\put(1.06,3.66){\spos{l}{u\!-}}\put(1.95,3.39){\spos{r}{(p\!-\!1)\l}}
\put(1.06,1.66){\spos{l}{u\!+}}\put(1.95,1.39){\spos{r}{(q\!-\!p)\l}}
\put(3.06,1.66){\spos{l}{u\!+}}
\put(3.95,1.39){\spos{r}{(q\!-\!1)\l}}\end{picture}}
\end{picture}}\eql{reorient1}\\
&=&A^p_{cd}\:A^q_{da}\;\sum_{g_1\ldots g_{p\!-\!1}}\;\sum_{h_1\ldots
h_{q\!-\!1}}\;
\raisebox{
-3.7\unitlength}[3.7\unitlength][3.1\unitlength]{\begin{picture}(5,6.8)
\put(2.5,1.3){\pos{b}{\displaystyle\frac{1}{\;
G^p_{\!a,e_1,\ldots,e_{p\!-\!1},b}\;}}}
\put(2.5,6.3){\pos{t}{G^p_{\!c,g_1,\ldots,g_{p\!-\!1},d}}}
\put(0,1.2){\begin{picture}(5,5)\multiput(1,1)(0,1){2}{\line(1,0){3}}
\multiput(1,3)(0,1){2}{\line(1,0){3}}
\multiput(1,1)(1,0){2}{\line(0,1){3}}\multiput(3,1)(1,0){2}{\line(0,1){3}}
\put(0.95,0.9){\pos{tr}{a}}\put(4.1,0.9){\pos{tl}{b}}\put(4.1,4.1){\pos{bl}{c}}
\put(0.9,4.1){\pos{br}{d}}
\put(2,0.92){\spos{t}{e_1}}\put(3,0.92){\spos{t}{e_{p\!-\!1}}}
\put(4.08,2){\spos{l}{f_1}}\put(4.08,3){\spos{l}{f_{\!q\!-\!1}}}
\put(3,4.12){\spos{b}{g_1}}\put(2,4.12){\spos{b}{g_{p\!-\!1}}}
\put(0.94,3){\spos{r}{h_1}}\put(0.72,2){\spos{b}{h_{q\!-\!1}}}
\d{2}{2}\d{3}{2}\d{2}{3}\d{3}{3}\d{1}{2}\d{1}{3}\d{2}{4}\d{3}{4}
\put(1.5,1.5){\spos{}{u}}
\put(3.06,1.66){\spos{l}{u\!-}}\put(3.95,1.39){\spos{r}{(p\!-\!1)\l}}
\put(3.06,3.66){\spos{l}{u\!+}}\put(3.95,3.39){\spos{r}{(q\!-\!p)\l}}
\put(1.06,3.66){\spos{l}{u\!+}}
\put(1.95,3.39){\spos{r}{(q\!-\!1)\l}}\end{picture}}
\end{picture}}\eql{reorient2}\\
&=&A^q_{bc}\:A^p_{cd}\;\sum_{f_1\ldots f_{\!q\!-\!1}}\;\sum_{g_1\ldots
g_{p\!-\!1}}\;
\raisebox{-3.7\unitlength}[3.7\unitlength][
3.1\unitlength]{\begin{picture}(8,6.8)
\put(4.2,1.3){\pos{b}{\displaystyle\frac{1}{\;G^p_{\!a,e_1,
\ldots,e_{p\!-\!1},b}\;}}}
\put(8,3.7){\pos{r}{G^q_{\!b,f_1,\ldots,f_{\!q\!-\!1},c}}}
\put(4.2,6.3){\pos{t}{G^p_{\!c,g_1,\ldots,g_{p\!-\!1},d}}}
\put(0,3.7){\pos{l}{\displaystyle
\frac{1}{\;G^q_{\!d,h_1,\ldots,h_{q\!-\!1},a}\;}}}
\put(1.7,1.2){\begin{picture}(5,5)
\multiput(1,1)(0,1){2}{\line(1,0){3}}\multiput(1,3)(0,1){2}{\line(1,0){3}}
\multiput(1,1)(1,0){2}{\line(0,1){3}}\multiput(3,1)(1,0){2}{\line(0,1){3}}
\put(0.95,0.9){\pos{tr}{a}}\put(4.1,0.9){\pos{tl}{b}}
\put(4.1,4.1){\pos{bl}{c}}\put(0.9,4.1){\pos{br}{d}}
\put(2,0.92){\spos{t}{e_1}}\put(3,0.92){\spos{t}{e_{p\!-\!1}}}
\put(4.08,2){\spos{l}{f_1}}\put(4.08,3){\spos{l}{f_{\!q\!-\!1}}}
\put(3,4.12){\spos{b}{g_1}}\put(2,4.12){\spos{b}{g_{p\!-\!1}}}
\put(0.94,3){\spos{r}{h_1}}\put(0.72,2){\spos{b}{h_{q\!-\!1}}}
\d{2}{2}\d{3}{2}\d{2}{3}\d{3}{3}\d{4}{2}\d{4}{3}\d{2}{4}\d{3}{4}
\put(1.5,3.5){\spos{}{u}}
\put(3.06,3.66){\spos{l}{u\!-}}\put(3.95,3.39){\spos{r}{(p\!-\!1)\l}}
\put(3.06,1.66){\spos{l}{u\!+}}\put(3.95,1.39){\spos{r}{(q\!-\!p)\l}}
\put(1.06,1.66){\spos{l}{u\!+}}\put(1.95,1.39){\spos{r}{(q\!-\!1)\l}}
\end{picture}}
\end{picture}}\eql{reorient3}
\eeqa
In these expressions, the external edge spins which are not summed are
arbitrary, as long as we have
$(e_1,\ldots,e_{p\!-\!1})\in P^p_{ab}$ in~\eqref{reorient2}
and~\eqref{reorient3},
$(f_1,\ldots,f_{\!q\!-\!1})\in P^q_{bc}$ in~\eqref{reorient0}
and~\eqref{reorient2},
$(g_1,\ldots,g_{p\!-\!1})\in P^p_{cd}$ in~\eqref{reorient0}
and~\eqref{reorient1}, and
$(h_1,\ldots,h_{q\!-\!1})\in P^q_{da}$ in~\eqref{reorient1}
and~\eqref{reorient3}.
One way to prove~\eqref{adjpushthrough}
and~\eqref{reorient1}--\eqref{reorient3} is to use the fusion projection
operators of~\cite{ZhoPea94a}, which satisfy a push-through
property relative to the fused face weights and whose entries are proportional
to the gauge factors $G$.

In~\cite{DatJimKunMiwOka88}, it is shown that for the ABF fused face weights,
the summation over multiple spins in~\eqref{fusFW} can always be reduced either
to a single term
or to a summation over a single index, and that
$\displaystyle\prod_{j=0}^{q\!-\!2}\T{p}{j}{u}$ always arises as a common
factor.
The resulting expressions for the weights are presented, and from these we find
that we have crossing symmetry
\beq\eql{fusFCS}\Wf{pq}{a}{b}{c}{d}{u}=(-1)^{p(q-1)}\;\epsilon_a\:\epsilon_b\:
\epsilon_c\:\epsilon_d\;\frac{\t(a\l)}{\t(d\l)}\;\frac{X^p_{cd}}{X^p_{ab}}\;
\Wf{pq}{d}{c}{b}{a}{\!-\!u\!+\!(p\!-\!q\!+\!1)\l}\eeq
partial height reversal symmetry
\beq\eql{fusFPHRS}\Wf{pq}{a}{b}{c}{d}{u}=(-1)^{pL}\;\epsilon_{a\!+\!d}\:
\epsilon_{b\!+\!c}\;\frac{X^q_{bc}}{X^q_{ad}}\;
\Wf{p,L\!-\!1\!-\!q}{L\!+\!1\!-\!a}{L\!+\!1\!-\!b}{c}{d}{u\!+\!(q\!+\!1)\l}
\eeq
full height reversal symmetry
\beq\eql{fusFFHRS}\Wf{pq}{a}{b}{c}{d}{u}=
\Wf{pq}{L\!+\!1\!-\!a}{L\!+\!1\!-\!b}{L\!+\!1\!-\!c}{L\!+\!1\!-\!d}{u}
\eeq
and an initial condition
\beq\eql{fusIC}\Wf{qq}{a}{b}{c}{d}{0}=\T{q}{q}{0}\;\delta_{ac}
\;A^q_{ab}\;A^q_{ad}\eeq
Properties~\eqref{fusFCS} and~\eqref{fusFFHRS} can be proved alternatively by
applying the corresponding properties of the level~1 weights directly
in~\eqref{fusFW}.

Using techniques similar to those used in~\cite{DatJimKunMiwOka88} to derive
explicit formulae for the ABF fused face weights, it can be also shown that,
for the boundary weights,
the summations over multiple spins in~\eqref{fusLBW} and \eqref{fusRBW} always
reduce to a single term, with
$\displaystyle\prod_{j=0}^{q\!-\!2}\T{j\!+\!1}{2j\!+\!1}{2u\!-\!\m}$
as a factor. This gives, for $A^q_{ab}=1$,
\beqa\rule[-3.5ex]{0ex}{3.5ex}\lefteqn{\BLf{q}{a}{b}{a}{u}\;=}
\eql{exfusLBW}\\
&&\prod_{j=1}^{\frac{a-b+q}{2}}\frac{\t(j\l)\;\t((a\!-\!j)\l)\;
\t(-\!u\!-\!(q\!-\!j)\l\!-\!\lm\!+\!\xiL(a))\;
\t(u\!+\!(q\!-\!j\!+\!a)\l\!+\!\lm\!+\!\xiL(a))}{\t(\l)^4}
\noeqno
&&\rule[-5ex]{0ex}{5ex}\quad\times\;
\prod_{j=1}^{\frac{b-a+q}{2}}
\frac{\t(j\l)\;\t((a\!+\!j)\l)\;\t(u\!+
\!(q\!-\!j)\l\!+\!\lm\!+\!\xiL(a))\;
\t(u\!+\!(q\!-\!j\!-\!a)\l\!+\!\lm\!-\!\xiL(a))}{\t(\l)^4}\nonumber
\eeqa
and
\beqa\BRf{q}{a}{b}{a}{u}&=&\prod_{j=1}^{\frac{a-b+q}{2}}
\frac{\t(-\!u\!-\!(q\!-\!j)\l\!-\!\lm\!+\!\xiR(a))\;
\t(u\!+\!(q\!-\!j\!+\!a)\l\!+
\!\lm\!+\!\xiR(a))}{\t(j\l)\;\t((b\!+\!j)\l)}
\eql{exfusRBW}\\
&&\quad\times\;\prod_{j=1}^{\frac{b-a+q}{2}}
\frac{\t(u\!+\!(q\!-\!j)\l\!+\!\lm\!+\!\xiR(a))\;
\t(u\!+\!(q\!-\!j\!-\!a)\l\!+\!\lm\!-
\!\xiR(a))}{\t(j\l)\;\t((b\!-\!j)\l)}
\nonumber\eeqa
It follows from these expressions that the ABF fused boundary weights satisfy
partial height reversal symmetry
\beqa\rule[-3.5ex]{0ex}{3.5ex}
\lefteqn{\T{L\!-\!q}{-\!2}{u\!+\!\lm\!-\!\xiL(a)}\;\T{L\!-
\!q}{-\!2}{u\!+\!\lm\!+\!\xiL(a)}\;
\BLf{q}{a}{b}{a}{u}\quad=}\noeqno
&&\rule[-4ex]{0ex}{4ex}-
\left(\frac{X^q_{ab}}{\T{L}{L}{0}}\right)^{\!\!2}\;
\T{q\!+\!1}{q\!-\!1\!-\!a}{u\!+\!\lm\!-\!\xiL(a)}\;
\T{q\!+\!1}{q\!-\!1\!+\!a}{u\!+\!\lm\!+\!\xiL(a)}\;
\BLf{L\!-\!1\!-\!q}{a}{L\!+\!1\!-\!b}{a}{u\!+\!(q\!
+\!1)\l}\noeqno
\eql{fusBPHRS}\\
\rule[-3.5ex]{0ex}{3.5ex}
\lefteqn{\T{L\!-\!q}{-\!2}{u\!
+\!\lm\!-\!\xiR(a)}\;\T{L\!-\!q}{-\!2}{u\!+\!\lm
\!+\!\xiR(a)}\;\BRf{q}{a}{b}{a}{u}\quad=}\noeqno
&&-\left(\frac{\T{L}{L}{0}}{X^q_{ab}}\right)^{\!\!2}\;
\T{q\!+\!1}{q\!-\!1\!-\!a}{u\!+\!\lm\!-\!\xiR(a)}\;
\T{q\!+\!1}{q\!-\!1\!+\!a}{u\!+\!\lm\!+\!\xiR(a)}\;
\BRf{L\!-\!1\!-\!q}{a}{L\!+\!1\!-\!b}{a}{u\!+\!(q\!+\!1)\l}
\nonumber\eeqa
and, provided that~\eqref{xicond} is satisfied, full height
reversal symmetry
\beqa\BLf{q}{a}{b}{c}{u}&=&
\BLf{q}{L\!+\!1\!-\!a}{L\!+\!1\!-\!b}{L\!+\!1\!-\!c}{u}\noeqno
\eql{fusBFHRS}\\
\BRf{q}{a}{b}{c}{u}&=&
\BRf{q}{L\!+\!1\!-\!a}{L\!+\!1\!-\!b}{L\!+\!1\!-\!c}{u}
\nonumber\eeqa

\subsub{Local Relations}
We now consider the fused local relations,~\eqref{genYBE}--\eqref{genRBC}. It
is shown in~\cite{DatJimKunMiwOka88} that the ABF fused face weights satisfy
the fused Yang-Baxter
equation~\eqref{genYBE}.  A proof proceeds as follows: if fusion level~$-1$ is
involved, then each side of~\eqref{genYBE} is zero,
if fusion level~$0$ is involved, then each side of~\eqref{genYBE} immediately
reduces to a product of the same terms,
and if higher fusion levels only are involved, then~\eqref{genYBE} can be
verified by seting internal arbitrary spins equal to adjoining summed spins,
using~\eqref{adjpushthrough} to push all explicit occurrences of the fused
adjacency condition to external edges, and applying the original Yang Baxter
equation,~\eqref{YBE}, $pqr$ times.

It can also be shown that the ABF fused face weights satisfy the fused
inversion relation~\eqref{genIR} with
\beq\rho^{qr}\!(u)=\t^q_{\!-\!1}(u)\eql{fusrho}\eeq
Again, if fusion level~$-1$ is involved, then each side of~\eqref{genIR} is
zero,
if fusion level~$0$ is involved, then the left side of~\eqref{genIR}
immediately reduces to the same product of terms
as the right side, and if higher fusion levels only are involved,
then~\eqref{genIR} can be verified by setting internal arbitrary spins equal to
adjoining summed spins,
using~\eqref{adjpushthrough} to push all explicit occurrences of the fused
adjacency condition to external edges, and applying the original inversion
relation,~\eqref{IR}, $qr$ times.

Finally, in Appendix B we show that the ABF fused face and boundary weights,
together with $\rho^{qr}$ given by~\eqref{fusrho}, also satisfy the fused
reflection
equations,~\eqref{genLRE} and~\eqref{genRRE}, and fused boundary crossing
equations,~\eqref{genLBC} and~\eqref{genRBC}, where $\l$ in these equations is
taken as the crossing parameter~\eqref{lambda} and $\m$ is arbitrary.

\subsub{Double-Row Transfer Matrices}
We now consider the ABF fused double-row transfer matrices $\D^{pq}\!(u)$,
which are defined by~\eqref{genDRTM}, with $\l$ given by~\eqref{lambda},
and the ABF fused double-row transfer matrices with fixed left and right
boundary spins $\aL$ and $\aR$, $\D^{pq}\!(\aL\aR|u)$, given by
\beq\langle
a_2,\ldots,a_\N\,|\,\D^{pq}\!(\aL\aR|u)\,|\,b_2,\ldots,b_\N\rangle=
\langle\aL,a_2,\ldots,a_\N,\aR\,|\,\D^{pq}\!(u)\,|\,\aL,b_2,
\ldots,b_\N,\aR\rangle\eeq
By re-configuring the fused face weights in the top row according
to~\eqref{reorient1}, using~\eqref{adjpushthrough} repeatedly to push all
explicit entries of $A^p$ to the lower edges, and all explicit entries of $A^q$
to a single internal edge, setting as many internal arbitrary spins as possible
equal to adjoining summed spins, and cancelling all of the gauge factors $G$
which appear along the top row, we find that $\D^{pq}\!(\aL\aR|u)$ can be
written as
%\newpage
\beq\langle
a_2,\ldots,a_\N\,|\,\D^{pq}\!(\aL\aR|u)\,|\,b_2,\ldots,b_\N\rangle\quad=\quad
\frac{A^p_{\aL a_2}\ldots A^p_{a_N\aR}}{\displaystyle
K^{pq}(u)}\;\;\sum_c\,A^q_{\aL c}\;\times\qquad\qquad\eql{ABFfusDRTM}\eeq
\setlength{\unitlength}{7.4mm}
\begin{center}$\quad$\begin{picture}(4,8.4)
\put(0,0){\line(0,1){0.7}}
\put(0,1.3){\line(0,1){5.4}}
\put(0,7.3){\line(0,1){0.7}}
\put(0,0){\line(1,1){4}}
\put(0,2){\line(1,1){3}}
\put(0,6){\line(1,1){1}}
\put(0,2){\line(1,-1){1}}
\put(0,6){\line(1,-1){3}}
\put(0,8){\line(1,-1){4}}
\dl{0}{0}{4}\dl{3}{5}{1}
\dl{1}{7}{3}\dl{0}{8}{4}
\d{2}{4}\d{1}{7}\d{3}{5}
\put(0,1){\um}\put(0,7){\uqlm}
\put(3,4){\luuql}
\put(0.1,-0.08){\spos{t}{\aL}}
\multiput(-0.06,2)(0,4){2}{\spos{r}{\aL}}
\put(0.1,8.18){\spos{b}{\aL}}
\put(1.03,0.97){\spos{tl}{c_1}}
\put(3.03,2.97){\spos{tl}{c_{\!q\!-\!1}}}
\put(3.94,3.65){\spos{r}{c}}
\end{picture}\begin{picture}(12,8)
\multiput(0,0)(0,1){2}{\line(1,0){12}}
\multiput(0,3)(0,1){3}{\line(1,0){12}}
\multiput(0,7)(0,1){2}{\line(1,0){12}}
\multiput(0,0)(1.5,0){2}{\line(0,1){8}}
\multiput(2.5,0)(1.5,0){2}{\line(0,1){8}}
\multiput(8,0)(1.5,0){2}{\line(0,1){8}}
\multiput(10.5,0)(1.5,0){2}{\line(0,1){8}}
\d{0}{1}\dd{0}{3}{3}\d{0}{7}\dd{1.5}{0}{2}
\dd{1.5}{3}{3}\d{1.5}{7}\dd{2.5}{0}{2}
\dd{2.5}{3}{3}\d{2.5}{7}\d{4}{1}\dd{4}{3}{3}
\d{4}{7}
\d{8}{1}\dd{8}{3}{3}\d{8}{7}\dd{9.5}{0}{2}
\dd{9.5}{3}{3}\d{9.5}{7}\dd{10.5}{0}{2}
\dd{10.5}{3}{3}\d{10.5}{7}\d{12}{1}
\dd{12}{3}{3}\d{12}{7}
\multiput(0,0)(8,0){2}{\pl}
\multiput(0,3)(8,0){2}{\qp}
\multiput(2.5,0)(8,0){2}{\fu}
\multiput(2.5,3)(8,0){2}{\ql}
\multiput(0,4)(8,0){2}{\umpl}
\multiput(0,7)(8,0){2}{\umqpll}
\multiput(2.5,4)(8,0){2}{\fum}
\multiput(2.5,7)(8,0){2}{\umql}
\put(0.1,-0.08){\spos{t}{\aL}}
\put(4.1,-0.08){\spos{t}{a_2}}
\put(8.1,-0.08){\spos{t}{a_{\!N}}}
\put(12.1,-0.08){\spos{t}{\aR}}
\put(0.1,8.18){\spos{b}{\aL}}
\put(1.6,8.18){\spos{b}{e_{11}}}
\put(2.8,8.18){\spos{b}{e_{1,p\!-\!1}}}
\put(4.1,8.18){\spos{b}{b_2}}
\put(8.1,8.18){\spos{b}{b_{\!N}}}
\put(9.6,8.18){\spos{b}{e_{\!N1}}}
\put(10.8,8.18){\spos{b}{e_{\!N,p\!-\!1}}}
\put(12.1,8.18){\spos{b}{\aR}}
\end{picture}\begin{picture}(4,8)
\put(4,0){\line(0,1){6.7}}
\put(4,7.3){\line(0,1){0.7}}
\put(4,0){\line(-1,1){4}}\put(4,2){\line(-1,1){3}}
\put(4,6){\line(-1,1){1}}
\put(4,2){\line(-1,-1){1}}
\put(4,6){\line(-1,-1){3}}
\put(4,8){\line(-1,-1){4}}
\dl{0}{0}{4}\dl{0}{1}{3}\dl{0}{3}{1}
\dl{0}{5}{1}\dl{0}{7}{3}\dl{0}{8}{4}
\d{3}{1}\d{1}{3}\d{1}{5}\d{2}{4}\d{3}{7}
\put(4,1){\u}\put(4,7){\uql}\put(1,4){\uuql}
\put(4.2,-0.08){\spos{t}{\aR}}
\multiput(4.06,2)(0,4){2}{\spos{l}{\aR}}
\put(4.2,8.18){\spos{b}{\aR}}
\end{picture}\end{center}
where
\beq K^{pq}(u)=\prod_{j=0}^{q\!-\!2}\left(\T{p}{j}{u}\;
\T{p}{-\!j\!-\!1}{-\!u\!+\!\m}\right)^N
\T{j\!+\!1}{2j\!+\!1}{2u\!-\!\m}\;
\T{j\!+\!1}{\!-\!j\!-\!1}{\!-\!2u\!+\!\m}
\eql{K}\eeq
and we must have
\[(e_{11},\ldots,e_{1,p\!-\!1})\in P^p_{\aL b_2},\;
\ldots\;,(e_{\!\N1},\ldots,e_{\!\N,p\!-\!1})\in P^p_{b_{\!N}\aR}\]
and, for each $c$ in the sum,
\[(c_1,\ldots,c_{q\!-\!1})\in P^q_{\aL c}\]
We note that the spins $c_1,\ldots,c_{q\!-\!1}$ can not be set equal to the
adjoining summed spins, as they only become arbitrary \e{after} the
summation has occured.

Since the required fused local relations are satisfied, we have,
from~\eqref{fuscomm}, commutativity
\beq\D^{pq}\!(\aL\aR|u)\:\D^{pr}\!(\aL\aR|v)=
\D^{pr}\!(\aL\aR|v)\:\D^{pq}\!(\aL\aR|u)\eeq
and, from~\eqref{fusCS}, crossing symmetry
\beq\eql{fusDCS}\D^{pq}\!(\aL\aR|u)=\D^{pq}\!(
\aL\aR|\!-\!u\!-\!(q\!-\!1)\l\!+\!\m)\eeq
It follows, from~\eqref{fusFPHRS} and~\eqref{fusBPHRS}, that the ABF fused
double-row transfer matrices also satisfy partial height reversal symmetry
\beq\alpha^{L\!-\!q}_{\!-\!2}(\aL\aR|u)\;
\D^{pq}\!(\aL\aR|u)\;=\;
(-1)^{pN}\;\beta^{q\!+\!1}_{q\!-\!1}(\aL\aR|u)
\;\D^{p,L\!-\!1\!-\!q}\!(\aL\aR|u\!+\!(q\!+\!1)\l)
\eql{fusPHRS1}\eeq
or, equivalently,
\beq\beta^{L\!-\!q}_{\!-\!2}(\aL\aR|u)\;
\D^{pq}\!(\aL\aR|u)\;=\;
(-1)^{pN}\;\alpha^{q\!+\!1}_{q\!-\!1}(\aL\aR|u)\;
\D^{p,L\!-\!1\!-\!q}\!(\aL\aR|u\!+\!(q\!+\!1)\l)
\eql{fusPHRS2}\eeq
where
\beqa\rule[-2ex]{0ex}{2ex}\alpha^r_k(\aL\aR|u)&=&
\T{r}{k}{u\!+\!\lm\!-\!\xiL(\aL)}\;\T{r}{k}{u\!+\!\lm\!
+\!\xiL(\aR)}\\
&&\rule[-4ex]{0ex}{4ex}\qquad\qquad\times\quad\T{r}{k}{u\!+
\!\lm\!-\!\xiR(\aR)}\;\T{r}{k}{u\!+\!\lm\!+\!\xiR(\aR)}\noeqno
\rule[-2ex]{0ex}{2ex}\beta^r_k(\aL\aR|u)&=&\T{r}{k\!-\!\aL}{u\!+
\!\lm\!-\!\xiL(\aL)}\;\T{r}{k\!+\!\aL}{u\!+\!\lm\!+\!\xiL(\aL)}\\
&&\qquad\qquad\times\quad\T{r}{k\!-\!\aR}{u\!+\!\lm\!
-\!\xiR(\aR)}
\;\T{r}{k\!+\!\aR}{u\!+\!\lm\!+\!\xiR(\aR)}\nonumber
\eeqa
Considering $q=-1$, $0$, $L\!-\!1$ and $L$ in~\eqref{fusPHRS1}
or~\eqref{fusPHRS2}, we have, using~\eqref{fus-1wt} and~\eqref{fus0wt},
\beqa\rule[-2ex]{0ex}{2ex}\D^{p,-1}\!(\aL\aR|u)&=&\mbox{\bf0}
\eql{Dp-1}\\
\rule[-2ex]{0ex}{2ex}\D^{p,L}\!(\aL\aR|u)&=&\mbox{\bf0}\\
\rule[-2ex]{0ex}{2ex}\D^{p,0}\!(\aL\aR|u)&=&f^p(u\!-\!\l)\;
\vec{I}^p\!(\aL\aR)\eql{Dp0}\\
\rule[-3.5ex]{0ex}{3.5ex}\D^{p,L\!-\!1}\!(\aL\aR|u)
&=&(-1)^{pN}\;
\frac{\alpha^L_{L\!-\!2}(\aL\aR|u)}{\beta^1_{\!-\!2}(\aL\aR|u)}
\;f^p(u\!-\!2\l)\;\vec{I}^p\!(\aL\aR)\\
&=&(-1)^{pN}\;
\frac{\beta^L_{L\!-\!2}(\aL\aR|u)}{\alpha^1_{\!-\!2}(\aL\aR|u)}
\;f^p(u\!-\!2\l)\;\vec{I}^p\!(\aL\aR)\eeqa
where
\beq f^p(u)=\left(\T{p}{0}{u}\:
\T{p}{\!-\!1}{\!-\!u\!+\!\m}\right)^{\!N}\eeq
and $\vec{I}^p\!(\aL\aR)$ is the adjacency-inclusive identity
\beqa\rule[-2ex]{0ex}{2ex}\lefteqn{\langle a_2,a_3,
\ldots,a_{\N\!-\!1},a_\N\,|\,\vec{I}^p\!(\aL\aR)\,|\,b_2,b_3,
\ldots,b_{\N\!-\!1},b_\N\rangle\quad=}\qquad\qquad\qquad\qquad
\qquad\qquad\\
&&\delta_{a_2b_2}\ldots\delta_{a_N b_N}\;A^p_{\aL a_2}
\,A^p_{a_2a_3}\ldots A^p_{a_{N\!-\!1}a_N}\,A^p_{a_N\aR}
\nonumber\eeqa
It follows from~\eqref{fusFFHRS} and~\eqref{fusBFHRS},
with~\eqref{xicond}, that the ABF fused double-row
transfer matrices satisfy full height reversal symmetry
\beq\D^{pq}\!(\aL\aR|u)=\vec{Y}\:\D^{pq}\!(L\!+\!1\!-\!
\aL,L\!+\!1\!-\!\aR|u)\:\vec{Y}\eeq
where
\beq\langle a_2,\ldots,a_\N\,|\,\vec{Y}\,|\,b_2,\ldots,b_\N
\rangle=\delta_{{\scriptscriptstyle L+1}\!-a_2\,,\,b_2}\:
\ldots\:\delta_{{\scriptscriptstyle L+1}\!-a_N\,,\,b_N}\eeq
For the ABF fused models, a natural choice for $\mu$ in
$\D^{pq}\!(\aL\aR|u)$ is
\beq\eql{mu}\m=p\l\eeq
We note that $p$-dependence, as opposed to $q$-dependence,
 of $\m$ in $\D^{pq}\!(\aL\aR|u)$ does not destroy commutativity or crossing
symmetry.
With this choice, crossing symmetry of the fused face weights,~\eqref{fusFCS},
implies that $\D^{pq}\!(\aL\aR|u)$ is similar to its transpose
\beq\eql{fusDT}\D^{pq}\!(\aL\aR|u)=\vec{S}^p\!(\aL\aR)^{-1}\;
\D^{pq}\!(\aL\aR|u)^t\;\vec{S}^p\!(\aL\aR)\eeq
where
\beqa\rule[-2.5ex]{0ex}{2.5ex}\lefteqn{\langle a_2,a_3,
\ldots,a_{\N\!-\!1},a_\N\,|\,\vec{S}^p\!(\aL\aR)\,|\,b_2,b_3,
\ldots,b_{\N\!-\!1},b_\N\rangle\quad=}\qquad\qquad\qquad\\
&&\delta_{a_2b_2}\ldots\delta_{a_N b_N}\;X^p_{\aL a_2}
\,X^p_{a_2a_3}\ldots X^p_{a_{N\!-\!1}a_N}\,X^p_{a_N\aR}\;
\frac{\t(\l)^{\N\!-\!1}}{\t(a_2\l)\ldots\t(a_\N\l)}
\nonumber\eeqa

\subsub{Functional Equations}
In Appendix~C, we show that the ABF fused double-row transfer matrices satisfy
functional equations
whose structure reflects that of the $su(2)$ fusion rule~\eqref{fusrule1}
satisfied by the adjacency matrices.  There are two families of functional
equations,
\beqa\rule[-2.5ex]{0ex}{2.5ex}\lefteqn{g^0_q(2u\!-\!\l)\;\D^{pq}\!(\aL\aR|u)
\;\D^{p1}\!(\aL\aR|u\!-\!\l)\quad=}\noeqno
&&\rule[-2.5ex]{0ex}{2.5ex}\alpha^1_{\!-\!1}(\aL\aR|u)\;
\beta^1_{\!-\!1}(\aL\aR|u)
\;g^{\!-\!1}_q(2u\!-\!\l)\;f^p(u\!-\!2\l)\;\D^{p,q\!-\!1}\!(\aL\aR|u\!+\!\l)
\eql{FE1}\\
&&\qquad\qquad\qquad\qquad\qquad\qquad\qquad+\quad g^1_q(2u\!-\!\l)
\;f^p(u\!-\!\l)\;\D^{p,q\!+\!1}\!(\aL\aR|u\!-\!\l)\nonumber\eeqa
and
\beqa\rule[-2.5ex]{0ex}{2.5ex}\lefteqn{g^0_q(2u\!+\!q\l)\;\D^{pq}
\!(\aL\aR|u)\;\D^{p1}\!(\aL\aR|u\!+\!q\l)\quad=}\noeqno
&&\rule[-2.5ex]{0ex}{2.5ex}\alpha^1_{q\!-\!1}(\aL\aR|u)\;
\beta^1_{q\!-\!1}(\aL\aR|u)\;g^{\!-\!1}_q(2u\!+\!q\l)\;f^p(u\!+\!q\l)
\;\D^{p,q\!-\!1}\!(\aL\aR|u)\eql{FE2}\\
&&\qquad\qquad\qquad\qquad\qquad\qquad+\quad g^1_q(2u\!+\!q\l)
\;f^p(u\!+\!(q\!-\!1)\l)\;\D^{p,q\!+\!1}\!(\aL\aR|u)\nonumber\eeqa
where $-1\leq p\leq L$, $0\leq q\leq L\!-\!1$, and
\beq g^k_q(u)=\frac{\t(u\!+\!(k\!-\!1)\l\!-\!\m)\:\t(u\!
+\!(q\!-\!k)\l\!-\!\m)}{\t(\l)^2}\eeq
The importance of these equations is that they describe the essential content
of the fusion hierarchy, since we see that either family, together with
$\D^{p0}\!(\aL\aR|u)$ and $\D^{p1}\!(\aL\aR|u)$, can be used to determine
recursively the higher fusion level double-row transfer matrices.

It can be shown, using induction as done in~\cite{KluPea92} for the
periodic-boundary case, that \eqref{Dp-1}, \eqref{Dp0}, and either~\eqref{FE1}
or~\eqref{FE2},
imply that the fused double-row transfer matrices also satisfy equations which
correspond to~\eqref{fusrule2}, and from which follow the generalised inversion
identity,
which corresponds to~\eqref{fusrule3},
\beqa\rule[-2ex]{0ex}{2ex}\vec{d}^{pq}(\aL\aR|u)\;
\vec{d}^{pq}(\aL\aR|u\!+\!\l)
&=&\left(\vec{I}^p\!(\aL\aR)+
\vec{d}^{p,q\!-\!1}(\aL\aR|u\!+\!\l)\right)\\
&&\qquad\qquad\qquad\times\quad\left(\vec{I}^p\!(\aL\aR)
+\vec{d}^{p,q\!+\!1}(\aL\aR|u)\right)\nonumber\eeqa
where $-1\leq p\leq L$, $1\leq q\leq L\!-\!2$, and
\beqa\rule[-2ex]{0ex}{2ex}\lefteqn{\vec{d}^{pq}(\aL\aR|u)
\quad=}\\
&&\frac{\t(2u\!+\!q\l\!-\!\m)^2\;\D^{p,q\!-\!1}\!(\aL\aR|u\!
+\!\l)\;\D^{p,q\!+\!1}\!(\aL\aR|u)}
{\alpha^q_{q\!-\!1}(\aL\aR|u)\;\beta^q_{q\!-\!1}(\aL\aR|u)\;
\t(2u\!-\!\l\!-\!\m)\;\t(2u\!+\!(2q\!+\!1)\l\!-\!\m)\;
f^p(u\!-\!\l)f^p(u\!+\!q\l)}\nonumber\eeqa
{}From~\eqref{fusDCS}, it follows that
\beq\vec{d}^{pq}(\aL\aR|u)=\vec{d}^{pq}(\aL\aR|\!-\!u\!
-\!q\l\!+\!\m)\eeq
and, if $\m$ is given by~\eqref{mu}, then~\eqref{fusDT} implies that
\beq\vec{d}^{pq}(\aL\aR|u)=\vec{S}^p\!(\aL\aR)^{-1}\;
\vec{d}^{pq}(\aL\aR|u)^t\;\vec{S}^p\!(\aL\aR)\eeq

\sect{Discussion}
We have presented a general formalism for applying fixed boundary
conditions to IRF models and have specialised to the case of ABF
models and their fusion hierarchy.  In future work, we intend both to
continue our study of ABF models and to proceed with the application
of fixed boundary conditions to other IRF models.

With regard to the ABF models, we note that the functional
equations~\eqref{FE1} and~\eqref{FE2} have the same $su(2)$ structure
as those satisfied by the ABF row transfer matrices with periodic
boundary conditions, and as first presented in~\cite{BazRes89}.  We
therefore plan to use the same approach as in~\cite{BazRes89}, to
obtain Bethe ansatz equations for the eigenspectra of the double-row
transfer matrices.  Subsequently, we hope to calculate the boundary
free energy of these models, and to use the technique
of~\cite{PeaKlu91,KluPea91,KluPea92} to calculate analytically the
central charges and conformal weights of the conformal field theories
associated with the models at criticality.  Other directions in which
our treatment of ABF models could be developed further would be to
investigate the existence of boundary weights of a non-diagonal form,
and to explore the connection between ABF boundary weights and known
$K$~matrices~\cite{DevGon94b,HouYue93,InaKon94} for the eight-vertex
model.

With regard to other models, the ABF models at criticality correspond to
the $A$ series within the classical $A$--$D$--$E$
models~\cite{Pas87a}, and we plan to study other members of this
group.  At criticality, the face weights of these models satisfy the
Yang-Baxter equation through the defining relations of the
Temperley-Lieb algebra alone and it would be valuable to find boundary
weights which similarly satisfy the IRF reflection equations through
this algebra.  The Temperley-Lieb-related $A$--$D$--$E$ models
are all based on the untwisted affine Lie algebra $A^{(1)}_1$, and we
also hope to consider the dilute $A$--$D$--$E$
models~\cite{Roc92,WarNieSea92} which are based on the twisted affine
Lie algebra $A^{(2)}_2$.  In these models, the unfused adjacency
matrices allow identical spins to be adjacent, which would therefore
enable us to consider the case of fixed boundaries of the form
$a,a,a,\ldots$, whereas for the level~1 ABF models only the form
$a,a\pm1,a,\ldots$ is possible.  Finally, it would be worthwhile to
study the application of fixed boundary conditions to higher-rank IRF
models, such as those based on the untwisted affine Lie algebras
$A^{(1)}_n$, $B^{(1)}_n$, $C^{(1)}_n$ and
$D^{(1)}_n$~\cite{JimMiwOka88}.

\rnc{\theequation}{A.\arabic{equation}}\setcounter{equation}{0}
\section*{Appendix A: Derivation of ABF Boundary Weights}
In this appendix, we find boundary weights which, together with the ABF face
weights~\eqref{FW}, satisfy the reflection equations~\eqref{LRE}
and~\eqref{RRE}.
We then show that these weights also satisfy the boundary crossing
equations~\eqref{LBC} and~\eqref{RBC}.

Since the ABF weights satisfy the symmetry~\eqref{refsym2}, the left and right
reflection equations are effectively the same, so it suffices to solve
them together.  We assume that there are solutions which have the diagonal form
\beq\BL{a}{b}{c}{u}=\BR{a}{b}{c}{u}=k(a,b|u)\:\delta_{ac}\eeq
The reflection equations then become
\beqa\lefteqn{\sum_f\;\W{b}{a}{f}{c}{u\!-\!v}\;
\W{c}{f}{a}{d}{\m\!-\!u\!-\!v}\;k(a,f|u)\;k(a,d|v)=}\qquad\noeqno
&&\sum_f\;\W{d}{c}{f}{a}{u\!-\!v}\;
\W{c}{b}{a}{f}{\m\!-\!u\!-\!v}\;k(a,f|u)\;k(a,b|v)\eql{DRE}\eeqa
This equation is trivially satisfied if $A_{ab}
\:A_{bc}\:A_{cd}\:A_{da}=0$. Proceeding to $A_{ab}
\:A_{bc}\:A_{cd}\:A_{da}=1$, we see that if $b=d=a\pm1$ and
$c=a$ or $c=a\pm2$, then both sides of~\eqref{DRE} are automatically equal,
since the ABF weights satisfy the symmetry~\eqref{refsym1}.  The only class of
assignments remaining is $b=a\pm1$, $c=a$ and $d=a\mp1$. This gives $L\!-\!2$
pairs of identical equations which we find, after substituting the explicit
face weights~\eqref{FW} into~\eqref{DRE}, are given by
\beq\eql{fundRE}\begin{array}{c}\displaystyle
\sqrt{\frac{\t((a\!+\!1)\l)}{\t((a\!-\!1)\l)}}\;
\t(u\!-\!v)\;\t(u\!+\!v\!-\!a\l\!+\!\l\!-\!\m)
\;k(a,a\!-\!1|u)\;k(a,a\!-\!1|v)\\
\qquad\qquad\qquad-\quad\t(u\!+\!v\!+\!\l\!-\!\m)
\;\t(u\!-\!v\!-\!a\l)\;k(a,a\!-\!1|u)\;k(a,a\!+\!1|v)\\
\rule{0ex}{3.5ex}\qquad\qquad\qquad-\quad\t(u\!+\!v\!+
\!\l\!-\!\m)\;\t(u\!-\!v\!+\!a\l)\;k(a,a\!+\!1|u)
\;k(a,a\!-\!1|v)\\
\displaystyle+\quad
\sqrt{\frac{\t((a\!-\!1)\l)}{\t((a\!+\!1)\l)}}\;\t(u\!-
\!v)\;\t(u\!+\!v\!+\!a\l\!+\!\l\!-\!\m)\;k(a,a\!+\!1|u)
\;k(a,a\!+\!1|v)\quad=\quad0\end{array}\eeq
where $a=2,\ldots,L\!-\!1$.
We note that the boundary weights $k(1,2|u)$ and
$k(L,L\!-\!1|u)$ do not appear in any of these equations and can therefore be
set to arbitrary
functions, $g_1(u)$ and $g_L(u)$ respectively.
Returning to $a=2,\ldots,L\!-\!1$, we now assume that there are constants
$\xi(a)$ for which $k(a,a\!-\!1|\xi(a)\!-\!\lm)=0$.
Taking $v=\xi(a)-\lm$ in~\eqref{fundRE}, we find that solutions must have the
form
\beq\eql{sol}\begin{array}{rcl}k(a,a\!-\!1|u)&=&
\sqrt{\t((a\!-\!1)\l)}\;\t(u\!+\!\lm\!-\!\xi(a))\;
\t(u\!+\!a\l\!+\!\lm\!+\!\xi(a))\;g_a(u)\\
k(a,a\!+\!1|u)&=&\sqrt{\t((a\!+\!1)\l)}\;\t(u\!+
\!\lm\!+\!\xi(a))\;\t(u\!-\!a\l\!+\!\lm\!-\!\xi(a))
\;g_a(u)\end{array}\eeq
for some functions $g_a$.

We now verify that these are in fact solutions for arbitrary constants $\xi(a)$
and arbitrary functions $g_a$. Substituting~\eqref{sol} into the left side
of~\eqref{fundRE} gives
\begin{eqnarray*}\lefteqn{\sqrt{\t((a\!+\!1)\l)\,
\t((a\!-\!1)\l)}\;g_a(u)\;g_a(v)\;\Bigl(Q_{\!a}(u+
\lm,v+\lm)-Q_{\!a}(u+\lm,-v-\lm)}\qquad\qquad\qquad
\qquad\qquad\qquad\qquad\\
&&\mbox{}-Q_{\!a}(-u-\lm,v+\lm)+Q_{\!a}(-u-\lm,-v-
\lm)\Bigr)\end{eqnarray*}
where
\beq Q_{\!a}(u,v)=\t(u\!-\!v)\;\t(u\!+\!v\!-\!a\l)\;
\t(u\!-\!\xi(a))\;\t(u\!+\!a\l\!+\!\xi(a))\;\t(v\!-
\!\xi(a))\;\t(v\!+\!a\l\!+\!\xi(a))\eeq
We now find that
\begin{eqnarray*}Q_{\!a}(u,v)-Q_{\!a}(u,-v)&=&
\t(u\!-\!\xi(a))\;\t(u\!+\!a\l\!+\!\xi(a))\times\\
&&\Bigl(\t(u\!-\!v)\;\t(u\!+\!v\!-\!a\l)\;\t(v\!-\!
\xi(a))\;\t(v\!+\!a\l\!+\!\xi(a))\\
&&\qquad\qquad-\quad\t(u\!+\!v)\;\t(u\!-\!v\!-\!a\l)
\;\t(\!-\!v\!-\!\xi(a))\;\t(\!-\!v\!+\!a\l\!+\!\xi(a))
\Bigr)\\
&=&\t(u\!-\!\xi(a))\;\t(u\!+\!a\l\!+\!\xi(a))\;\t(u\!-
\!a\l\!-\!\xi(a))\;\t(u\!+\!\xi(a))\;\t(a\l)\;\t(2v)
\end{eqnarray*}
where we have used the identity~\eqref{thetaid}
with
\beq s=u-\frac{a\l}{2},\quad x=-v+\frac{a\l}{2},\quad
t=v+\frac{a\l}{2},\quad y=-\frac{a\l}{2}-\xi(a)\eeq
We can now see that $Q_{\!a}(u,v)-Q_{\!a}(u,-v)$ is even in $u$ and therefore
that the left side of~\eqref{fundRE} vanishes as required.
The solution~\eqref{sol} obtained here matches that of~\eqref{BW} if we take
\beqa g_1(u)&=&\sqrt{\frac{\t(2\l)}{\t(\l)}}\;\frac{\t(u\!+
\!\lm\!+\!\xi(1))\;\t(u\!-\l\!+\!\lm\!-\!\xi(1))}{\t(\l)^2}
\noeqno
g_a(u)&=&\frac{1}{\sqrt{\t(a\l)}\;\t(\l)^2}\\
g_L(u)&=&\sqrt{\frac{\t((L\!-\!1)\l)}{\t(L\l)}}\;\frac{\t(u\!+
\!\lm\!-\!\xi(L))\;\t(u\!+\!L\l\!+\!\lm\!+\!\xi(L))}{\t(\l)^2}
\nonumber\eeqa
and set $\xi\mapsto\xiL$ and $\xi\mapsto\xiR$ for the left and right boundary
weights respectively.

Finally, we consider the boundary crossing equations,~\eqref{LBC}
and~\eqref{RBC}, with the ABF face weights, the boundary weights found here,
and $\rho$
given by~\eqref{rho}.  These equations are satisfied since if
$\delta_{ac}\:A_{ab}=0$, then both sides of the equations are
zero, if $a=c=1,\:L$ and $b=2,\:L\!-\!1$, then the left sides are single terms
which we immediately find are equal to the terms on the right side,
and if $2\leq a=c\leq L\!-\!1$ and $b=a\pm1$, then the left sides are sums of
two terms which we find can be reduced to the terms on the right side using a
single application of~\eqref{thetaid}.

\rnc{\theequation}{B.\arabic{equation}}\setcounter{equation}{0}
\section*{Appendix B: ABF Fused Reflection and Boundary\\Crossing Equations}
In this appendix, we show that the fused right reflection and boundary crossing
equations,~\eqref{genRRE} and~\eqref{genRBC}, are satisfied
by the ABF fused weights.  The proofs for the fused left reflection and
boundary crossing equations,~\eqref{genLRE} and~\eqref{genLBC}, are similar.

We begin with~\eqref{genRRE}. If $q=-1$ or $r=-1$ then, due to~\eqref{fus-1wt},
each side of~\eqref{genRRE} is zero, and if
$q=0$ or $r=0$ then, using~\eqref{fus-1wt} and~\eqref{fusrho}, we find that
each side of~\eqref{genRRE} reduces to a product of the same terms.

We now proceed to the case $q\geq1$ and $r\geq1$.  Having substituted the ABF
fused weights,~\eqref{fusFW} and~\eqref{fusRBW}, and $\rho$ given
by~\eqref{fusrho}, into~\eqref{genRRE},
we then re-configure the central fused face weights on each side according
to~\eqref{reorient1} and the upper fused face weight on the right side
according to~\eqref{reorient3},
set internal arbitrary spins equal to adjoining summed spins,
use~\eqref{adjpushthrough} to push all explicit occurrences of the fused
adjacency condition to external
edges, cancel internal gauge factors $G$, and take external arbitrary spins to
be the same on each side of the equation.
After these steps, we find that the left side of~\eqref{genRRE} is given by
\begin{eqnarray*}\lefteqn{\frac{\T{q}{\!-\!1}{u\!-\!v\!+
\!(q\!-\!r)\l}\:\T{q}{\!-\!1}{\!-\!u\!-\!v
\!-(r\!-\!1)\l\!+\!\m}}{\;
\displaystyle\prod_{j=0}^{q\!-\!2}\T{r}{j}{u
\!-\!v}\:\T{r}{j}{\!-\!u\!-\!v\!-(q\!-\!1)\l\!
+\!\m}\:\T{j\!+\!1}{2j\!+\!1\;}{2u\!-\!\m}
\;\prod_{j=0}^{r\!-\!2}\T{j\!+\!1}{2j\!+
\!1\;}{2v\!-\!\m}\;}\quad\times}\\
&&\rule{0ex}{4ex}\qquad\frac{\delta_{ae}
\:A^r_{ab}\:A^q_{bc}}{G^q_{\!c,h_1,\ldots,h_{q\!
-\!1},d}\;G^r_{\!d,i_1,\ldots,i_{r\!-\!1},a}}\;\;
\sum_{f_1\ldots f_{\!r\!-\!1}}\;\sum_{g_1\ldots
g_{q\!-\!1}}\;
{\cal L}^{qr}\!(u,v)_{a,b,c,d,a,f_1,\ldots,f_{\!r
\!-\!1},g_1,\ldots,g_{q\!-\!1},h_1,\ldots,h_{q\!-
\!1},i_1,\ldots,i_{r\!-\!1}}\end{eqnarray*}
\rule{0ex}{2ex}and that the right side of~\eqref{genRRE} is given by
\begin{eqnarray*}\lefteqn{\frac{\T{r}{\!-\!1}{u\!-\!v}
\:\T{r}{\!-\!1}{\!-\!u\!-\!v\!-(q\!-\!1)\l\!+\!\m}}{\;
\displaystyle\prod_{j=0}^{r\!-\!2}\T{q}{j}{u\!-
\!v\!+\!(q\!-\!r)\l}\:\T{q}{j}{\!-\!u\!-\!v
\!-(r\!-\!1)\l\!+\!\m}\:\T{j\!+\!1}{2j\!+
\!1\;}{2v\!-\!\m}
\;\prod_{j=0}^{q\!-\!2}\T{j\!+\!1}{2j
\!+\!1\;}{2u\!-\!\m}\;}\quad\times}\\
&&\rule{0ex}{4ex}\qquad\frac{\delta_{ae}\:A^r_{ab}
\:A^q_{bc}}{G^q_{\!c,h_1,\ldots,h_{q\!-
\!1},d}\;G^r_{\!d,i_1,\ldots,i_{r\!-\!1},a}}\;\;
\sum_{f_1\ldots f_{\!r\!-\!1}}\;
\sum_{g_1\ldots g_{q\!-\!1}}\;
{\cal R}^{qr}\!(u,v)_{a,b,c,d,a,f_1,
\ldots,f_{\!r\!-\!1},g_1,\ldots,g_{q\!-\!1},h_1,
\ldots,h_{q\!-\!1},i_1,\ldots,i_{r\!-\!1}}\end{eqnarray*}
where we must have $(h_1,\ldots,h_{q\!-\!1})
\in P^q_{cd}$ and $(i_1,\ldots,i_{r\!-\!1})
\in P^r_{da}$, and where
\newpage
\setlength{\unitlength}{7.4mm}
\begin{center}
\text{6}{18}{0}{13}{l}{{\cal L}^{qr}\!(u,v)_{a,b,c,d,e,f_1,
\ldots,f_{\!r\!-\!1},g_1,\ldots,g_{q\!-\!1},h_1,
\ldots,h_{q\!-\!1},i_1,\ldots,i_{r\!-\!1}}\quad=}
\begin{picture}(14,18)
\put(13,1){\line(0,1){6.7}}
\put(13,8.3){\line(0,1){7.4}}
\put(13,16.3){\line(0,1){0.7}}
\put(1,5){\line(1,1){12}}
\put(2,4){\line(1,1){11}}
\put(4,2){\line(1,1){9}}
\put(5,1){\line(1,1){8}}
\put(10,4){\line(1,1){3}}
\put(12,2){\line(1,1){1}}
\put(1,5){\line(1,-1){4}}
\put(2,6){\line(1,-1){4}}
\put(4,8){\line(1,-1){4}}
\put(5,9){\line(1,-1){8}}
\put(6,10){\line(1,-1){7}}
\put(8,12){\line(1,-1){5}}
\put(9,13){\line(1,-1){4}}
\put(10,14){\line(1,-1){3}}
\put(12,16){\line(1,-1){1}}
\d{6}{2}\d{12}{2}\d{5}{3}
\d{8}{4}\d{10}{4}\d{3}{5}
\d{7}{5}\d{9}{5}\d{11}{5}\d{8}{6}
\d{10}{6}\d{5}{7}\d{9}{7}\d{6}{8}
\d{12}{8}\d{7}{9}\d{11}{9}\d{12}{10}
\d{9}{11}\d{10}{12}\d{11}{13}
\d{13}{3}\d{13}{7}\d{13}{9}\d{13}{11}
\d{13}{15}
\dl{5}{1}{8}\dl{6}{2}{6}\dl{8}{4}{2}
\put(13,2){\u}\put(13,8){\uql}
\put(10,5){\uuql}\put(13,10){\v}
\put(13,16){\vrl}\put(10,13){\vvrl}
\put(5,2){\uv}\put(8,5){\uvq}
\put(2,5){\uvr}\put(5,8){\uvqr}
\put(9,6){\uvm}\put(12,9){\uvqm}
\put(6,9){\uvrm}\put(9,12){\uvqrm}
\put(5,0.9){\pos{t}{a}}\put(13,0.9){\pos{t}{a}}
\put(0.9,5){\pos{r}{b}}
\put(4.93,9.07){\pos{br}{c}}
\put(8.93,13.07){\pos{br}{d}}
\put(13,17.1){\pos{b}{e}}
\put(3.97,1.97){\spos{tr}{f_1}}
\put(1.97,3.97){\spos{tr}{f_{\!r\!-\!1}}}
\put(1.95,6.05){\spos{br}{g_1}}
\put(3.95,8.05){\spos{br}{g_{q\!-\!1}}}
\put(5.95,10.05){\spos{br}{h_1}}
\put(7.95,12.05){\spos{br}{h_{q\!-\!1}}}
\put(9.95,14.05){\spos{br}{i_1}}
\put(11.95,16.05){\spos{br}{i_{r\!-\!1}}}
\end{picture}\end{center}
\raisebox{-3ex}[0ex][0ex]{and}
\newpage
\begin{center}
\text{6}{18}{0}{6}{l}{{\cal R}^{qr}\!(u,v)_{a,b,c,d,e,f_1,
\ldots,f_{\!r\!-\!1},g_1,\ldots,g_{q\!-\!1},h_1,
\ldots,h_{q\!-\!1},i_1,\ldots,i_{r\!-\!1}}\quad=}
\begin{picture}(14,18)
\put(13,1){\line(0,1){6.7}}\put(13,8.3){\line(0,1){7.4}}
\put(13,16.3){\line(0,1){0.7}}
\put(1,13){\line(1,1){4}}\put(2,12){\line(1,1){4}}
\put(4,10){\line(1,1){4}}\put(5,9){\line(1,1){8}}
\put(6,8){\line(1,1){7}}\put(8,6){\line(1,1){5}}
\put(9,5){\line(1,1){4}}\put(10,4){\line(1,1){3}}
\put(12,2){\line(1,1){1}}
\put(1,13){\line(1,-1){12}}\put(2,14){\line(1,-1){11}}
\put(4,16){\line(1,-1){9}}\put(5,17){\line(1,-1){8}}
\put(10,14){\line(1,-1){3}}\put(12,16){\line(1,-1){1}}
\d{11}{5}\d{10}{6}\d{9}{7}\d{12}{8}\d{7}{9}
\d{11}{9}\d{6}{10}\d{12}{10}\d{5}{11}
\d{9}{11}\d{8}{12}\d{10}{12}
\d{3}{13}\d{7}{13}\d{9}{13}\d{11}{13}
\d{8}{14}\d{10}{14}\d{5}{15}\d{6}{16}\d{12}{16}
\dl{8}{14}{2}\dl{6}{16}{6}\dl{5}{17}{8}
\d{13}{3}\d{13}{7}\d{13}{9}\d{13}{11}\d{13}{15}
\put(13,2){\v}\put(13,8){\vrl}\put(10,5){\vvrl}
\put(13,10){\u}\put(13,16){\uql}\put(10,13){\uuql}
\put(5,10){\uv}\put(2,13){\uvq}\put(8,13){\uvr}
\put(5,16){\uvqr}
\put(9,6){\uvm}\put(6,9){\uvqm}\put(12,9){\uvrm}
\put(9,12){\uvqrm}
\put(13,0.9){\pos{t}{a}}\put(8.93,4.93){\pos{tr}{b}}
\put(4.91,8.91){\pos{tr}{c}}\put(0.9,13){\pos{r}{d}}
\put(5,17.1){\pos{b}{e}}\put(13,17.1){\pos{b}{e}}
\put(11.97,1.97){\spos{tr}{f_1}}
\put(9.97,3.97){\spos{tr}{f_{\!r\!-\!1}}}
\put(7.97,5.97){\spos{tr}{g_1}}
\put(5.97,7.97){\spos{tr}{g_{q\!-\!1}}}
\put(3.97,9.97){\spos{tr}{h_1}}
\put(1.97,11.97){\spos{tr}{h_{q\!-\!1}}}
\put(1.95,14.05){\spos{br}{i_1}}
\put(3.95,16.05){\spos{br}{i_{r\!-\!1}}}
\end{picture}\end{center}
We now claim that, for any model in which the original Yang Baxter
equation,~\eqref{YBE}, and right reflection equation,~\eqref{RRE}, are
satisfied, and
for arbitrary $\l$, we in fact have
\beqa\lefteqn{{\cal L}^{qr}\!(u,v)_{a,b,c,d,e,f_1,
\ldots,f_{\!r\!-\!1},g_1,\ldots,g_{q\!-\!1},h_1,
\ldots,h_{q\!-\!1},i_1,\ldots,i_{r\!-\!1}}\;=}
\eql{fusRRE}\\
&&\qquad\qquad\qquad\qquad\qquad\qquad\qquad\qquad\qquad
{\cal R}^{qr}\!(u,v)_{a,b,c,d,e,f_1,
\ldots,f_{\!r\!-\!1},g_1,\ldots,g_{q\!-\!1},h_1,
\ldots,h_{q\!-\!1},i_1,\ldots,i_{r\!-\!1}}\nonumber
\eeqa
This can be proved by induction, which consists of showing that
\[{\cal L}^{1,1}\!(u,v)={\cal R}^{1,1}\!(u,v)\]
that
\[{\cal L}^{1,1}\!(u,v)={\cal R}^{1,1}\!(u,v)\quad\mbox{and}
\quad{\cal L}^{q\!-\!1,1}\!(u,v)={\cal R}^{q\!-\!1,1}\!(u,v)
\quad\mbox{imply that}
\quad{\cal L}^{q,1}\!(u,v)={\cal R}^{q,1}\!(u,v)\]
and, finally, that
\[{\cal L}^{q,1}\!(u,v)={\cal R}^{q,1}\!(u,v)\quad\mbox{and}
\quad{\cal L}^{q,r\!-\!1}\!(u,v)={\cal R}^{q,r\!-\!1}\!(u,v)
\quad\mbox{imply that}
\quad{\cal L}^{q,r}\!(u,v)={\cal R}^{q,r}\!(u,v)\]
We know the first statement holds, since it is simply the original right
reflection equation~\eqref{RRE}.
We shall only explicitly demonstrate the second statement, since the third can
be demonstrated similarly.  We have, for $q\geq2$,
\newpage
\rnc{\equals}{\text{2}{12}{1}{6}{}{=}}
\setlength{\unitlength}{7.2mm}
\begin{flushleft}\text{4}{12}{0}{9}{l}{{\cal L}^{q,1}\!(u,v)
\quad=}\begin{picture}(11,12)
\put(11,0){\line(0,1){6.7}}
\put(11,7.3){\line(0,1){1.4}}
\put(11,9.3){\line(0,1){2.7}}
\put(0,1){\line(1,1){11}}
\put(1,0){\line(1,1){10}}
\put(7,4){\line(1,1){4}}
\put(8,3){\line(1,1){3}}
\put(10,1){\line(1,1){1}}
\put(1,0){\line(-1,1){1}}
\put(2,1){\line(-1,1){1}}
\put(4,3){\line(-1,1){1}}
\put(5,4){\line(-1,1){1}}
\put(11,0){\line(-1,1){6}}
\put(11,2){\line(-1,1){5}}
\put(11,6){\line(-1,1){3}}
\put(11,8){\line(-1,1){2}}
\put(11,10){\line(-1,1){1}}
\d{2}{1}\d{10}{1}\d{11}{2}\d{4}{3}
\d{8}{3}\d{5}{4}\d{7}{4}\d{9}{4}\d{6}{5}
\d{8}{5}\d{7}{6}\d{11}{6}\d{10}{7}\d{9}{8}
\d{11}{8}\d{10}{9}\d{11}{10}
\dl{1}{0}{10}\dl{2}{1}{8}\dl{4}{3}{4}
\dl{5}{4}{2}
\put(11,1){\u}\put(11,7){\uqll}
\put(11,9){\uql}\put(11,11){\v}
\put(8,4){\uuqll}\put(7,5){\uuql}
\put(10,8){\uuqqlll}
\put(1,1){\uv}\put(4,4){\uvll}
\put(5,5){\uvq}\put(6,6){\uvm}
\put(9,9){\uvllm}\put(10,10){\uvqm}
\end{picture}\end{flushleft}
\begin{flushleft}\equals\begin{picture}(13,12)
\put(9,0){\line(0,1){6.7}}\put(9,7.3){\line(0,1){0.7}}
\put(11,8){\line(0,1){0.7}}
\put(11,9.3){\line(0,1){2.7}}
\put(3,6){\line(1,1){4}}\put(0,1){\line(1,1){11}}
\put(1,0){\line(1,1){10}}\put(6,3){\line(1,1){3}}
\put(8,1){\line(1,1){1}}\put(1,0){\line(-1,1){1}}
\put(2,1){\line(-1,1){1}}
\put(4,3){\line(-1,1){1}}\put(9,0){\line(-1,1){6}}
\put(9,2){\line(-1,1){5}}\put(9,6){\line(-1,1){3}}
\put(9,8){\line(-1,1){2}}\put(11,8){\line(-1,1){2}}
\put(11,10){\line(-1,1){1}}
\d{2}{1}\d{8}{1}\d{9}{2}\d{4}{3}\d{6}{3}\d{5}{4}
\d{7}{4}\d{6}{5}\d{5}{6}\d{9}{6}\d{8}{7}\d{7}{8}
\d{9}{8}\d{11}{8}\d{8}{9}\d{10}{9}\d{11}{10}
\dl{1}{0}{8}\dl{2}{1}{6}\dl{4}{3}{2}\dl{9}{8}{2}
\dl{7}{10}{2}
\put(9,1){\u}\put(9,7){\uqll}\put(11,9){\uql}
\put(11,11){\v}\put(6,4){\uuqll}\put(4,6){\uuql}
\put(7,9){\uuqqlll}
\put(1,1){\uv}\put(4,4){\uvll}\put(9,9){\uvq}
\put(5,5){\uvm}\put(8,8){\uvllm}
\put(10,10){\uvqm}
\end{picture}\raisebox{
6\unitlength}{\parbox{7\unitlength}{\raggedright
using the Yang-Baxter
equation $q\!-\!1$ times}}\end{flushleft}
\newpage
\begin{flushleft}\equals\begin{picture}(11,12)
\put(9,0){\line(0,1){6.7}}\put(9,7.3){\line(0,1){3.4}}
\put(9,11.3){\line(0,1){0.7}}
\put(6,11){\line(1,1){1}}\put(3,6){\line(1,1){6}}
\put(0,1){\line(1,1){9}}\put(1,0){\line(1,1){8}}
\put(6,3){\line(1,1){3}}\put(8,1){\line(1,1){1}}
\put(1,0){\line(-1,1){1}}
\put(2,1){\line(-1,1){1}}\put(4,3){\line(-1,1){1}}
\put(9,0){\line(-1,1){6}}\put(9,2){\line(-1,1){5}}
\put(9,6){\line(-1,1){3}}\put(9,8){\line(-1,1){3}}
\put(9,10){\line(-1,1){2}}
\d{2}{1}\d{8}{1}\d{9}{2}\d{4}{3}\d{6}{3}\d{5}{4}
\d{7}{4}\d{6}{5}\d{5}{6}\d{9}{6}\d{8}{7}\d{7}{8}
\d{9}{8}\d{8}{9}\d{9}{10}\d{8}{11}
\dl{1}{0}{8}\dl{2}{1}{6}\dl{4}{3}{2}\dl{7}{12}{2}
\put(9,1){\u}\put(9,7){\uqll}\put(9,9){\v}
\put(9,11){\uql}\put(6,4){\uuqll}\put(4,6){\uuql}
\put(7,9){\uuqqlll}
\put(1,1){\uv}\put(4,4){\uvll}\put(7,11){\uvq}
\put(5,5){\uvm}\put(8,8){\uvllm}\put(8,10){\uvqm}
\end{picture}
\raisebox{6\unitlength}{\parbox{7\unitlength}{\raggedright
applying\\ \rule{0ex}{3ex}${\cal L}^{1,1}
\!(u,v)=$\\ \rule{0ex}{3ex}${\cal R}^{1,1}
\!(u,v)$}}\end{flushleft}
\begin{flushleft}\equals\begin{picture}(16,12)
\put(14,0){\line(0,1){8.7}}\put(14,9.3){\line(0,1){0.7}}
\put(6,10){\line(0,1){0.7}}\put(6,11.3){\line(0,1){0.7}}
\put(3,11){\line(1,1){1}}\put(0,6){\line(1,1){6}}
\put(1,5){\line(1,1){5}}\put(6,8){\line(1,1){1}}
\put(8,6){\line(1,1){1}}\put(9,5){\line(1,1){5}}
\put(10,4){\line(1,1){4}}
\put(12,2){\line(1,1){2}}\put(13,1){\line(1,1){1}}
\put(1,5){\line(-1,1){1}}\put(2,6){\line(-1,1){1}}
\put(4,8){\line(-1,1){1}}\put(14,0){\line(-1,1){11}}
\put(14,2){\line(-1,1){10}}
\put(14,4){\line(-1,1){3}}\put(14,8){\line(-1,1){1}}
\d{14}{2}\d{13}{3}\d{14}{4}\d{11}{5}\d{2}{6}\d{8}{6}
\d{10}{6}\d{12}{6}\d{9}{7}\d{11}{7}\d{4}{8}\d{6}{8}
\d{14}{8}\d{5}{9}\d{7}{9}\d{13}{9}\d{6}{10}\d{14}{10}
\d{5}{11}
\dl{1}{5}{8}\dl{2}{6}{6}\dl{9}{7}{2}\dl{4}{8}{2}
\dl{7}{9}{6}\dl{6}{10}{8}\dl{4}{12}{2}
\put(14,1){\v}\put(14,3){\u}\put(14,9){\uqll}
\put(6,11){\uql}\put(11,6){\uuqll}\put(1,6){\uuql}
\put(4,9){\uuqqlll}
\put(9,6){\uv}\put(6,9){\uvll}\put(4,11){\uvq}\put(13,2){\uvm}
\put(10,5){\uvllm}\put(5,10){\uvqm}
\end{picture}
\raisebox{6\unitlength}{\parbox{4\unitlength}{\raggedright
applying\\ \rule{0ex}{3ex}${\cal L}^{q\!-\!1,1}\!(u,v)
=$\\ \rule{0ex}{3ex}${\cal R}^{q\!-\!1,1}\!(u,v)$}}
\end{flushleft}
\newpage
\begin{flushleft}\equals\begin{picture}(13,12)
\put(11,0){\line(0,1){8.7}}
\put(11,9.3){\line(0,1){1.4}}
\put(11,11.3){\line(0,1){0.7}}
\put(0,11){\line(1,1){1}}
\put(1,10){\line(1,1){1}}
\put(2,9){\line(1,1){1}}
\put(4,7){\line(1,1){1}}
\put(5,6){\line(1,1){6}}
\put(6,5){\line(1,1){5}}
\put(7,4){\line(1,1){4}}
\put(9,2){\line(1,1){2}}
\put(10,1){\line(1,1){1}}
\put(11,0){\line(-1,1){11}}
\put(11,2){\line(-1,1){10}}
\put(11,4){\line(-1,1){4}}
\put(11,8){\line(-1,1){2}}
\put(11,10){\line(-1,1){1}}
\d{11}{2}\d{10}{3}\d{11}{4}
\d{8}{5}\d{7}{6}\d{9}{6}\d{6}{7}
\d{8}{7}\d{5}{8}\d{7}{8}\d{11}{8}
\d{10}{9}\d{3}{10}\d{9}{10}\d{11}{10}
\d{2}{11}\d{10}{11}
\dl{5}{8}{2}\dl{3}{10}{6}\dl{2}{11}{8}
\dl{1}{12}{10}
\put(11,1){\v}\put(11,3){\u}\put(11,9){\uqll}
\put(11,11){\uql}\put(8,6){\uuqll}\put(7,7){\uuql}
\put(10,10){\uuqqlll}
\put(5,7){\uv}\put(2,10){\uvll}\put(1,11){\uvq}
\put(10,2){\uvm}\put(7,5){\uvllm}\put(6,6){\uvqm}
\end{picture}\raisebox{
6\unitlength}{\parbox{7\unitlength}{\raggedright
using  the Yang-Baxter equation $q\!-\!1$ times}}
\end{flushleft}
\begin{flushleft}\makebox[2\unitlength]{$=$}$\displaystyle
\quad{\cal R}^{q,1}\!(u,v)$\end{flushleft}
Having established~\eqref{fusRRE}, it is straightforward to
verify that the $u$, $v$ dependent factors on each side of~\eqref{genRRE} are
equal,
which completes our proof that~\eqref{genRRE} is satisfied.

The proof that the fused right boundary crossing equation,~\eqref{genRBC}, is
also satisfied by the ABF weights corresponds closely to that for the fused
right reflection equation.
If $q=-1$ then each side of~\eqref{genRBC} is zero, and if $q=0$, then each
side of~\eqref{genRBC} is given by $\delta_{ab}\:\delta_{bc}$.
For $q\geq1$, we re-configure the fused face weight on the left side
of~\eqref{genRBC} according to~\eqref{reorient2},
set internal arbitrary spins equal to adjoining summed spins,
use~\eqref{adjpushthrough} to push explicit occurrences of the fused adjacency
condition to external
edges, cancel internal gauge factors $G$, and take the external arbitrary spins
to be the same on each side of the equation.
After these steps, we find that each side of~\eqref{genRRE} is proportional to
a sum of products of level~1 face and boundary weights, and that
the proof can be completed by using induction on $q$ to show that face and
boundary weight components of each side are proportional,
and then verifying that the overall proportionality factors on each side are
the same.
The induction argument here is valid for any model in which the original
inversion relation,~\eqref{IR}, and right boundary crossing
equation,~\eqref{RBC}, are satisfied.

\rnc{\theequation}{C.\arabic{equation}}\setcounter{equation}{0}
\section*{Appendix C: Proof of ABF Functional Equations}
In this appendix, we prove that the ABF fused double-row transfer matrices
satisfy the functional equations~\eqref{FE2}. The proof of~\eqref{FE1} is
similar.

We note that for $q=0$,~\eqref{FE2} is immediately satisfied  due
to~\eqref{Dp-1} and~\eqref{Dp0}.
We therefore proceed to the case $q\geq1$ and begin by considering an entry of
$\D^{pq}\!(\aL\aR|u)$ $\D^{p1}\!(\aL\aR|u\!+\!q\l)$. Using,~\eqref{ABFfusDRTM},
we find that
\newpage
\beqa\langle a_2,\ldots,a_\N\,|\,\D^{pq}\!(\aL\aR|u)\;
\D^{p1}\!(\aL\aR|u\!+\!q\l)\,|\,b_2,
\ldots,b_\N\rangle\;\;=\;\;
\frac{A^p_{\aL a_2}\ldots
A^p_{a_N\aR}}{\displaystyle
K^{pq}(u)}\;\;\sum_c\,A^q_{\aL c}\:
\times&&\noeqno&&\noeqno&&\noeqno&&
\eql{FEder1}\eeqa
\setlength{\unitlength}{7.4mm}
\begin{center}\raisebox{0ex}[
10.3\unitlength][-2ex]{\begin{picture}(5,12)
\put(0,0){\line(0,1){0.7}}\put(0,1.3){\line(0,1){5.4}}
\put(0,7.3){\line(0,1){1.4}}\put(0,9.3){\line(0,1){0.7}}
\put(4,10){\line(0,1){0.7}}\put(4,11.3){\line(0,1){0.7}}
\put(0,0){\line(1,1){5}}\put(0,2){\line(1,1){4}}
\put(0,6){\line(1,1){2}}\put(0,8){\line(1,1){1}}
\put(4,10){\line(1,1){1}}
\put(0,2){\line(1,-1){1}}\put(0,6){\line(1,-1){3}}
\put(0,8){\line(1,-1){4}}\put(0,10){\line(1,-1){5}}
\put(4,12){\line(1,-1){1}}
\dl{0}{0}{5}\dl{4}{6}{1}\dl{2}{8}{3}\dl{1}{9}{4}
\dl{0}{10}{5}\dl{4}{12}{1}
\d{2}{4}\d{3}{5}\d{4}{6}\d{1}{7}\d{2}{8}\d{1}{9}
\put(0,1){\um}\put(0,7){\uqllm}\put(0,9){\uqlm}
\put(4,11){\uqm}
\put(1,8){\luuqqlll}\put(3,4){\luuqll}
\put(4,5){\luuql}
\put(0.1,-0.08){\spos{t}{\aL}}
\put(-0.06,2){\spos{r}{\aL}}
\multiput(-0.06,6)(0,2){3}{\spos{r}{\aL}}
\put(4.1,12.15){\spos{b}{\aL}}
\put(1.03,0.97){\spos{tl}{c_1}}
\put(3.03,2.97){\spos{tl}{c_{\!q\!-\!2}}}
\put(4.03,3.97){\spos{tl}{c_{\!q\!-\!1}}}
\put(4.94,4.65){\spos{r}{c}}
\end{picture}\begin{picture}(12,12)
\multiput(0,0)(0,1){2}{\line(1,0){12}}
\multiput(0,3)(0,1){4}{\line(1,0){12}}
\multiput(0,8)(0,1){5}{\line(1,0){12}}
\multiput(0,0)(1.5,0){2}{\line(0,1){12}}
\multiput(2.5,0)(1.5,0){2}{\line(0,1){12}}
\multiput(8,0)(1.5,0){2}{\line(0,1){12}}
\multiput(10.5,0)(1.5,0){2}{\line(0,1){12}}
\d{0}{1}\dd{0}{3}{4}\dd{0}{8}{2}\d{0}{11}
\dd{1.5}{0}{2}\dd{1.5}{3}{4}\dd{1.5}{8}{4}
\dd{2.5}{0}{2}\dd{2.5}{3}{4}\dd{2.5}{8}{4}
\d{4}{1}\dd{4}{3}{4}\dd{4}{8}{4}
\d{8}{1}\dd{8}{3}{4}\dd{8}{8}{4}\dd{9.5}{0}{2}
\dd{9.5}{3}{4}\dd{9.5}{8}{4}\dd{10.5}{0}{2}
\dd{10.5}{3}{4}\dd{10.5}{8}{4}\d{12}{1}
\dd{12}{3}{4}\dd{12}{8}{2}\d{12}{11}
\multiput(0,0)(8,0){2}{\pl}\multiput(0,3)(8,0){2}{\qpl}
\multiput(0,4)(8,0){2}{\qp}
\multiput(2.5,0)(8,0){2}{\fu}\multiput(2.5,3)(8,0){2}{\qll}
\multiput(2.5,4)(8,0){2}{\ql}
\multiput(0,5)(8,0){2}{\umpl}\multiput(0,8)(8,0){2}{\umqplll}
\multiput(0,9)(8,0){2}{\umqpll}\multiput(0,10)(8,0){2}{\uqpl}
\multiput(0,11)(8,0){2}{\umqpl}
\multiput(2.5,5)(8,0){2}{\fum}\multiput(2.5,8)(8,0){2}{\umqll}
\multiput(2.5,9)(8,0){2}{\umql}\multiput(2.5,10)(8,0){2}{\q}
\multiput(2.5,11)(8,0){2}{\umq}
\put(0.1,-0.08){\spos{t}{\aL}}\put(4.1,-0.08){\spos{t}{a_2}}
\put(8.1,-0.08){\spos{t}{a_{\!N}}}\put(12.1,-0.08){\spos{t}{\aR}}
\put(0.1,12.18){\spos{b}{\aL}}\put(1.6,12.18){\spos{b}{f_{11}}}
\put(2.8,12.18){\spos{b}{f_{1,p\!-\!1}}}
\put(4.1,12.18){\spos{b}{b_2}}
\put(8.1,12.18){\spos{b}{b_{\!N}}}
\put(9.6,12.18){\spos{b}{f_{\!N1}}}
\put(10.8,12.18){\spos{b}{f_{\!N,p\!-\!1}}}
\put(12.1,12.18){\spos{b}{\aR}}
\end{picture}\begin{picture}(5,12)
\put(5,0){\line(0,1){6.7}}\put(5,7.3){\line(0,1){1.4}}
\put(5,9.3){\line(0,1){0.7}}\put(1,10){\line(0,1){0.7}}
\put(1,11.3){\line(0,1){0.7}}
\put(5,0){\line(-1,1){5}}\put(5,2){\line(-1,1){4}}
\put(5,6){\line(-1,1){2}}\put(5,8){\line(-1,1){1}}
\put(1,10){\line(-1,1){1}}
\put(5,2){\line(-1,-1){1}}\put(5,6){\line(-1,-1){3}}
\put(5,8){\line(-1,-1){4}}\put(5,10){\line(-1,-1){5}}
\put(1,12){\line(-1,-1){1}}
\dl{0}{0}{5}\dl{0}{1}{4}\dl{0}{3}{2}\dl{0}{4}{1}
\dl{0}{6}{1}\dl{0}{8}{3}\dl{0}{9}{4}\dl{0}{10}{5}
\dl{0}{12}{1}
\d{4}{1}\d{2}{3}\d{3}{4}\d{1}{4}\d{2}{5}\d{4}{7}
\d{1}{6}\d{3}{8}\d{4}{9}
\put(5,1){\u}\put(5,7){\uqll}\put(5,9){\uql}\put(1,11){\uq}
\put(1,5){\uuql}\put(2,4){\uuqll}\put(4,8){\uuqqlll}
\put(5.2,-0.08){\spos{t}{\aR}}\put(5.06,2){\spos{l}{\aR}}
\multiput(5.06,6)(0,2){3}{\spos{l}{\aR}}\put(1.2,12.18){\spos{b}{\aR}}
\end{picture}}\end{center}
\rule{0ex}{5ex}where we must have $(c_1,\ldots,c_{q\!-\!1})\in P^q_{\aL c}$,
for each $c$ in the sum, and
$(f_{11},\ldots,f_{1,p\!-\!1})\in P^p_{\aL b_2}$, \ldots,
$\rule{0ex}{2.5ex}(f_{\!\N1},\ldots,f_{\!\N,p\!-\!1})\in P^p_{b_{\!N}\aR}$.
We now use the identity
\beq\raisebox{-4\unitlength}[
3.5\unitlength][4\unitlength]{\begin{picture}(14,8)
\put(1,6){\line(1,1){1}}\put(2,5){\line(1,1){1}}
\put(3,4){\line(1,1){1}}\put(5,2){\line(1,1){1}}
\put(6,1){\line(1,1){6}}\put(8,1){\line(1,1){5}}
\put(1,6){\line(1,-1){5}}\put(2,7){\line(1,-1){6}}
\put(8,3){\line(1,-1){1}}\put(10,5){\line(1,-1){1}}
\put(11,6){\line(1,-1){1}}\put(12,7){\line(1,-1){1}}
\dl{2}{7}{10}\dl{3}{6}{8}\dl{4}{5}{6}\dl{6}{3}{2}\dl{6}{1}{2}
\d{3}{6}\d{11}{6}\d{4}{5}\d{10}{5}\d{6}{3}\d{8}{3}\d{7}{2}
\put(2,6){\luuqql}\put(3,5){\luuqqll}\put(6,2){\luuq}
\put(8,2){\uuq}\put(11,5){\uuqqll}\put(12,6){\uuqql}
\put(1.99,4.98){\spos{tr}{a_2}}
\put(2.99,3.98){\spos{tr}{a_3}}
\put(4.99,1.98){\spos{tr}{a_{\!q}}}
\put(9.03,1.97){\spos{tl}{c_{\!q}}}
\put(11.03,3.97){\spos{tl}{c_3}}\put(12.03,4.97){\spos{tl}{c_2}}
\put(0.98,6){\spos{r}{a_1}}\put(6,0.96){\spos{t}{b}}
\put(8,0.96){\spos{t}{b}}\put(13.03,6){\spos{l}{c_1}}
\put(12,7.05){\spos{b}{d}}\put(2,7.05){\spos{b}{d}}
\end{picture}\text{0}{14}{-2}{3}{bl}{=
\quad(-1)^q\:\T{q}{2q\!-
\!2}{2u\!-\!\m}\:\T{q}{2q}{2u\!-\!\m}}
\text{6}{14}{0.5}{2.5}{tl}{\times\;\delta_{a_1c_1}\ldots
\delta_{a_qc_q}\;A_{da_1}\ldots A_{a_qb}}}\eql{multinv}\eeq
which can be obtained by using the inversion relation $q$ times.
By inserting~\eqref{multinv} just below the top row of faces in~\eqref{FEder1},
and then using the Yang-Baxter equation $qpN$ times,
we find that
\beqa\rule[-2.5ex]{0ex}{2.5ex}\lefteqn{(-1)^q\:
\T{q}{2q\!-\!2}{2u\!-\!\m}\:\T{q}{2q}{2u\!-\!\m}\:
\langle a_2,\ldots,a_\N\,|\,\D^{pq}\!(\aL\aR|u)\;
\D^{p1}\!(\aL\aR|u\!+\!q\l)\,|\,b_2,\ldots,
b_\N\rangle\;=}\qquad\qquad\qquad\noeqno
&&\frac{1}{\displaystyle K^{pq}(u)}\;\sum_d\;\sum_c
\,A^q_{\aL c}\;\DD(\aL,c_1,\ldots,c_{q\!-\!2},c_{q\!-\!1},c,d,c)
\Big|_{(c_1,\ldots,c_{q\!-\!1})\in P^q_{\aL c}}\eql{FEder2}\eeqa
\newpage
\noindent where we define
\beq\rule[-2ex]{0ex}{2ex}\DD(\aL,c_1,\ldots,c_{q\!-\!2},
c_{q\!-\!1},c,d,e)\quad=\quad A^p_{\aL a_2}\ldots
A^p_{a_N\aR}\;\times\qquad\qquad\qquad\qquad\qquad\qquad\eql{DD}\eeq
\setlength{\unitlength}{7.4mm}
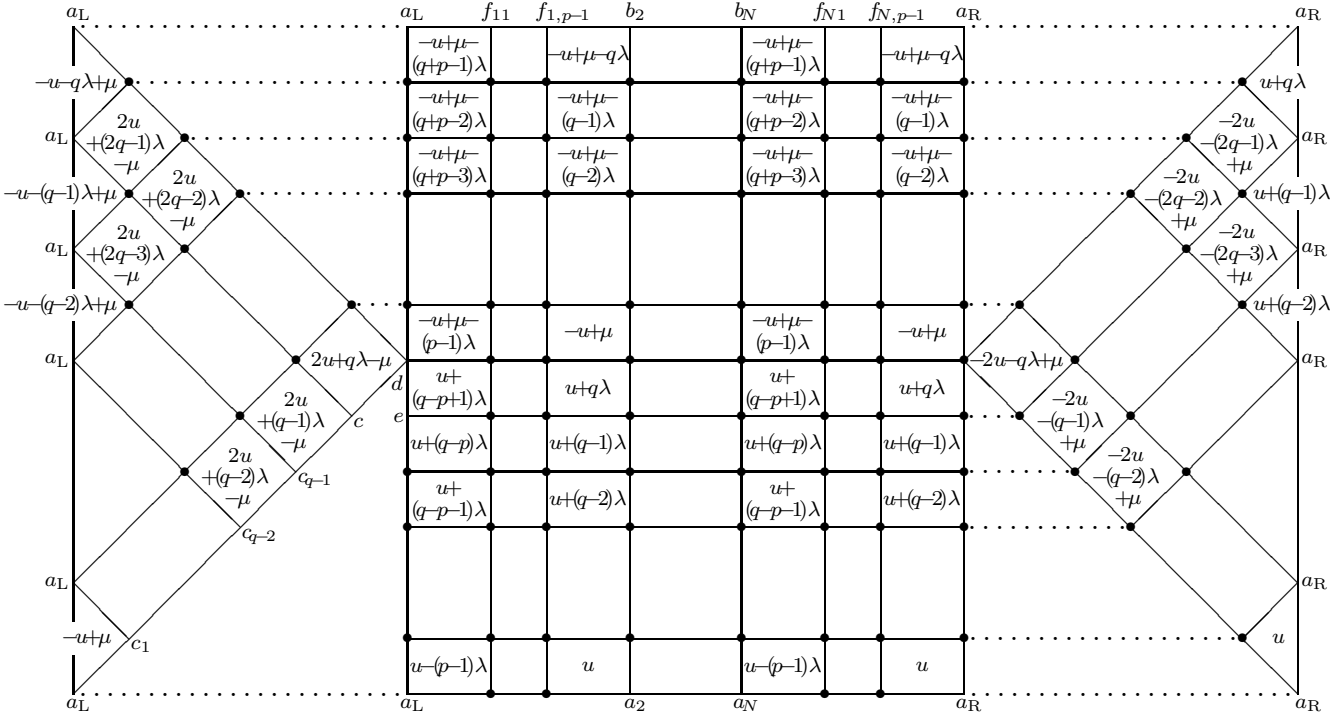
\begin{center}\begin{picture}(6,12)
\put(0,0){\line(0,1){0.7}}\put(0,1.3){\line(0,1){5.4}}
\put(0,7.3){\line(0,1){1.4}}\put(0,9.3){\line(0,1){1.4}}
\put(0,11.3){\line(0,1){0.7}}
\put(0,0){\line(1,1){6}}\put(0,2){\line(1,1){5}}
\put(0,6){\line(1,1){3}}\put(0,8){\line(1,1){2}}
\put(0,10){\line(1,1){1}}
\put(0,2){\line(1,-1){1}}\put(0,6){\line(1,-1){3}}
\put(0,8){\line(1,-1){4}}\put(0,10){\line(1,-1){5}}
\put(0,12){\line(1,-1){6}}
\dl{0}{0}{6}\dl{5}{7}{1}\dl{3}{9}{3}\dl{2}{10}{4}
\dl{1}{11}{5}\dl{0}{12}{6}
\d{2}{4}\d{1}{7}\d{3}{5}\d{1}{9}\d{2}{8}
\d{4}{6}\d{1}{11}\d{2}{10}\d{3}{9}\d{5}{7}
\put(0,1){\um}\put(0,7){\uqllm}
\put(0,9){\uqlm}\put(0,11){\uqm}
\put(1,8){\luuqqlll}\put(1,10){\luuqql}
\put(2,9){\luuqqll}\put(3,4){\luuqll}
\put(4,5){\luuql}\put(5,6){\luuq}
\put(0.1,-0.08){\spos{t}{\aL}}
\put(-0.06,2){\spos{r}{\aL}}
\multiput(-0.06,6)(0,2){3}{\spos{r}{\aL}}
\put(0.1,12.18){\spos{b}{\aL}}
\put(1.03,0.97){\spos{tl}{c_1}}
\put(3.03,2.97){\spos{tl}{c_{\!q\!-\!2}}}
\put(4.03,3.97){\spos{tl}{c_{\!q\!-\!1}}}
\put(5.03,4.97){\spos{tl}{c}}
\put(5.92,5.6){\spos{r}{d}}
\put(5.94,4.98){\spos{r}{e}}
\end{picture}\begin{picture}(10,12)
\multiput(0,0)(0,1){2}{\line(1,0){10}}
\multiput(0,3)(0,1){5}{\line(1,0){10}}
\multiput(0,9)(0,1){4}{\line(1,0){10}}
\multiput(0,0)(1.5,0){2}{\line(0,1){12}}
\multiput(2.5,0)(1.5,0){2}{\line(0,1){12}}
\multiput(6,0)(1.5,0){2}{\line(0,1){12}}
\multiput(8.5,0)(1.5,0){2}{\line(0,1){12}}
\d{0}{1}\dd{0}{3}{2}\d{0}{7}\dd{0}{9}{3}
\dd{1.5}{0}{2}\dd{1.5}{3}{5}\dd{1.5}{9}{3}
\dd{2.5}{0}{2}\dd{2.5}{3}{5}\dd{2.5}{9}{3}
\d{4}{1}\dd{4}{3}{5}\dd{4}{9}{3}
\d{6}{1}\dd{6}{3}{5}\dd{6}{9}{3}\dd{7.5}{0}{2}
\dd{7.5}{3}{5}\dd{7.5}{9}{3}\dd{8.5}{0}{2}
\dd{8.5}{3}{5}\dd{8.5}{9}{3}\d{10}{1}
\dd{10}{3}{5}\dd{10}{9}{3}
\multiput(0,0)(6,0){2}{\pl}
\multiput(0,3)(6,0){2}{\qpl}
\multiput(0,4)(6,0){2}{\qp}
\multiput(0,5)(6,0){2}{\uqpl}
\multiput(2.5,0)(6,0){2}{\fu}
\multiput(2.5,3)(6,0){2}{\qll}
\multiput(2.5,4)(6,0){2}{\ql}
\multiput(2.5,5)(6,0){2}{\q}
\multiput(0,6)(6,0){2}{\umpl}
\multiput(0,9)(6,0){2}{\umqplll}
\multiput(0,10)(6,0){2}{\umqpll}
\multiput(0,11)(6,0){2}{\umqpl}
\multiput(2.5,6)(6,0){2}{\fum}
\multiput(2.5,9)(6,0){2}{\umqll}
\multiput(2.5,10)(6,0){2}{\umql}
\multiput(2.5,11)(6,0){2}{\umq}
\put(0.1,-0.08){\spos{t}{\aL}}
\put(4.1,-0.08){\spos{t}{a_2}}
\put(6.1,-0.08){\spos{t}{a_{\!N}}}
\put(10.1,-0.08){\spos{t}{\aR}}
\put(0.1,12.18){\spos{b}{\aL}}
\put(1.6,12.18){\spos{b}{f_{11}}}
\put(2.8,12.18){\spos{b}{f_{1,p\!-\!1}}}
\put(4.1,12.18){\spos{b}{b_2}}
\put(6.1,12.18){\spos{b}{b_{\!N}}}
\put(7.6,12.18){\spos{b}{f_{\!N1}}}
\put(8.8,12.18){\spos{b}{f_{\!N,p\!-\!1}}}
\put(10.1,12.18){\spos{b}{\aR}}
\end{picture}\begin{picture}(6,12)
\put(6,0){\line(0,1){6.7}}
\put(6,7.3){\line(0,1){1.4}}
\put(6,9.3){\line(0,1){1.4}}
\put(6,11.3){\line(0,1){0.7}}
\put(6,0){\line(-1,1){6}}\put(6,2){\line(-1,1){5}}
\put(6,6){\line(-1,1){3}}\put(6,8){\line(-1,1){2}}
\put(6,10){\line(-1,1){1}}
\put(6,2){\line(-1,-1){1}}\put(6,6){\line(-1,-1){3}}
\put(6,8){\line(-1,-1){4}}\put(6,10){\line(-1,-1){5}}
\put(6,12){\line(-1,-1){6}}
\dl{0}{0}{6}\dl{0}{1}{5}\dl{0}{3}{3}\dl{0}{4}{2}
\dl{0}{5}{1}\dl{0}{7}{1}\dl{0}{9}{3}\dl{0}{10}{4}
\dl{0}{11}{5}\dl{0}{12}{6}
\d{5}{1}\d{3}{3}\d{4}{4}\d{2}{4}\d{3}{5}\d{5}{7}
\d{1}{5}\d{2}{6}\d{4}{8}\d{5}{9}\d{1}{7}\d{3}{9}
\d{4}{10}\d{5}{11}
\put(6,1){\u}\put(6,7){\uqll}\put(6,9){\uql}
\put(6,11){\uq}
\put(1,6){\uuq}\put(2,5){\uuql}\put(3,4){\uuqll}
\put(4,9){\uuqqll}\put(5,8){\uuqqlll}\put(5,10){\uuqql}
\put(6.2,-0.08){\spos{t}{\aR}}\put(6.06,2){\spos{l}{\aR}}
\multiput(6.06,6)(0,2){3}{\spos{l}{\aR}}\put(6.2,12.18){\spos{b}{\aR}}
\end{picture}\end{center}
\rule{0ex}{5ex}We note that the dependence of $\DD$ on $p$, $q$, $u$ and all of
the external spins except $\aL$ has been suppressed.
The next step in the proof of~\eqref{FE2}, will be to decompose the sum over
$c$, in~\eqref{FEder2}, into antisymmetric and symmetric sums.  However, in
order to do so,
we shall need several subsidiary results.  We begin with the following local
identities:
\newcounter{subeq}\rnc{\theequation}{C.\arabic{equation}\alph{subeq}}
\setlength{\unitlength}{9mm}
\beqa\rule[-1.8\unitlength]{0ex}{1.8\unitlength}
\sum_c\;\epsilon_b\:
\epsilon_e\;\raisebox{
-1.2\unitlength}{\bdoubface{b}{a}{c}{a}{d}{e}{u}{u\!+\!\l}}
\;\epsilon_a\:\epsilon_c&\;\;=\;\;&
\frac{\t(a\l)}{\t(b\l)}\;\frac{\t(\l\!-\!u)\,\t(\l\!
+\!u)}{\t(\l)^2}\;\delta_{bd}\:A_{ab}\:A_{be}
\stepcounter{subeq}\eql{coll1}\\
\sum_c\;\epsilon_a\:\epsilon_c\;\frac{\t(c\l)}{\t(a\l)}
\;\raisebox{-1.2\unitlength}{\tdoubface{a}{b}{e}{d}{a}{c}{u\!
+\!\l}{u}}\;\epsilon_d\:\epsilon_e\;\frac{\t(d\l)}{\t(e\l)}&=&
\frac{\t(d\l)}{\t(a\l)}\;
\frac{\t(\l\!-\!u)\,\t(\l\!+\!u)}{\t(\l)^2}
\;\delta_{bd}\:A_{ab}\:A_{be}
\stepcounter{subeq}\addtocounter{equation}{-1}\eql{coll2}\\
\sum_c\;\raisebox{
-1.4\unitlength}{\begin{picture}(2.6,3.2)
\put(1.3,0.55){\pos{t}{\epsilon_d\:\epsilon_e}}
\put(0,0.7){\sdoubface{b}{e}{d}{a}{c}{a}{u}{u\!+\!\l}}
\put(1.3,2.5){\pos{b}{\epsilon_a\:
\epsilon_c}}\end{picture}}&=&\frac{\t(a\l)}{\t(d\l)}
\;\frac{\t(\l\!-\!u)\,\t(\l\!+\!u)}{\t(\l)^2}
\;\delta_{bd}\:A_{ab}\:A_{be}
\stepcounter{subeq}\addtocounter{equation}{-1}\eql{coll3}
\eeqa
\beqa\sum_c\;\;\raisebox{-1.8\unitlength}{\begin{picture}(4,4.2)
\put(0,0){\leftdoubtri{a}{e}{a}{c}{u\!
+\!\l}{\;u}{\!\!-\!2u\!-\!\l\!+\!\m}}
\put(1.3,0.9){\pos{tl}{\epsilon_a\:\epsilon_c}}
\put(1.4,3){\pos{bl}{\displaystyle\epsilon_a\:
\epsilon_e\;\frac{\t(a\l)}{\t(e\l)}}}
\end{picture}}&=&
\left(\!\frac{\t(a\l)}{\t(\l)}\!\right)^{\!\!2}
\;\frac{\t(2u\!+\!3\l\!-\!\m)}{\t(\l)}\;
\gamma_{\mbox{\tiny L}}(a|u)\;A_{ae}
\stepcounter{subeq}\addtocounter{equation}{-1}\eql{coll4}\\
\sum_c\raisebox{
-1.8\unitlength}{\begin{picture}(4,4.3)\put(1.8,0){
\rightdoubtri{a}{c}{a}{e}{u\;}{u\!+
\!\l}{\!\!-\!2u\!-\!\l\!+\!\m}}
\put(2.7,0.9){\pos{tr}{\epsilon_a\:
\epsilon_e}}\put(2.6,3){\pos{br}{\displaystyle
\epsilon_a\:\epsilon_c\;\frac{\t(c\l)}{\t(a\l)}}}
\end{picture}}&=&
\left(\!\frac{\t(\l)}{\t(a\l)}\!\right)^{\!\!2}\;
\frac{\t(2u\!+\!3\l\!-\!\m)}{\t(\l)}\;
\gamma_{\mbox{\tiny R}}(a|u)\;A_{ae}
\stepcounter{subeq}\addtocounter{equation}{-1}\eql{coll5}
\eeqa
\rnc{\theequation}{C.\arabic{equation}}
where
\beqa\rule[-2ex]{0ex}{2ex}\lefteqn{\gamma_{\mbox{\tiny L/R}}(a|u)\quad=}\\
&&\frac{\t(u\!+\!\lm\!-\!\xi_{\mbox{\tiny L/R}}\!(a))
\;\t(u\!+\!\lm\!+\!\xi_{\mbox{\tiny L/R}}\!(a))\;
\t(u\!-\!a\l\!+\!\lm\!-\!\xi_{\mbox{\tiny L/R}}\!(a))
\;\t(u\!+\!a\l\!+\!\lm\!+\!\xi_{\mbox{\tiny L/R}}\!(a))}{\t(\l)^4}
\nonumber\eeqa
Identities~\eqref{coll1}--~\eqref{coll3} can each be proved as follows: if the
external spins do not satisfy the adjacency conditions, then both sides of the
equation are zero;
if $b=d=a\pm1$ and $e=a\pm2$, or else $a=1$ or $L$, then the left side is a
single term, which we immediately find is equal to the term on the right side;
if $b=a\pm1$, $d=a\mp1$ and $e=a$, then the left side is a sum of two terms,
which immediately cancel, as required by the delta function on the right;
finally, if $2\leq a\leq L\!-\!1$, $b=d=a\pm1$ and  $e=a$, then the left side
is a sum of two terms which can be reduced to the term on the right side
using a single application of~\eqref{thetaid}.

The proofs of~\eqref{coll4} and~\eqref{coll5} are similar: if $a$ and $e$ do
not satisfy the adjacency condition, then both sides of the equation are zero;
if $a=1$ or $L$ then the left side is a single term; and, if $2\leq a\leq
L\!-\!1$ and $e=a\pm1$, then the left side is a sum of two terms which can be
reduced to the
term on the right side using~\eqref{thetaid}.

We now use these identities repeatedly in~\eqref{DD}.  By starting at $c$ and
proceeding in a clockwise loop, using~\eqref{coll1} $q\!-\!1$
times,~\eqref{coll4} once,~\eqref{coll2} $pN$
times,~\eqref{coll5} once,~\eqref{coll3} $q\!-\!1$ times, and~\eqref{coll1}
$pN$ times, we find that, for $q\geq2$,
\newpage
\rnc{\theequation}{C.\arabic{equation}\alph{subeq}}\setcounter{subeq}{0}
\stepcounter{subeq}\beq\epsilon_e\;\sum_c\;\epsilon_c\;\DD(\aL,c_1,
\ldots,c_{q\!-\!2},d,c,d,e)\quad=\quad A_{de}\;M^{pq}(\aL\aR|u)\;
\times\qquad\qquad\qquad\qquad\eql{Dcoll1}\eeq
\setlength{\unitlength}{7.4mm}
\begin{center}\begin{picture}(4,8)
\put(0,0){\line(0,1){0.7}}
\put(0,1.3){\line(0,1){5.4}}
\put(0,7.3){\line(0,1){0.7}}
\put(0,0){\line(1,1){4}}
\put(0,2){\line(1,1){3}}
\put(0,6){\line(1,1){1}}
\put(0,2){\line(1,-1){1}}
\put(0,6){\line(1,-1){3}}
\put(0,8){\line(1,-1){4}}
\dl{0}{0}{4}\dl{3}{5}{1}
\dl{1}{7}{3}\dl{0}{8}{4}
\d{2}{4}\d{1}{7}\d{3}{5}
\put(0,1){\um}\put(0,7){\uqllm}
\put(3,4){\luuqll}
\put(0.1,-0.08){\spos{t}{\aL}}
\multiput(-0.06,2)(0,4){2}{\spos{r}{\aL}}
\put(0.1,8.18){\spos{b}{\aL}}
\put(1.03,0.97){\spos{tl}{c_1}}
\put(3.03,2.97){\spos{tl}{c_{\!q\!-\!2}}}
\put(3.92,3.6){\spos{r}{d}}
\end{picture}\begin{picture}(14,8)
\multiput(0,0)(0,1){2}{\line(1,0){14}}
\multiput(0,3)(0,1){3}{\line(1,0){14}}
\multiput(0,7)(0,1){2}{\line(1,0){14}}
\multiput(0,0)(1.5,0){2}{\line(0,1){8}}
\multiput(2.5,0)(1.5,0){2}{\line(0,1){8}}
\multiput(10,0)(1.5,0){2}{\line(0,1){8}}
\multiput(12.5,0)(1.5,0){2}{\line(0,1){8}}
\d{0}{1}\d{0}{3}\d{0}{5}\d{0}{7}\dd{1.5}{0}{2}
\dd{1.5}{3}{3}\d{1.5}{7}\dd{2.5}{0}{2}
\dd{2.5}{3}{3}\d{2.5}{7}\d{4}{1}
\dd{4}{3}{3}\d{4}{7}
\d{10}{1}\dd{10}{3}{3}\d{10}{7}\dd{11.5}{0}{2}
\dd{11.5}{3}{3}\d{11.5}{7}\dd{12.5}{0}{2}
\dd{12.5}{3}{3}\d{12.5}{7}\d{14}{1}
\dd{14}{3}{3}\d{14}{7}
\multiput(0,0)(10,0){2}{\pl}\multiput(0,3)(10,0){2}{\qpl}
\multiput(2.5,0)(10,0){2}{\fu}\multiput(2.5,3)(10,0){2}{\qll}
\multiput(0,4)(10,0){2}{\umpl}\multiput(0,7)(10,0){2}{\umqplll}
\multiput(2.5,4)(10,0){2}{\fum}\multiput(2.5,7)(10,0){2}{\umqll}
\put(0.1,-0.08){\spos{t}{\aL}}
\put(4.1,-0.08){\spos{t}{a_2}}
\put(10.1,-0.08){\spos{t}{a_{\!N}}}\put(14.1,-0.08){\spos{t}{\aR}}
\put(0.1,8.18){\spos{b}{\aL}}\put(1.6,8.18){\spos{b}{f_{11}}}
\put(2.8,8.18){\spos{b}{f_{1,p\!-\!1}}}\put(4.1,8.18){\spos{b}{b_2}}
\put(10.1,8.18){\spos{b}{b_{\!N}}}
\put(11.6,8.18){\spos{b}{f_{\!N1}}}
\put(12.8,8.18){\spos{b}{f_{\!N,p\!-\!1}}}\put(14.1,8.18){\spos{b}{\aR}}
\end{picture}\begin{picture}(4,8)
\put(4,0){\line(0,1){6.7}}\put(4,7.3){\line(0,1){0.7}}
\put(4,0){\line(-1,1){4}}
\put(4,2){\line(-1,1){3}}
\put(4,6){\line(-1,1){1}}
\put(4,2){\line(-1,-1){1}}
\put(4,6){\line(-1,-1){3}}\put(4,8){\line(-1,-1){4}}
\dl{0}{0}{4}\dl{0}{1}{3}\dl{0}{3}{1}
\dl{0}{5}{1}\dl{0}{7}{3}\dl{0}{8}{4}
\d{3}{1}\d{1}{3}\d{1}{5}\d{2}{4}\d{3}{7}
\put(4,1){\u}\put(4,7){\uqll}\put(1,4){\uuqll}
\put(4.2,-0.08){\spos{t}{\aR}}
\multiput(4.06,2)(0,4){2}{\spos{l}{\aR}}
\put(4.2,8.18){\spos{b}{\aR}}
\end{picture}\end{center}
while, for $q=1$,
\stepcounter{subeq}\addtocounter{equation}{-1}\beq
\epsilon_e\;\sum_c\;\epsilon_c\;\DD(\aL,c,\aL,e)=
A_{\aL e}\;M^{p1}(\aL\aR|u)\;
\langle
a_2,\ldots,a_\N\,|\,\vec{I}^p\!(\aL\aR)\,|\,b_2,\ldots,b_\N\rangle
\eql{Dcoll2}\eeq
\rnc{\theequation}{C.\arabic{equation}}
where
\beqa\rule[-2ex]{0ex}{2ex}
\lefteqn{M^{pq}(\aL\aR|u)\quad=}\eql{M}\\
&&\rule[-2ex]{0ex}{2ex}\T{q\!-\!1}{\!-\!1}{2u\!+
\!(2q\!-\!3)\l\!-\!\m}\;\T{q\!-\!1}{1}{2u\!+
\!(2q\!-\!3)\l\!-\!\m}\;
\T{1}{0}{\!-\!2u\!-\!(2q\!-\!3)\l\!+\!\m}\;
\gamma_{\mbox{\tiny L}}(\aL|\!-\!u\!-q\l\!+
\!\m)\noeqno
&&\times\;\rule[-2ex]{0ex}{2ex}\Big(
\T{p}{\!-\!1}{\!-\!u\!-\!q\l\!+\!\m}\;
\T{p}{1}{\!-\!u\!-\!q\l\!+\!\m}\Big)^{\!N}
\noeqno
&&\times\;\rule[-2ex]{0ex}{2ex}\T{1}{0}{2u\!
+\!(2q\!+\!1)\l\!-\!\m}\;
\gamma_{\mbox{\tiny R}}(\aR|u\!+(q\!-\!1)\l)\;
\T{q\!-\!1}{\!-\!1}{\!-\!2u\!-\!q\l\!+\!\m}\;
\T{q\!-\!1}{1}{\!-\!2u\!-\!q\l\!+\!\m}\noeqno
&&\times\;\Big(\T{p}{\!-\!1}{u\!+\!(q\!-\!1)\l}
\;\T{p}{1}{u\!+\!(q\!-\!1)\l}\Big)^{\!N}\nonumber
\eeqa
\rnc{\theequation}{C.\arabic{equation}\alph{subeq}}
\setcounter{subeq}{0}
We now assert the following properties of $\DD$:
\stepcounter{subeq}\beq\begin{array}{lcr}
\rule[-1.5ex]{0ex}{1.5ex}
A^q_{\aL e}\;\DD(\aL,c_1,\ldots,c_{q\!-\!2},c_{q\!-\!1},c,d,e)
&=&A^q_{\aL e}A^q_{\aL c}\;\DD(\aL,c_1,\ldots,c_{q\!-
\!2},c_{q\!-\!1},c,d,e),\\
&&\rule[-1.5ex]{0ex}{1.5ex}\mbox{for }(c_1,\ldots,c_{q\!-
\!1})\in P^q_{\aL c}\end{array}\eql{prop1a}\eeq
\stepcounter{subeq}\addtocounter{equation}{-1}\beq
\rule[-2ex]{0ex}{2ex}A^q_{\aL e}\;\DD(\aL,c_1,\ldots,c_{q\!-
\!2},c_{q\!-\!1},c,d,e)\mbox{ is independent of }(c_1,\ldots
,c_{q\!-\!1})\in P^q_{\aL c}\eql{prop1b}\eeq
\stepcounter{subeq}\addtocounter{equation}{-1}\beq
\epsilon_e\;\sum_c\;\epsilon_c\;\DD(\aL,c'_1,
\ldots,c'_{q\!-\!2},d,c,d,e)
\mbox{ is independent of }e\mbox{, for }
A_{de}=1\eql{prop2a}\eeq
\stepcounter{subeq}\addtocounter{equation}{-1}\beq A^{q
\!-\!1}_{\aL d}\:\;\epsilon_e\;\sum_c\;
\epsilon_c\;\DD(\aL,c'_1,\ldots,
c'_{q\!-\!2},d,c,d,e)
\mbox{ is independent of }(c'_1,\ldots,
c'_{q\!-\!2})\in P^{q\!-\!1}_{\aL d}
\eql{prop2b}\eeq
\stepcounter{subeq}\addtocounter{equation}{-1}
\beq A^{q\!+\!1}_{\aL d}\;\sum_e\;\DD(\aL,c''_1,
\ldots,c''_{q\!-\!2},c''_{q\!-\!1},c''_q,d,e)
\mbox{ is independent of }(c''_1,\ldots,c''_q)
\in P^{q\!+\!1}_{\aL d}\eql{prop3}\eeq
\rnc{\theequation}{C.\arabic{equation}}
Properties~\eqref{prop1a} and~\eqref{prop1b} follow by considering
$\rule[-1.5ex]{0ex}{1.5ex}\displaystyle A^q_{\aL
e}\;\DD(\aL,c_1,\ldots,c_{q\!-\!2},c_{q\!-\!1},c,d,e)$, with
$(c_1,\ldots,c_{q\!-\!1})\in P^q_{\aL c}$, as a linear combination of terms of
the form
\begin{eqnarray*}\lefteqn{\rule[-3.5ex]{0ex}{3.5ex}
\Wf{pq}{\aL}{a_2}{g_2}{e}{u}\:\ldots\:
\Wf{pq}{a_\N}{\aR}{g_{\N\!+\!1}}{g_\N}{u}
\:\BRf{q}{\aR}{g_{\N\!+\!1}}{\aR}{u}}\\
&\times&\rule[-3.5ex]{0ex}{3.5ex}
\Wf{1q}{h_{\N\!+\!1}}{g_{\N\!+
\!1}}{\aR}{i_{\N\!+\!1}}{\!\!-
\!2u\!-\!(2q\!-\!1)\l\!+\!\m}\\
&\times&\rule[-3.5ex]{0ex}{3.5ex}
\Wf{pq}{h_\N}{h_{\N\!+\!1}}{i_{\N\!+
\!1}}{i_\N}{\!\!-\!u\!-\!(q\!-\!1)\l\!+
\!\m}\:\ldots\:\Wf{pq}{d}{h_2}{i_2}{i_1}{\!\!-
\!u\!-\!(q\!-\!1)\l\!+\!\m}\\
&\times&\Wf{q1}{\aL}{c}{d}{i_1}{2u\!+\!(2q\!-
\!1)\l\!-\!\m}\:\BLf{q}{\aL}{c}{\aL}{\!-\!u\!-
\!(q\!-\!1)\l\!+\!\m}\end{eqnarray*}
Property~\eqref{prop2a} follows immediately from~\eqref{Dcoll1}
and~\eqref{Dcoll2}, while property~\eqref{prop2b} follows from~\eqref{Dcoll1}
by considering $\rule{0ex}{3ex}\displaystyle
A^{q\!-\!1}_{\aL d}\:\;\epsilon_e\;\sum_c\;
\epsilon_c\;\DD(\aL,c'_1,\ldots,c'_{q\!-\!2},d,c,d,e)$,
with $A_{de}=1$ and $(c'_1,\ldots,c'_{q\!-\!2})\in P^{q\!-\!1}_{\aL d}$, as
proportional to a sum of terms of the form
\begin{eqnarray*}\lefteqn{\rule[-3.5ex]{0ex}{3.5ex}
\Wf{p,q\!-\!1}{\aL}{a_2}{g_2}{g_1}{u}\:\ldots\:
\Wf{p,q\!-\!1}{a_\N}{\aR}{g_{\N\!+\!1}}{g_\N}{u}\,
\BRf{q\!-\!1}{\aR}{g_{\N\!+\!1}}{\aR}{u}}\\
&\times&\rule[-3.5ex]{0ex}{3.5ex}
\Wf{p,q\!-\!1}{g_\N}{g_{\N\!+\!1}}{\aR}{b_\N}{\!\!-\!u\!-\!(q\!-\!2)\l\!+\!\m}
\:\ldots\:\Wf{p,q\!-\!1}{g_1}{g_2}{b_2}{\aL}{\!\!-\!u\!-\!(q\!-\!2)\l\!+\!\m}\\
&\times&\BLf{q\!-\!1}{\aL}{g_1}{\aL}{\!-\!u\!-\!(q\!-\!2)
\l\!+\!\m}\end{eqnarray*}
Finally, property~\eqref{prop3} follows by considering $\displaystyle
A^{q\!+\!1}_{\aL d}
\;\sum_e\;\DD(\aL,c''_1,\ldots,c''_{q\!-\!2},c''_{q\!-\!1},c''_q,d,e)$, with
$(c''_1,\ldots,c''_q)\in P^{q\!+\!1}_{\aL d}$, as a sum of terms of the form
\begin{eqnarray*}\lefteqn{\rule[-3.5ex]{0ex}{3.5ex}
\Wf{p,q\!+\!1}{\aL}{a_2}{g_2}{g_1}{u}\:\ldots\:
\Wf{p,q\!+\!1}{a_\N}{\aR}{g_{\N\!+\!1}}{g_\N}{u}\,
\BRf{q\!+\!1}{\aR}{g_{\N\!+\!1}}{\aR}{u}}\\
&\times&\rule[-3.5ex]{0ex}{3.5ex}
\Wf{p,q\!+\!1}{g_\N}{g_{\N\!+\!1}}{\aR}{b_\N}{\!\!
-\!u\!-\!q\l\!+\!\m}
\:\ldots\:\Wf{p,q\!+\!1}{g_1}{g_2}{b_2}{\aL}{\!\!
-\!u\!-\!q\l\!+\!\m}\\
&\times&\BLf{q\!+\!1}{\aL}{g_1}{\aL}{\!-
\!u\!-\!q\l\!+\!\m}
\end{eqnarray*}
We now return to the sum over $c$ in~\eqref{FEder2}, which we claim can be
decomposed into antisymmetric and symmetric sums,
\beqa\rule[-3.5ex]{0ex}{3.5ex}\lefteqn{\sum_c\,A^q_{\aL c}\;
\DD(\aL,c_1,\ldots,c_{q\!-\!2},c_{q\!-\!1},c,d,c)
\quad=}\eql{FEder3}\\
&&A^{q\!-\!1}_{\aL d}\:\;\epsilon_e\;\sum_c\;
\epsilon_c\;\DD(\aL,c'_1,\ldots,c'_{q\!-\!2},d,c,d,e)
\quad+\quad A^{q\!+\!1}_{\aL d}\;\sum_e\;
\DD(\aL,c''_1,\ldots,c''_{q\!-\!2},c''_{q\!-\!1},c''_q,d,e)
\nonumber\eeqa
In this decomposition we assume the following: that $(c_1,
\ldots,c_{q\!-\!1})\in P^q_{\aL c}$ for each $c$ in the
 sum of the left side;
that $e$ satisfies $\rule{0ex}{2.5ex}A_{de}=1$ and that
$(c'_1,\ldots,c'_{q\!-\!2})\in P^{q\!-\!1}_{\aL d}$ in the antisymmetric sum;
and that
$(c''_1,\ldots,c''_q)\in P^{q\!+\!1}_{\aL d}$ in the symmetric sum.
Therefore, due to\rule{0ex}{2.5ex}~\eqref{prop1b}--\eqref{prop3}, all of these
spins are arbitrary\rule{0ex}{2.5ex}.

We now proceed to prove~\eqref{FEder3}. We begin by constructing the following
table of values of the adjacency matrix entries
which appear in~\eqref{FEder3} (as well as in the $su(2)$ fusion
rule~\eqref{fusrule1}):
\[\begin{array}{|@{}*{16}{c@{}|@{}}c@{}|}\cline{2-17}
\multicolumn{1}{@{}c@{}|@{}}{}&
\multicolumn{3}{@{}l@{}|@{}}{
\raisebox{-1ex}[0ex]{$\,\sm{d\!-
\!a\not\in}$}}&\multicolumn{2}{@{}c@{}|@{}}{}&
\multicolumn{2}{@{}c@{}|@{}}{}&
\multicolumn{9}{@{}c@{}|}{\rule[-2ex]{0ex}{5ex}
\sm{d\!-\!a\in\{\!-q\!+\!1,\!-\!q\!+\!3,
\ldots,q\!-\!3,q\!-\!1\}}}\\\cline{9-17}
\multicolumn{1}{@{}c@{}|@{}}{}&
\multicolumn{3}{@{}l@{}|@{}}{
\raisebox{3ex}[0ex]{$\:\sm{\{\!-
\!q\!-\!1,-\!q\!+\!1,-\!q\!+\!3,}$}}&
\multicolumn{2}{@{}c@{}|@{}}{\raisebox{
3ex}[0ex]{$\sm{d\!-\!a=-\!q\!-\!1}$}}&
\multicolumn{2}{@{}c@{}|@{}}{
\raisebox{3ex}[0ex]{$\sm{d\!-\!a=q\!+\!1}$}}&
\multicolumn{2}{@{}c@{}|@{}}{\sm{d\!+\!a\leq q\!-\!1}}&
\multicolumn{2}{@{}c@{}|@{}}{\sm{d\!+\!a=q\!+\!1}}&
\multicolumn{1}{@{}l@{}|@{}}{
\raisebox{1ex}[4.2ex]{$\sm{\,q\!
+\!3\leq d\!+\!a}$}}&
\multicolumn{2}{@{}c@{}|@{}}{\sm{d\!+\!a=2L\!-\!q\!+\!1}}&
\multicolumn{2}{@{}c@{}|}{\sm{d\!+\!a\geq 2L\!-\!q\!+\!3}}\\
\multicolumn{1}{@{}c@{}|@{}}{}&
\multicolumn{3}{@{}r@{}|@{}}{
\raisebox{3.7ex}[0ex]{$\sm{\ldots,q\!-
\!3,q\!-\!1,q\!+\!1\}}\,$}}&
\multicolumn{2}{@{}c@{}|@{}}{}&
\multicolumn{2}{@{}c@{}|@{}}{}&
\multicolumn{2}{@{}c@{}|@{}}{}&
\multicolumn{2}{@{}c@{}|@{}}{}&
\multicolumn{1}{@{}r@{}|@{}}{
\raisebox{1.6ex}[0ex]{$\sm{\leq2L\!-
\!q\!-\!1\,}$}}&\multicolumn{2}{@{}c@{}|@{}}{}&
\multicolumn{2}{@{}c@{}|}{}\\\cline{2-17}
\multicolumn{1}{@{}c@{}|@{}}{}&
\raisebox{0ex}[3ex][2ex]{$\sm{\,d=
1}$}&\sm{\,2\!\leq d\leq L\!-\!1}&\sm{\,d=L\,}&
\sm{\,d=1}&\sm{\,2\!\leq d\leq L\!-\!1}&
\sm{\,2\!\leq d\leq L\!-\!1}&\sm{\,d=L\,}&\sm{\,d=1}&
\sm{\,2\!\leq d\leq L\!-\!1}&\sm{\,d=1}&\sm{\,2\!\leq
d\leq L\!-\!1}&\sm{\,2\!\leq d\leq L\!-\!1}&
\sm{\,2\!\leq d\leq L\!-\!1}&\sm{\,d=L\,}&\sm{\,2\!\leq
d\leq L\!-\!1}&\sm{\,d=L\,}\\\hline
\rule[-2ex]{0ex}{5.5ex}\,A^q_{a,d\!-
\!1}\,&-&0&0&-&0&1&1&-&0&-&0&1&1&1&0&0
\rule[-2ex]{0ex}{3ex}\\\hline
\rule[-2ex]{0ex}{5.5ex}\,A^q_{a,d\!+\!1}
\,&0&0&-&1&1&0&-&0&0&1&1&1&0&-&0&-
\rule[-2ex]{0ex}{3ex}\\\hline
\rule[-2ex]{0ex}{5.5ex}
A^{q\!-\!1}_{a,d}&0&0&0&0&0&0&0&0&0&1&1&1&1&1&0&0
\rule[-2ex]{0ex}{3ex}\\\hline
\rule[-2ex]{0ex}{5.5ex}A^{q\!+\!1}_{a,d}
&0&0&0&1&1&1&1&0&0&0&0&1&0&0&0&0
\rule[-2ex]{0ex}{3ex}\\\hline
\end{array}\]
%\begin{center}Table 1: Values of each term in the $su(2)$ fusion
%rule\end{center}
The entries in this table can all be obtained directly from the fused adjacency
conditions,~\eqref{AC1} and ~\eqref{AC2}.
We now denote the left side of~\eqref{FEder3} by $\L$ and the right side
of~\eqref{FEder3} by $\R$, and consider cases
corresponding to those listed in the table.\\
\rule[-2ex]{0ex}{6ex}{\bf(I)} \ $d\!-\!\aL\not\in\{\!-
\!q\!-\!1,-\!q\!+\!1,\ldots,q\!-\!1,q\!+\!1\}$, \ $d\!+
\!\aL\leq q\!-\!1$ \ or \ $d\!+\!\aL\geq 2L\!-
\!q\!+\!3$\\
In these cases, $\L$ and $\R$ are each zero.\\
\rule[-2ex]{0ex}{6ex}{\bf(II)} \ $d\!-\!\aL=\pm(q\!+
\!1)$\\
In these cases, $\L$ and $\R$ are each given by the single term
\[\L=\R=\DD(\aL,\aL\!\pm\!1,\ldots,\aL\!\pm\!q,\aL
\!\pm\!(q\!+\!1),\aL\!\pm\!q)\]
\rule[-2ex]{0ex}{2ex}{\bf(III)} \ $d\!-\!\aL
\in\{\!-\!q\!+\!1,\ldots,q\!-\!1\}$\\
In these cases, we can satisfy $(c_1,
\ldots,c_{q\!-\!1})\in P^q_{\aL c}$, for each $c$ in the sum in $\L$, by taking
$c_{q\!-\!1}=d$ and
$\rule{0ex}{2.5ex}(c_1,\ldots,c_{q\!-\!2})\in P^{q\!-\!1}_{\aL,d}$.
We use this choice in each of the following subcases:\\
\rule[-2ex]{0ex}{6ex}{\bf(i)} \ $d\!+\!\aL=q\!+\!1$\\
In this case, $\L$ is comprised of the single term
\[\L=\DD(\aL,c_1,\ldots,c_{q\!-
\!2},d,d\!+\!1,d,d\!+\!1)\]
Meanwhile, $\R$ is comprised of the antisymmetric sum only, for which we choose
$e=d\!+\!1$.
If $d=1$, we have a single term, which immediately matches $\L$.  For
$\rule{0ex}{2.5ex}d\geq2$, we have
\[\R=\DD(\aL,c'_1,\ldots,c'_{q\!-\!2},d,d
\!+\!1,d,d\!+\!1)-\DD(\aL,c'_1,\ldots,c'_{q
\!-\!2},d,d\!-\!1,d,d\!+\!1)\]
but, by taking $c=d\!-\!1$ and $e=d\!+\!1$ in~\eqref{prop1a}, we find that the
second of these terms vanishes and, therefore,
we again have a single term which matches $\rule{0ex}{2.5ex}\L$.\\
\rule[-2ex]{0ex}{6ex}{\bf(ii)} \ $d\!+\!\aL=2L\!-\!q\!+\!1$\\
This case is similar to the previous one, with $\L$ now comprised of the single
term
\[\L=\DD(\aL,c_1,\ldots,c_{q\!-\!2},d,d\!-\!1,d,d\!-\!1)\]
Again, $\rule{0ex}{2.5ex}\R$ is comprised of the antisymmetric sum only, for
which we now choose $e=d\!-\!1$.
If $d=L$ we immediately have a term which matches $\rule{0ex}{2.5ex}\L$, while
for $d\leq L\!-\!1$, we have
\[\R=\DD(\aL,c'_1,\ldots,c'_{q\!-\!2},d,d\!-
\!1,d,d\!-\!1)-\DD(\aL,c'_1,\ldots,c'_{q\!-\!2},d,d\!+\!1,d,d\!-\!1)\]
but, as before, we find that the second of these terms vanishes by taking
$c=d\!+\!1$ and $\rule{0ex}{2.5ex}e=d\!-\!1$ in~\eqref{prop1a}.\\
\rule[-2ex]{0ex}{6ex}{\bf(iii)} \ $q\!+
\!3\leq d\!+\!\aL\leq2L\!-\!q\!-\!1$\\
In this case, we have
\[\L=\DD(\aL,c_1,\ldots,c_{q\!-\!2},d,d\!-
\!1,d,d\!-\!1)+\DD(\aL,c_1,\ldots,c_{q
\!-\!2},d,d\!+\!1,d,d\!+\!1)\]
For $\R$, we choose, in the antisymmetric sum,
$e=d\pm1$, and, in the symmetric sum, $c''_q=d\mp1$, \
$c''_{q\!-\!1}=d$, \ and
$(c''_1,\ldots,c''_{q\!-\!2})\in P^{q\!-\!1}_{\aL d}$,
\rule{0ex}{2.5ex}giving
\begin{eqnarray*}\R&=&\DD(\aL,c'_1,\ldots,c'_{q
\!-\!2},d,d\!\pm\!1,d,d\!\pm\!1)-\DD(\aL,c'_1,
\ldots,c'_{q\!-\!2},d,d\!\mp\!1,d,d\!\pm\!1)\\
&&\mbox{}+\rule{0ex}{3ex}\DD(\aL,c''_1,\ldots,c''_{q
\!-\!2},d,d\!\mp\!1,d,d\!\pm\!1)+\DD(\aL,c''_1,\ldots,c''_{q\!-
\!2},d,d\!\mp\!1,d,d\!\mp\!1)\end{eqnarray*}
We see that the two middle terms of $\R$ cancel, while the two outer terms
match those of $\L$.
This completes our proof of~\eqref{FEder3}

We now substitute~\eqref{FEder3} and~\eqref{Dcoll1} or~\eqref{Dcoll2}
into~\eqref{FEder2}, and use~\eqref{DD}, \eqref{ABFfusDRTM} and~\eqref{K} to
give
\beqa\lefteqn{(-1)^q\:
\T{q}{2q\!-\!2}{2u\!-\!\m}\:\T{q}{2q}{2u\!-\!\m}\:
\langle a_2,\ldots,a_\N\,|\,\D^{pq}\!(\aL\aR|u)\;
\D^{p1}\!(\aL\aR|u\!+\!q\l)\,|\,b_2,\ldots,b_\N\rangle\;=}\noeqno
\eql{FEder4}\\
&&\frac{M^{pq}(\aL\aR|u)}{\rule[-3ex]{0ex}{6ex}\left(
\T{p}{q\!-\!2}{u}\;\T{p}{-\!q\!+\!1}{\!-\!u\!+\!\m}\right)^{\!N}
\T{q\!-\!1}{2q\!-\!3}{2u\!-\!\m}\;
\T{q\!-\!1}{\!-\!q\!+\!1}{\!-\!2u\!+\!\m}}
\;\;\langle a_2,\ldots,a_\N\,|\,\D^{p,q\!-\!1}
\!(\aL\aR|u)\,|\,b_2,\ldots,b_\N\rangle\noeqno
&&\mbox{}+\left(\T{p}{q\!-\!1}{u}\;\T{p}{-\!q}{\!-
\!u\!+\!\m}\right)^{\!N}\T{q}{2q\!-\!1}{2u\!-\!\m}\;
\T{q}{\!-\!q}{\!-\!2u\!+\!\m}\;\;
\langle a_2,\ldots,a_\N\,|\,\D^{p,q\!+\!1}\!(\aL\aR|u)
\,|\,b_2,\ldots,b_\N\rangle\nonumber\eeqa
By using~\eqref{M} and then cancelling the common factor
$(-1)^q\:\T{q\!-\!1}{2q\!-\!2}{2u\!-\!\m}\;
\T{q\!-\!1}{2q\!-\!1}{2u\!-\!\m}$ from each
side of~\eqref{FEder4}\rule{0ex}{2.5ex},
it is relatively straightforward to show that the
coefficients of each term in~\eqref{FEder4} match those in~\eqref{FE2}, which
completes our proof of~\eqref{FE2}.

\end{document}